\documentclass[prl,showpacs,twocolumn,floats,10pt,aps,citeautoscript,longbibliography,superscriptaddress]{revtex4-2}
\usepackage{dcolumn}
\usepackage{anyfontsize} 
\usepackage{microtype} 
\usepackage{graphicx} 
\usepackage{xspace} 

\usepackage[usenames,dvipsnames]{xcolor} 

\usepackage{xparse} 

\usepackage{algorithm}
\usepackage{algpseudocode}

\usepackage[normalem]{ulem} 
\definecolor{darkgreen}{rgb}{0,0.5,0}
\definecolor{purple}{rgb}{0.6,0,0.5}
\definecolor{orange}{rgb}{1,0.5,0}
\definecolor{darkred}{rgb}{.7,0,0}
\definecolor{darkblue}{rgb}{0,0,.6}
\definecolor{grey}{rgb}{.6,.6,.6}
\definecolor{dimgreen}{rgb}{0.2,0.7,0.2}

\newcommand{\EllipseWhiteLambda}{\protect\raisebox{0mm}{\protect\includegraphics[width=0.0333\linewidth]{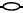}}}
\newcommand{\CircleWhiteC}{\protect\raisebox{-0.5mm}{\protect\includegraphics[width=0.037\linewidth]{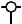}}}
\newcommand{\CircleGreenC}{\protect\raisebox{-0.5mm}{\protect\includegraphics[width=0.037\linewidth]{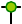}}}
\newcommand{\QuaterCircleBlackA}{\protect\raisebox{-0.5mm}{\protect\includegraphics[width=0.037\linewidth]{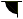}}}
\newcommand{\QuaterCircleBlackB}{\protect\raisebox{-0.5mm}{\protect\includegraphics[width=0.037\linewidth]{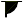}}}
\newcommand{\QuaterCircleBlackThreeS}{\protect\raisebox{-0.8mm}{\protect\includegraphics[width=0.037\linewidth]{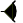}}}
\newcommand{\TriangleWhiteA}{\protect\raisebox{-0.5mm}{\protect\includegraphics[width=0.037\linewidth]{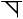}}}
\newcommand{\TriangleWhiteB}{\protect\raisebox{-0.5mm}{\protect\includegraphics[width=0.037\linewidth]{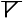}}}
\newcommand{\TriangleGreyA}{\protect\raisebox{-0.5mm}{\protect\includegraphics[width=0.037\linewidth]{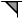}}}
\newcommand{\TriangleGreyB}{\protect\raisebox{-0.5mm}{\protect\includegraphics[width=0.037\linewidth]{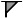}}}
\newcommand{\TriangleGreyAdagger}{\protect\raisebox{-0.25mm}{\protect\includegraphics[width=0.037\linewidth]{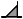}}}
\newcommand{\TriangleGreyBdagger}{\protect\raisebox{-0.25mm}{\protect\includegraphics[width=0.037\linewidth]{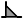}}}
\newcommand{\TriangleRedA}{\protect\raisebox{-0.5mm}{\protect\includegraphics[width=0.037\linewidth]{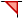}}}
\newcommand{\TriangleRedAdagger}{\protect\scalebox{-1}[1]{\protect\rotatebox[origin=c]{180}
{\raisebox{-0.25mm}{\protect\includegraphics[width=0.037\linewidth]{Eq/TriangleRedA}}}}}
\newcommand{\TriangleOrangeA}{\protect\raisebox{-0.5mm}{\protect\includegraphics[width=0.037\linewidth]{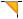}}}
\newcommand{\TriangleOrangeB}{\protect\raisebox{-0.5mm}{\protect\includegraphics[width=0.037\linewidth]{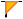}}}
\newcommand{\TriangleGreenA}{\protect\raisebox{-0.5mm}{\protect\includegraphics[width=0.037\linewidth]{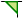}}}
\newcommand{\TriangleYellowA}{\protect\raisebox{-0.5mm}{\protect\includegraphics[width=0.037\linewidth]{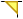}}}
\newcommand{\TrianglePinkA}{\protect\raisebox{-0.5mm}{\protect\includegraphics[width=0.037\linewidth]{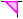}}}
\newcommand{\DiamondYellowS}{\protect\raisebox{-1mm}{\protect\includegraphics[width=0.037\linewidth]{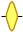}}}
\newcommand{\DiamondOrangeS}{\protect\raisebox{-0.75mm}{\protect\includegraphics[width=0.037\linewidth]{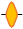}}}
\newcommand{\SlopingRectangleYellowU}{\protect\raisebox{-1mm}{\protect\includegraphics[width=0.037\linewidth]{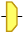}}}
\newcommand{\SlopingRectangleOrangeU}{\protect\raisebox{-1mm}{\protect\includegraphics[width=0.037\linewidth]{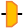}}}
\newcommand{\jvdomit}[1]{}

\usepackage{amsmath} 
\usepackage{amssymb}
\usepackage{mathrsfs} 
\usepackage{bbm}  
\renewcommand{\vec}[1]{{\boldsymbol{#1}}} 
\newcommand{\mr}[1]{\ensuremath{\mathrm{#1}}}

\allowdisplaybreaks 
\usepackage{enumitem} 

\usepackage{scalefnt} 

\usepackage{tikz}
\usepackage{algorithm}
\usepackage{algpseudocode}

\usepackage[colorlinks, citecolor={blue!50!black}, urlcolor={blue!50!black}, linkcolor={red!50!black}]{hyperref}

\usepackage{scalerel,stackengine}  

\newcommand{\Eq}[1]{Eq.~\eqref{#1}}
\newcommand{\Eqs}[1]{Eqs.~\eqref{#1}}

\newcommand{\Fig}[1]{Fig.~\ref{#1}}

\def\Cc{\mathcal{C}}

\newcommand{\ellminusone}{{\ell \hspace{-0.2mm}-\hspace{-0.2mm} 1}}
\newcommand{\ellplusone}{{\ell  \hspace{-0.2mm}+\hspace{-0.2mm} 1}}

\newcommand{\eLL}{{\mbox{\small$\mathscr{L}$}}}

\newcommand{\scripteLL}{{\scriptstyle \! \mathscr{L}}}
\newcommand{\sseLL}{{{\mbox{\tiny$\!\mathscr{L}$}}}}
\newcommand{\boldell}{{\boldsymbol{\ell}}}
\newcommand{\boldellp}{{\boldsymbol{\ell}}'}

\def\Sc{\mathcal{S}}

\def\Ab{{\overline{A}}} 
\def\Ah{{\widehat{A}}}

\def\Bb{{\overline{B}}}

\def\Vt{{\widetilde{V}}} 
 
\def\Vh{{\widehat{V}}}

\def\St{{\widetilde{S}}} 
\def\Sb{{\overline{S}}} 
\def\Sh{{\widehat{S}}}

\def\dast{d^\ast}
\def\wast{w^\ast}
\def\Dast{D^\ast}
\def\Dt{\widetilde{D}} 
\def\Dtast{\widetilde{D}{}^\ast} 
\def\Db{\overline{D}} 
\def\Dbast{\overline{D}{}^\ast} 
\def\Dh{\widehat{D}} 
\def\Dhast{\widehat{D}{}^\ast} 
 
\def\Dpast{D^{\prime\ast}}  
 
\def\Dfast{D_\fin^\ast} 
 
\def\Diast{D_\init^\ast} 
 
\def\Dmax{D_{\mathrm{max}}}
\def\Dmaxast{D_\mathrm{max}^\ast}

\def\Mh{\widehat{M}} 
\def\Mb{\overline{M}}

\def\Uh{\widehat{U}}
\def\Ut{\widetilde{U}}
\def\uh{\widehat{u}}
\def\ut{\widetilde{u}}
\def\ubar{\overline{u}}
\def\vh{\widehat{v}}
\def\vt{\widetilde{v}}

\def\wh{\widehat{w}}
\def\wt{\widetilde{w}}
\def\Sc{\scalebox{0.8}{$\mathcal{S}$}}
\def\Scb{{\overline{\Sc}}}
\def\Sch{{\widehat{\Sc}}}
\def\Sct{{\widetilde{\Sc}}}

\def\Figureone{1}

\def\Figurethree{3}
\def\Figurefour{4}

\newcommand*{\ndots}{\kern-0.075em.\kern-0.05em.\kern-0.05em.}  
\newcommand*{\nidots}{.\kern-0.05em.\kern-0.05em.} 
\newcommand*{\ncdots}{\kern-0.15em\cdot\kern-0.2em\cdot\kern-0.2em\cdot\kern-0.15em}  

\newcommand{\ket}[1]{\ensuremath{| #1 \rangle}}

\newcommand{\pdag}{{\protect\vphantom{dagger}}}

\newcommand{\Nphmax}{N_{\mr{ph}}^\maximum}

\newcommand{\trunc}{{\mathrm{tr}}}
\newcommand{\pruned}{{\mathrm{pr}}}
\newcommand{\full}{{\mathrm{full}}}
\newcommand{\init}{{\mathrm{i}}}
\newcommand{\fin}{{\mathrm{f}}}

\newcommand{\expand}{{\mathrm{ex}}}

\newcommand{\maximum}{{\mathrm{max}}}

\newcommand{\onesite}{\textrm{1s}}
\newcommand{\twosite}{\textrm{2s}}

\newcommand{\monesite}{\mathrm{1s}}
\newcommand{\mtwosite}{\mathrm{2s}}

\newcommand{\shrewd}{shrewd}
\newcommand{\Shrewd}{Shrewd}

\newcommand{\Atrunc}{{\widetilde A}{}^\trunc}
\newcommand{\Btrunc}{{\widetilde B}{}^\trunc}
\newcommand{\Apruned}{{\widehat A}{}^\pruned}

\newcommand{\Abtrunc}{{\overline A}{}^\trunc}
\newcommand{\Ahtrunc}{{\widehat A}{}^\trunc}

\NewDocumentCommand{\doubleI}{O{}}{\mathbbm{1}_{#1}}
\NewDocumentCommand{\doubleIb}{O{}}{{\overline{\mathbbm{1}}_{#1}}}
\NewDocumentCommand{\doubleIk}{O{}}{\mathbbm{1}^\ks_{\! #1}}
\NewDocumentCommand{\doubleId}{O{}}{\mathbbm{1}^\ds_{\! #1}}
\NewDocumentCommand{\doubleIp}{O{}}{\mathbbm{1}^\ps_{\! #1}}
\NewDocumentCommand{\doubleV}{O{}}{\mathbbm{V}_{\! #1}}
\NewDocumentCommand{\doubleVk}{O{}}{\mathbbm{V}^\ks_{\! #1}}
\NewDocumentCommand{\doubleVd}{O{}}{\mathbbm{V}^\ds_{\! #1}}
\NewDocumentCommand{\doubleVp}{O{}}{\mathbbm{V}^\ps_{\! #1}}
\NewDocumentCommand{\doublev}{o}{{\mathbbm{v}_{#1}}}
\NewDocumentCommand{\doubleVb}{o}{{\overline{\mathbbm{V}}_{\! #1}}}
\NewDocumentCommand{\doubleVt}{o}{{\widetilde{\mathbbm{V}}_{\! #1}}}
\NewDocumentCommand{\doubleVh}{o}{\widehat{{\mathbbm{V}}_{\! #1}}}
\NewDocumentCommand{\doubleW}{o}{\mathbbm{W}_{\! #1}}
\NewDocumentCommand{\doubleWk}{o}{\mathbbm{W}^\ks_{\! #1}}
\NewDocumentCommand{\doubleWd}{o}{\mathbbm{W}^\ds_{\! #1}}
\NewDocumentCommand{\doubleWb}{o}{{\overline{\mathbbm{W}}_{\! #1}}}
\NewDocumentCommand{\doubleWt}{o}{{\widetilde{\mathbbm{V}}_{\! #1}}}
\NewDocumentCommand{\doubleWh}{o}{{\widehat{\mathbbm{V}}_{\! #1}}}

\def\D{{\scriptstyle {\rm D}}} 
\def\DD{{\scriptstyle {\rm DD}}} 
\def\rDD{\mathrm{r}{\scriptstyle {\rm DD}}} 
\def\K{{\scriptstyle {\rm K}}} 
\def\B{{\scriptstyle {\rm B}}} 
\def\P{{\scriptstyle {\rm P}}} 

\def\ds{{\scriptscriptstyle {\rm D}}}
\def\ks{{\scriptscriptstyle {\rm K}}}
\def\ps{{\scriptscriptstyle {\rm P}}}

\def\ps{{\scriptscriptstyle {\rm P}}}

\def\st{{\widetilde{s}}}

\def\sh{{\widehat{s}}}

\def\vh{{\widehat{v}}}

\def\variance{\Delta_{E}}

\NewDocumentCommand{\cor}{mod()}
{
	#1\IfValueTF{#2}{[#2]}{}\IfValueTF{#3}{(#3)}{}
}

\usepackage{mathtools}

\usepackage{xparse} 

\NewDocumentCommand{\xA}{
         O{fill=black}mm D<|{0} D|>{0} D//{0} O{}O{0}O{0}}  
{
	\draw [#1] (#2-00.135,#3) -- (#2,#3) -- (#2,-00.135+#3) -- cycle;
	\draw (#2-00.25-#4,#3) -- (#2-00.135,#3); 
	\draw (#2,#3) -- (#2+00.25+#5,#3); 
	\draw (#2,#3-00.135) -- (#2,#3-00.25-#6);
	\draw (#2+#8,#3+#9) node (X) {#7};
}

\NewDocumentCommand{\xAd}{
         O{fill=black}mm D<|{0} D|>{0} D//{0} O{}O{0}O{0}}  
{
	\draw [#1] (#2-00.135,#3) -- (#2,#3) -- (#2,#3+00.135) -- cycle;
	\draw (#2-00.25-#4,#3) -- (#2-00.135,#3); 
	\draw (#2,#3) -- (#2+00.25+#5,#3); 
	\draw (#2,#3+00.135) -- (#2,#3+00.25+#6);
	\draw (#2+#8,#3+#9) node (X) {#7};
}

\NewDocumentCommand{\xB}{
         O{fill=black}mm D<|{0} D|>{0} D//{0} O{}O{0}O{0}} 
{
	\draw [#1] (#2,#3) -- (#2+0.135,#3) -- (#2,-0.135+#3) -- cycle;
	\draw (#2-00.25-#4,#3) -- (#2,#3); 
	\draw (#2+0.135,#3) -- (#2+00.25+#5,#3); 
	\draw (#2,#3-0.135) -- (#2,#3-00.25-#6);
	\draw (#2+#8,#3+#9) node (X) {#7};
}

\NewDocumentCommand{\xBd}{
         O{fill=black}mm D<|{0} D|>{0} D//{0} O{}O{0}O{0}}  
{
	\draw [#1] (#2,#3) -- (#2+0.135,#3) -- (#2,#3+0.135) -- cycle;
	\draw (#2-00.25-#4,#3) -- (#2,#3); 
	\draw (#2+0.135,#3) -- (#2+00.25+#5,#3); 
	\draw (#2,#3+0.135) -- (#2,#3+00.25+#6);
	\draw (#2+#8,#3+#9) node (X) {#7};
}

\NewDocumentCommand{\xC}{
         O{fill=black}mm D<|{0} D|>{0} D//{0} O{}O{0}O{0}}  
{
	\draw [#1] (#2,#3) circle (0.065);
	\draw (#2-00.25-#4,#3) -- (#2-0.065,#3); 
	\draw (#2+0.065,#3) -- (#2+00.25+#5,#3); 
	\draw (#2,#3-0.065) -- (#2,#3-00.25-#6);
	\draw (#2+#8,#3+#9) node (X) {#7};
}

\NewDocumentCommand{\xCd}{
         O{fill=black}mm D<|{0} D|>{0} D//{0} O{}O{0}O{0}}  
{
	\draw [#1] (#2,#3) circle (0.065);
	\draw (#2-00.25-#4,#3) -- (#2-0.065,#3); 
	\draw (#2+0.065,#3) -- (#2+00.25+#5,#3); 
	\draw (#2,#3+0.065) -- (#2,#3+00.25+#6);
	\draw (#2+#8,#3+#9) node (X) {#7};
}

\NewDocumentCommand{\xW}{
         O{fill=black}mm D<|{0} D|>{0} D//{0} O{}O{0}O{0}}  
{
	\draw [#1] (#2-0.065,#3-0.065) rectangle (#2+0.065,#3+0.065);
	\draw (#2-00.25-#4,#3) -- (#2-0.065,#3); 
	\draw (#2+0.065,#3) -- (#2+00.25+#5,#3); 
	\draw (#2,#3-0.065) -- (#2,#3-00.25-#6);
	\draw (#2,#3+0.065) -- (#2,#3+00.25+#6);
	\draw (#2+#8,#3+#9) node (X) {#7};
}

\NewDocumentCommand{\lcurl}{mmm  O{}O{0}O{0} D<>{0}}   
{
	\draw (#1,#3) edge[out=180,in=-90] (#1-0.1+#7,#2*0.5+#3*0.5);
	\draw (#1-0.1+#7,#2*0.5+#3*0.5) edge[out=90,in=180] (#1,#2);
	\draw (#1+#5,0.5*#3+0.5*2+#6) node (X) {#4};
}

\NewDocumentCommand{\rcurl}{mmm  O{}O{0}O{0} D<>{0}}   
{
	\draw (#1,#3) edge[out=0,in=-90] (#1+0.1+#7,#2*0.5+#3*0.5);
	\draw (#1+0.1+#7,#2*0.5+#3*0.5) edge[out=90,in=0] (#1,#2);
	\draw (#1+#5,0.5*#2+0.5*#3+#6) node (X) {#4};
}

\NewDocumentCommand{\effL}{O{fill=black} mmm  O{$\,$}O{0}O{0} D<>{0}}   
{
	\draw (#2,#4) edge[out=180,in=-90] (#2-00.2+#8,#3*0.5+#4*0.5-0.1);
	\draw (#2-00.2+#8,#3*0.5+#4*0.5+0.1) edge[out=90,in=180] (#2,#3);
	\draw [#1](#2-00.2+#8,#3*0.5+#4*0.5+0.1) -- 
	(#2-00.2+#8,#3*0.5+#4*0.5-0.1) -- (#2-00.1+#8,#3*0.5+#4*0.5) -- cycle;
	\draw (#2-00.2+#8,#3*0.5+#4*0.5) -- (#2,#3*0.5+#4*0.5);
	\draw (#2+#6,#4+#7) node (X) {#5};
}

\NewDocumentCommand{\effR}{O{fill=black} mmm  O{$\,$}O{0}O{0} D<>{0}}   
{
	\draw (#2,#4) edge[out=0,in=-90] (#2+00.2+#8,#3*0.5+#4*0.5-0.1);
	\draw (#2+00.2+#8,#3*0.5+#4*0.5+0.1) edge[out=90,in=0] (#2,#3);
	\draw [#1] (#2+00.2+#8,#3*0.5+#4*0.5+0.1) -- 
	(#2+00.2+#8,#3*0.5+#4*0.5-0.1) -- (#2+00.1+#8,#3*0.5+#4*0.5) -- cycle;
	\draw (#2+00.2+#8,#3*0.5+#4*0.5) -- (#2,#3*0.5+#4*0.5);
	\draw (#2+#6,#4+#7) node (X) {#5};
}

\makeatletter
\def\maketitle{
\@author@finish
\title@column\titleblock@produce
\suppressfloats[t]}
\makeatother

\begin{document} 

\title{Controlled bond expansion for 
DMRG ground state search at single-site costs
}
\author{Andreas Gleis}
\affiliation{Arnold Sommerfeld Center for Theoretical Physics, 
Center for NanoScience,\looseness=-1\,  and 
Munich Center for \\ Quantum Science and Technology,\looseness=-2\, 
Ludwig-Maximilians-Universit\"at M\"unchen, 80333 Munich, Germany}
\author{Jheng-Wei Li}
\affiliation{Arnold Sommerfeld Center for Theoretical Physics, 
Center for NanoScience,\looseness=-1\,  and 
Munich Center for \\ Quantum Science and Technology,\looseness=-2\, 
Ludwig-Maximilians-Universit\"at M\"unchen, 80333 Munich, Germany}
\author{Jan von Delft}
\affiliation{Arnold Sommerfeld Center for Theoretical Physics, 
Center for NanoScience,\looseness=-1\,  and 
Munich Center for \\ Quantum Science and Technology,\looseness=-2\, 
Ludwig-Maximilians-Universit\"at M\"unchen, 80333 Munich, Germany}

\begin{abstract}
\begin{center}
(Dated: \today)
\end{center}

DMRG ground state search algorithms employing symmetries must be able to expand virtual bond spaces 
by adding or changing symmetry sectors if these lower the energy. Traditional single-site DMRG does not allow
bond expansion; two-site DMRG does, but at much higher computational costs. We present a controlled bond expansion (CBE) algorithm that yields two-site accuracy and convergence per sweep, at single-site costs. Given a matrix product state $\Psi$ defining a variational space, CBE identifies  parts of the orthogonal space carrying significant weight in $H\Psi$ and expands bonds to 
include only these. CBE--DMRG uses no mixing parameters and is fully variational.
Using CBE--DMRG,
we show that the Kondo--Heisenberg model on a width 4 cylinder features two distinct phases differing 
in their Fermi surface volumes.
\\
\\
\noindent
\vspace{-2mm}
DOI: \vspace{-2mm}

\end{abstract}

\maketitle
\textit{Introduction.---}
A powerful tool for studying ground state properties
of one- and two-dimensional quantum systems is the  density martrix renormalization group (DMRG) \cite{White1992,White1993,Verstraete2004,Schollwoeck2005,Schollwoeck2011,White1996,Stoudenmire2012}. 
Prominent two-dimensional applications include the $t$-$J$  \cite{White1998,White2004,White2009a,Jiang2021} and Hubbard \cite{LeBlanc2015,Ehlers2017,Zheng2017,Huang2018,Qin2020,Jiang2020,Jiang2022} models, and quantum magnets \cite{Yan2011,Depenbrock2012,Kolley2015,He2017a}. Due to their
high numerical costs, such studies are currently limited to either small finite-sized systems or cylinders with small circumference. 
Progress towards computationally cheaper DMRG ground state search algorithms would clearly be welcome.
In this paper, we address this challenge. A DMRG ground state search explores a variational space spanned by matrix product states~\cite{Oestlund1995,Rommer1997}. If symmetries are exploited, the  algorithm must be able to expand the auxiliary spaces associated with virtual bonds by adjusting symmetry sectors if  this  lowers the energy. Traditional single-site 
(\onesite) DMRG, which variationally updates one site at a time, does not allow such bond expansions. As a result, it often gets stuck in metastable configurations having quantum numbers different from the actual ground state. Two-site 
(\twosite) DMRG naturally leads to bond expansion, but carries much higher computational costs. 
Hence, schemes have been proposed for achieving
bond expansions at sub-\twosite\ costs,
 such as density matrix perturbation \cite{White2005}
or strictly single-site DMRG (DMRG3S)~\cite{Hubig2015}.
However, in these schemes, the degree of subspace expansion per local update
is controlled by  a heuristic mixing factor. Depending on its value,
some subspace expansion updates increase, rather than decrease, the energy.  
Here, we present a controlled bond expansion (CBE) algorithm which lowers the energy with each step and yields 
\twosite\ accu\-racy and convergence per sweep, at \onesite\ costs. Given a matrix product state $\Psi$ defining a vari\-ational space, our key idea is to identify parts of the \twosite\ orthogonal space that carry significant weight in $H\Psi$,  and to include only these parts when expanding the virtual bonds of a \onesite\ Hamiltonian. Remarkably, these parts can be found via a projector that can be \mbox{constructed at \onesite\ costs.}
Using CBE--DMRG we study the Kondo--Heisenberg model on a width 4 cylinder and show that it features two phases differing in their Fermi surface volumes. We thereby further advance the understanding of this highly debated model using a controlled method.
\textit{MPS basics.---} 
We briefly recall some standard MPS concepts 
\cite{Schollwoeck2011}, 
adopting the diagrammatic conventions of Ref.~\onlinecite{Gleis2022a}.
Consider an $\eLL$-site system with an open boundary MPS wavefunction $\Psi$ having dimensions $d$ for physical sites 
and $D$ for virtual bonds.
$\Psi$ can be written in bond-canonical form w.r.t.\ to any bond $\ell$, 
\vspace{-2mm}
\begin{align}
	\Psi &  = \,  
	\raisebox{-3mm}{\includegraphics[width=0.713\linewidth]{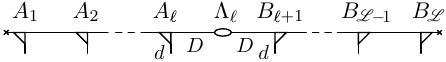}} \, .
\end{align}
The tensors $\Lambda_\ell \, (\EllipseWhiteLambda)$, $A_\ell \,  (\TriangleWhiteA)$ and $B_{\ell} \, (\TriangleWhiteB)$ are variational parameters. They are linked by gauge relations, $A_\ell \Lambda_\ell = \Lambda_{\ellminusone} B_\ell$,
useful for shifting the bond tensor $\Lambda_\ell$ to neighboring bonds. 
 $A_\ell$ and $B_\ell$ are  left and right-sided isometries, respectively, projecting $Dd$-dimensional \textit{parent} ($\P$) spaces to $D$-dimensional \textit{kept} ($\K$) image spaces \cite{Gleis2022a}; they satisfy
\vspace{-2mm}
\begin{flalign}
\label{eq:IsometricConditions}
& \raisebox{-6.5mm}{
 \includegraphics[width=0.866\linewidth]{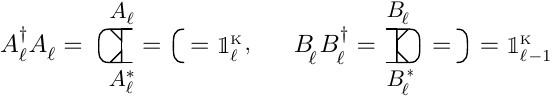}}.  \hspace{-1cm} &
 \end{flalign}
 
 \vspace{-2mm}
 \noindent 
The Hamiltonian can similarly be expressed as a matrix product operator (MPO) with virtual bond dimension $w$,
\hspace{-4mm}
\begin{align}
H =
     \raisebox{-3.0mm}{\includegraphics[width=0.75\linewidth]{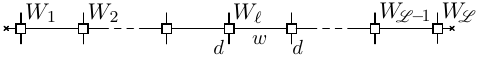}}  \, . &
\end{align}

\vspace{-2mm}
 \noindent 
For \twosite\ 
or \onesite\ DMRG, the energy of $\Psi$  is lowered by projecting $H$ to a local variational space
associated with sites  $(\ell,\ell+1)$ or $\ell$, respectively, and using its 
ground state (GS) within that space to locally update $\Psi$. 
The effective \twosite\ 
and \onesite\  Hamiltonians 
can be computed recursively using
\vspace{-3mm}
\begin{subequations}
\begin{flalign}
H^\mtwosite_\ell \! & = \!\!\!\!
\raisebox{-5.3mm}{
 \includegraphics[width=0.783\linewidth]{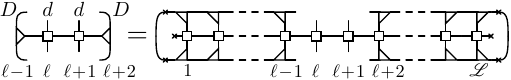}}  ,
 \hspace{-1cm} 
\\
H^\monesite_\ell \! & =  \!\!\!\!
\raisebox{-5.3mm}{
 \includegraphics[width=0.7\linewidth]{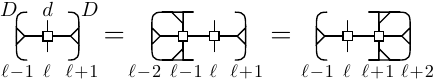}}  \!\! . 
 \hspace{-1cm} & 
\end{flalign}
\end{subequations}
To perform \twosite\  
or \onesite\ updates, one replaces $\psi_\ell^\mtwosite \!=\! 
A_{\ell} \Lambda_\ell  B_{\ellplusone}  $ or $\psi_\ell^\monesite \!=\! C_\ell
= \! A_\ell \Lambda_\ell \, (\CircleWhiteC)$ by the GS solutions of
\vspace{-2mm}
\begin{subequations}
\label{subeq:SchroedingerMain}
\begin{flalign}
\label{eq:SchroedingerMain-two}
 \quad   (H_\ell^\mtwosite \!-\! E) \psi^\mtwosite_\ell & = 0 \, ,  \quad
	\raisebox{-5.3mm}{
 \includegraphics[width=0.44\linewidth]{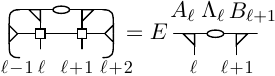}} \!\! , \hspace{-1cm} &
\\ 
\label{eq:SchroedingerMain-one}
\quad		(H_\ell^\monesite \!-\! E) \psi^\monesite_\ell
		& =  0  \, , \; \quad
\raisebox{-5.8mm}{
 \includegraphics[width=0.316\linewidth]{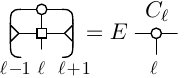}} . \hspace{-1cm} & 
\end{flalign}
\end{subequations}
Updating site by site, one sweeps back and forth through the MPS until the GS energy converges.
The local variational space is 
larger for \twosite\ than \onesite\ DMRG by a factor $d$,
$\mathcal{O}(D^2d^2)$ vs.\  $\mathcal{O}(D^2d)$.
This enables \twosite\ DMRG to increase (``expand'') the bond dimension 
during updates by including new states (and 
symmetry sectors!) from the \twosite\ space. 
\onesite\ DMRG  cannot do this, and hence often fails
to yield accurate GS energies. 
The better performance of \twosite\ vs.\ \onesite\
has its price: much higher numerical costs, 
$\mathcal{O}\bigl(D^3 d^3 + D^3 d^2 w\bigr) $ vs.\ $\mathcal{O}\bigl(D^3 d w\bigr)$  \cite{Schollwoeck2011}. 
\textit{Discarded spaces.---}
To track those parts of
\twosite\ spaces not contained in 
\onesite\ spaces, we introduce orthogonal complements of
$A^\pdag_{\ell}$ and $B_{\ell}^\pdag$, denoted $\Ab^\pdag_{\ell} (\TriangleGreyA)$ 
and $\Bb_{\ell}^\pdag (\TriangleGreyB)$. 
These isometries have  image 
spaces, called \textit{discarded} $(\D)$ spaces \cite{Gleis2022a},
of dimension $\Db \!=\! D(d\!-\! 1)$, orthogonal to the 
kept images of  $A^\pdag_{\ell}$ and $B_{\ell}^\pdag$. Thus 
$A^\mathbbm{1}_\ell (\QuaterCircleBlackA) \!=\! A^\pdag_\ell \!\oplus\! \Ab^\pdag_\ell$ 
and $B^\mathbbm{1}_\ell (\QuaterCircleBlackB)\!=\! B^\pdag_\ell \!\oplus\! \Bb^\pdag_\ell$ are unitaries on their parent spaces, with
\vspace{-5mm}
\begin{flalign}
\label{eq:full-unitaries}
 & \!\raisebox{-4,7mm}{
 \includegraphics[width=0.9\linewidth]{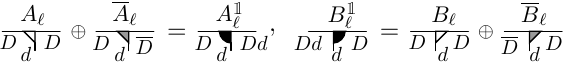}} \, . 
 \hspace{-1cm} & 
\end{flalign}

\vspace{-1mm}
\noindent
The unitarity conditions for $A^\mathbbm{1}_\ell$ and $B^\mathbbm{1}_\ell$ imply 
orthonormality and completeness relations complementing Eq.~\eqref{eq:IsometricConditions}, 
\vspace{-5mm}
\begin{subequations}
\label{subeq:AdditionalOrthonormalityRelations}
\begin{flalign}
 & \hspace{-1mm} 
 \raisebox{-4mm}{
 \includegraphics[width=0.89\linewidth]{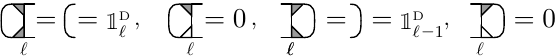}}
 \hspace{-1cm} &
\\
\label{eq:CompletenessMain}
 & \hspace{-2mm} \raisebox{-3.5mm}{
 \includegraphics[width=0.886\linewidth]{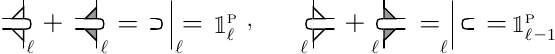}} .
 \hspace{-2cm} &
\end{flalign}
\end{subequations}

\vspace{-1mm}

\noindent
If the unitary maps $A^{\mathbbm{1}\dagger}_\ell$ and $B_\ellplusone^{\mathbbm{1}\dagger}$ of \Eq{eq:full-unitaries} are applied to some of the open indices of 
$H_\ell^\monesite \psi_\ell^\monesite$, $H_\ellplusone^\monesite \psi_\ellplusone^\monesite$ and $H_\ell^\mtwosite \psi^\mtwosite_\ell$ as indicated below, they map the diagrams of \Eqs{subeq:SchroedingerMain} to
\vspace{-2mm}
\begin{align}
\nonumber
 & \!\! \raisebox{-3.5mm}{
 \includegraphics[width=\linewidth]{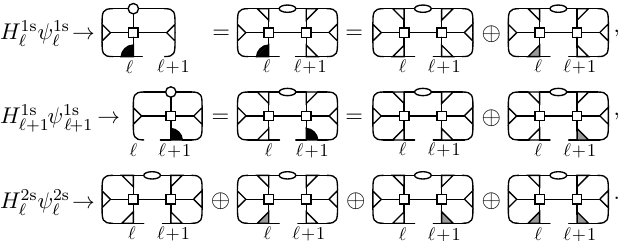}} 
\end{align} 

\vspace{-1mm}

\noindent
The first three terms from the third line
also appear in the first two lines, 
but the fourth, involving $\TriangleGreyAdagger \TriangleGreyBdagger$,
does not. Let $\DD$ denote the image of the orthogonal complements $\Ab_\ell \otimes \Bb_\ellplusone \, 
(\TriangleGreyA \otimes \TriangleGreyB)$,  then $\DD$
is orthogonal to the variational space explored by \onesite\ DMRG on sites $(\ell,\ell\!+\!1)$.
$\DD$ is much larger than the latter, of dimension $\Db{}^2 = D^2(d\!-\!1)^2$ vs.\ $2D^2d$, and (importantly!) may contain new symmetry sectors. Thus $\DD$ is the \twosite\ ingredient lacking in \onesite\ schemes. 
This can also be seen considering the energy variance
$\variance \!=\! \|(H \! - \! E)\Psi\|^2$.  
By expanding it into contributions involving orthogonal
projections on one, two, or more sites \cite{Hubig2018}, 
$\variance \! = \variance^{1\perp} \!+\! \variance^{2\perp} \!+\! \dots$, one obtains \cite{Gleis2022a}
\vspace{-2mm}
\begin{align}
	\label{eq:variance}
\variance^{1\perp} = \raisebox{-5.3mm}{
 \includegraphics[width=0.233\linewidth]{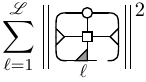}} , 
 \quad \variance^{2\perp} = 
\raisebox{-5.3mm}{
 \includegraphics[width=0.3\linewidth]{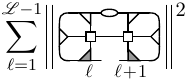}} .
\end{align}

\vspace{-2mm}

\noindent
1s DMRG minimizes only $\variance^{1\perp}$, 2s minimizes 
$\variance^{1\perp}$ \textit{and} $\variance^{2\perp}$.
We thus seek to expand
the $\K$ image of \TriangleWhiteA\ or \TriangleWhiteB at the expense of the $\D$ image of $\TriangleGreyA$ or $\TriangleGreyB$. This transfers weight from $\variance^{2\perp}$ to $\variance^{1\perp}$, making it accessible to \onesite\ minimization.

\begin{figure} 
\includegraphics[width=\linewidth]{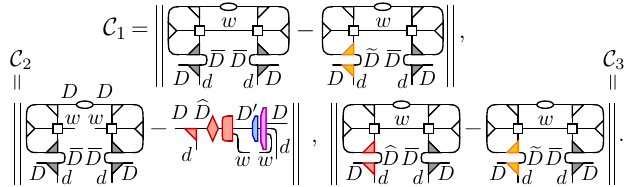}
\caption{Shrewd selection for a right-to-left sweep:
Ideally, the truncated complement $\Atrunc_\ell (\TriangleOrangeA)$ should be found by minimizing the cost
function $\Cc_1$, but that would involve \twosite\ cost, $\mathcal{O}(D^3 d^2 w)$. To achieve \onesite\ cost,  $\mathcal{O}(D^3 d w)$,
we instead use \textit{shrewd selection}, involving two separate truncations: The first truncation (\textit{preselection}) truncates
$\Ab{}_\ell (\TriangleGreyA)$ to $\Apruned_\ell (\TriangleRedA)$ by minimizing the cost function $\Cc_2$.
The second truncation (\textit{final selection}) further truncates $\Apruned_\ell (\TriangleRedA) \to \Atrunc_\ell (\TriangleOrangeA)$
by minimizing the cost function $\Cc_3$. For details, see Fig.~S-2 in Sec.~S-1 of the supplement \cite{supplement}. 
} \vspace{-4.5mm}
\label{fig:cost-function} 
\end{figure}

\textit{Controlled bond expansion.---}
The CBE algorithm rests on two new insights, substantiated by the quality of its results. The first insight is that the subspace of $\DD$ relevant for lowering the GS energy is relatively small:
 it is the subspace on which $H^\mtwosite_\ell \psi^\mtwosite_\ell$ and hence $\variance^{2\perp}$ have significant weight.  
When expanding a bond, it thus suffices to add only this small
subspace (hence the moniker \textit{controlled} bond expansion), or only part of it,
to be called relevant $\DD$ ($\rDD$)~\cite{RelevantSubspace}. Since $\DD$ is the image of $\Ab_\ell \otimes \Bb_\ellplusone (\TriangleGreyA \otimes \TriangleGreyB)$, 
$\rDD$ can be viewed as the image of $\Atrunc_\ell 
\otimes \Bb{}_\ellplusone (\TriangleOrangeA \otimes \TriangleGreyB)$ or $\Ab{}_\ell \otimes \Btrunc_\ellplusone (\TriangleGreyA \otimes \TriangleOrangeB)$, where the isometries $\Atrunc_\ell\,
(\TriangleOrangeA)$
or $\Btrunc_\ellplusone\, (\TriangleOrangeB)$ are \textit{truncated} versions of $\Ab{}_\ell$ or $\Bb{}_\ellplusone$ and have image dimensions $\Dt$, say.  
It turns out that one may choose $\Dt < D$, independent of $d$, thus 
$\rDD$, of dimension $\Dt \Db$, is indeed much smaller than $\DD$.
The second insight is that $\Atrunc_\ell$ or $\Btrunc_\ellplusone$ can be constructed at \onesite\ costs
using a novel scheme explained in Fig.~\ref{fig:cost-function}.
We call it \textit{\shrewd\ selection} since it is cheap, efficient
and practical, though not strictly optimal (that would
require \twosite\ costs). 
\textit{Shrewd selection.---}
Ideally, $\Atrunc_\ell $ should minimize the cost function 
$\Cc_1$ (Fig.~\ref{fig:cost-function}, top), 
the difference between applying the projectors $\Ab{}^\pdag_\ell \Ab{}^{\dag}_\ell$ or $\Atrunc_\ell \Atrunc_\ell{}^\dag$ to $ H^\mtwosite_\ell \psi^\mtwosite_\ell \Bb{}^{\dag}_\ellplusone \Bb{}^{\pdag}_\ellplusone$.
However,  exact minimization of $\Cc_1$ 
would involve \twosite\ costs (feasible if $d$, $w$ and $D$ are comparatively
small, but in general undesirable). To maintain \onesite\ costs, $\mathcal{O}(D^3 d w)$, we instead  use \textit{\shrewd\ selection},
involving two separate truncations,
depicted schematically in Fig.~\ref{fig:PreselectionConcept}
and explained in detail in Sec.~S-1 of the supplement~\cite{supplement}.
The first truncation (\textit{preselection}) truncates
the central MPS bond from 
$D \!\to \! D'$ (specified below) in the presence of its environment by minimizing the cost function $\Cc_2$ (Fig.~\ref{fig:cost-function}, bottom left);
this replaces the full complement 
by a preselected complement, $\Ab_\ell 
\TriangleGreyA \!\to \!\! \Apruned_\ell \TriangleRedA$, 
with reduced image dimension,
$\Db \! \to \! \Dh \!=\! D' w$~\cite{FirstTruncation}. The second truncation
(\textit{final selection}) minimizes the cost function $\Cc_3$ (Fig.~\ref{fig:cost-function}, bottom right)
with central MPO bond closed as appropriate for $H^\mtwosite_\ell\psi^\mtwosite_\ell$: it further truncates  $\Apruned_\ell$ to yield the final truncated complement, $\Atrunc_\ell$, $\TriangleRedA \!\to \!\!
\TriangleOrangeA$, 
$\Dh \! \to \! \Dt < \! D$. 
To ensure \onesite\ costs for final selection we need $\Dh \!=\! D$, 
and thus choose $D' \!=\! D/w$ for preselection. 
\textit{CBE update.---}
A CBE update of bond $\ell$ proceeds in four substeps.
We describe them for a right-to-left sweep for building $\Atrunc_\ell$ 
and updating $C_\ellplusone$ (left-to-right sweeps, building 
$\Btrunc_\ellplusone$ and updating $C_\ell$, are analogous).

(i) Compute $\Atrunc_\ell \, (\TriangleOrangeA)$
using shrewd selection.

(ii) Expand bond $\ell$ from dimension 
$D$ to $D \!+\! \Dt$ by replacing 
$A_\ell$ by an expanded isometry
$A_\ell^\expand (\TriangleGreenA) = A_\ell \oplus \Atrunc_\ell$, 
and $C_\ellplusone$ by an expanded tensor initialized as  
$C_\ellplusone^{\expand,\init} \, (\CircleGreenC)$,
defined such that $A^\expand_\ell C_\ellplusone^{\expand,\init} = A_\ell C_\ellplusone $:
\vspace{-3mm}
\begin{flalign}
&  
\raisebox{-4mm}{
 \includegraphics[width=0.9\linewidth]{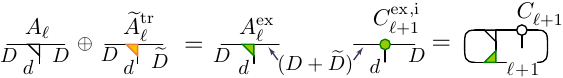}} .
\hspace{-1cm} & 
\label{eq:expand-bond}
\end{flalign}

\vspace{-1.5mm}
\noindent
Also construct an expanded \textit{one}-site Hamiltonian, defined
in a variational space of dimension $D (D+\Dt) d$:
\vspace{-2mm}
\begin{align}
\label{eq:H1expand}
H^{\monesite,\expand}_\ellplusone & =
\raisebox{-4.7mm}{
\includegraphics[width=0.516\linewidth]{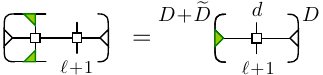}} \, .
\end{align}

\vspace{-1mm}

(iii) Update $C_{\ellplusone}^\expand$ variationally by using an iterative eigensolver,
as usual in DMRG, to find the GS solution of $(H^{\monesite,\expand}_\ellplusone \!-\! E) C^\expand_\ellplusone= 0$, starting from $C^{\expand,\init}_\ellplusone$. 
(We employ a Lanczos eigensolver.)
This has costs of $\mathcal{O}\bigl(D^3dw\bigr)$. 
Thus, $C^\expand_\ellplusone$ can be updated at \onesite\ costs, while including only the most relevant \twosite\ information via the contribution of $\Atrunc_\ell$.

(iv) Shift the isometry center from
site $\ell +1$ to site $\ell$ using a singular value decomposition (SVD) 
and truncate (\textit{trim}) bond $\ell$ from dimension $D + \Dt$ back to $D$, removing low-weight states. The discarded 
weight, say $\xi$, of this bond trimming serves as an error measure~\cite{supplement}.
\nocite{Stoudenmire2012,Hubig2018,White2005,Jeckelmann1998,Tezuka2007,Fehske2008, Ejima2010,Reinhard2019,Coleman2007,Ye2022,Coleman2001,Luttinger1960,Senechal1999,Oshikawa2000,Si2014,Nishikawa2018}
 \begin{figure}[t!] 
 \includegraphics[width=\linewidth]{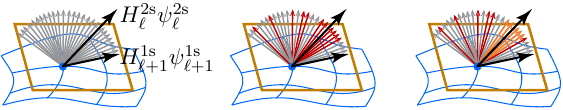} 
  \vspace{-7mm}
 \caption{
The projection $H^{\mtwosite}_{\ell} {\psi^\mtwosite_\ell} 
 \overset{A_{\ell}^{\dagger}}{\longmapsto} H^{\monesite}_{\ell+1}
 {\psi^\monesite_\ellplusone}$ to the tangent space (yellow) of the MPS manifold (blue) discards information from $\DD$ (depicted by grey arrows for $\DD$ basis vectors).  \textit{Relevant} information is recovered at \onesite\ cost by constructing  $\rDD$ through preselection (red), then final selection (orange). \vspace{-3mm}
 \label{fig:PreselectionConcept}} 
 \end{figure}
The energy minimization based on $H^{\monesite,\expand}_\ellplusone$ is variational, hence 
each CBE update strictly lowers the GS energy.
Though 
\shrewd\ selection involves severe bond reductions, 
it yields $\rDD$s suitable for efficiently lowering the GS energy (in step (iii)).
Moreover, although CBE explores a much smaller variational space than
\twosite\ DMRG, it converges at the same rate 
and accuracy (see below and Ref.~\cite{supplement}),
since it focuses on the subspace that really matters for energy reduction.
Section~S-1 in \cite{supplement} illustrates this 
by analysing singular value spectra. All in all,
CBE is
a \onesite\ cost version of the \twosite\ update,
compatible with established DMRG parallelization schemes
\cite{Stoudenmire2013}.
Similar to \twosite~\cite{Stoudenmire2012}, CBE can also be combined with mixing during
the initial few sweeps (see Ref.~\onlinecite{supplement}, Sec.~S-3).

We note
that bond expansion using a truncated $\DD$ has been proposed before \cite{Hubig2015,ZaunerStauber2018}.
But our
$A_\ell^\expand (\TriangleGreenA)$ 
outperforms that of  
DMRG3S \cite{Hubig2015} (see below and Ref.~\cite{supplement});
and we find $A_\ell^\expand (\TriangleGreenA)$ 
at \onesite\ costs, whereas Ref.~\onlinecite{ZaunerStauber2018} (on variational uniform MPS~\cite{Vanderstraeten2019}) uses an SVD requiring \twosite\ costs.

\textit{Sweeping.---}
Our computations exploit 
$\mathrm{U}(1)_{\mathrm{ch}}\otimes \mathrm{SU}(2)_{\mathrm{sp}}$ charge and spin symmetries using QSpace~\cite{Weichselbaum2012,Weichselbaum2020}, 
with bond dimensions  $\Dast$ (or $D$) counting symmetry multiplets (or states). 
Usually, 
$\Dast$ is increased 
with each update during sweeping, 
 from an initial $\Diast$
to a final $\Dfast \!= \! \alpha \Diast$, with $\alpha \! > \! 1$.
To achieve this with  CBE we (i,ii) use
$\Dpast \!\simeq\! \Dfast /\wast$, $\Dhast \!=\! \Dfast$ (cf.~\Fig{fig:cost-function}) and expand from $\Diast$ to $\Diast +  \Dtast =  \Dfast(1  +   \delta)$, 
(iii) call the iterative eigensolver, and (iv) truncate back to $\Dfast$ when shifting the isometry center. 
We use $\delta \!=\! 0.1$ for CBE, unless stated otherwise.
\textit{Benchmarks.---}
As a first benchmark, we consider the 1D Hubbard-Holstein~(HH) model~\cite{Jeckelmann1998,Tezuka2007,Fehske2008,Ejima2010,Reinhard2019}, 
described by
\begin{align}
H_{\mr{HH}} 
&= -\sum_{\ell\sigma} \bigl(c^{\dagger}_{\ell\sigma}c^{\phantom{\dagger}}_{\ell+1\sigma} + \textrm{h.c.}\bigr) 
+ 0.8 \sum_{\ell} n_{\ell\uparrow}n_{\ell\downarrow} 
\\ \nonumber
&+0.5 \sum_{\ell} b^{\dagger}_\ell b^{\phantom{\dagger}}_\ell
+ \sqrt{0.2}\sum_{\ell} \bigl(n_{\ell\uparrow}+n_{\ell\downarrow} - 1\bigr)\bigl(b^{\dagger}_\ell + b^{\phantom{\dagger}}_\ell\bigr)
\,  . 
\end{align}
Here, $c^\dagger_{\ell \sigma}$ creates an electron 
and $b^\dagger_\ell$ a phonon at site $\ell$, and $n_{\ell\sigma} = c^\dagger_{\ell \sigma} c^\pdag_{\ell \sigma}$. We search for the GS with
$N\! =\! \eLL\! =\! 50$, total spin $S\!=\!0$, and restrict the maximum local number of excited phonons to $\Nphmax$. Then, $d^\ast[d] = 3(\Nphmax\!+\!1) 
\, [4(\Nphmax\!+\!1)]$. 
Figure~\ref{fig:HubbardHolstein_HubbardCyl}(a) shows the relative error in energy vs.\ number of half-sweeps $n_s$ for different $\Dmax^{\ast}$ at fixed $d^\ast=12$, 
comparing CBE and \twosite\ DMRG schemes. The convergence with $n_s$ is similar for CBE and \twosite.  
Figure~\ref{fig:HubbardHolstein_HubbardCyl}(b) compares the CPU time
(measured on a single core of an Intel Core i7-9750H CPU)
per sweep 
for CBE and \twosite\ for different $d^\ast$ at fixed $\Dmax^{\ast}$. 
 Linear and quadratic fits 
confirm the expected $d^\ast$~(\onesite) or $d^{\ast2}$~(\twosite) scaling, respectively, highlighting the speedup from CBE.
Next, we consider $\eLL_{\!x} \! \times\! \eLL_{\!y}\! = 10 \!\times \!4$ and $10 \! \times \! 6$ Hubbard cylinders~(HC),
described by (following Ref.~\onlinecite{Hubig2018})
\begin{align}
H_{\mr{HC}} \!= \!-\sum_{\langle \boldell, \boldellp \rangle,\sigma} (c^{\dagger}_{\boldell\sigma}c^{\phantom{\dagger}}_{\boldellp\sigma} \! + \! \mathrm{h.c.}) 
+  8 \sum_{\boldell} n_{\boldell\uparrow} n_{\boldell\downarrow}\, . 
\end{align}
Here, 
$\boldell = (x,y)$ is a 2D site index 
and $\sum_{\langle \boldell, \boldellp \rangle}$ a nearest-neighbor sum.
We search for the GS with total filling $N=0.9\eLL_{\!x}\eLL_{\!y}$
and spin $S\!=\!0$.
We use a real-space MPO, not the hybrid-space MPO~\cite{Motruk2016,Ehlers2017} used in Ref.~\onlinecite{Hubig2018}.
Figure~\ref{fig:HubbardHolstein_HubbardCyl}(c,d) benchmarks CBE (black) against \twosite\ DMRG (red);
their accuracies match (same GS energy for given $\Dast$). 
CBE--DMRG yields controlled convergence for sufficiently large $\Dast$, where the energy error decreases linearly with $\xi$. 
DMRG3S does not reach \twosite\ accuracy for this model, as 
is clear
from the data shown in Ref.~\onlinecite{Hubig2018} Sec.~V~E.
%

%
 \begin{figure}[bt!]
 \includegraphics[width=\linewidth]{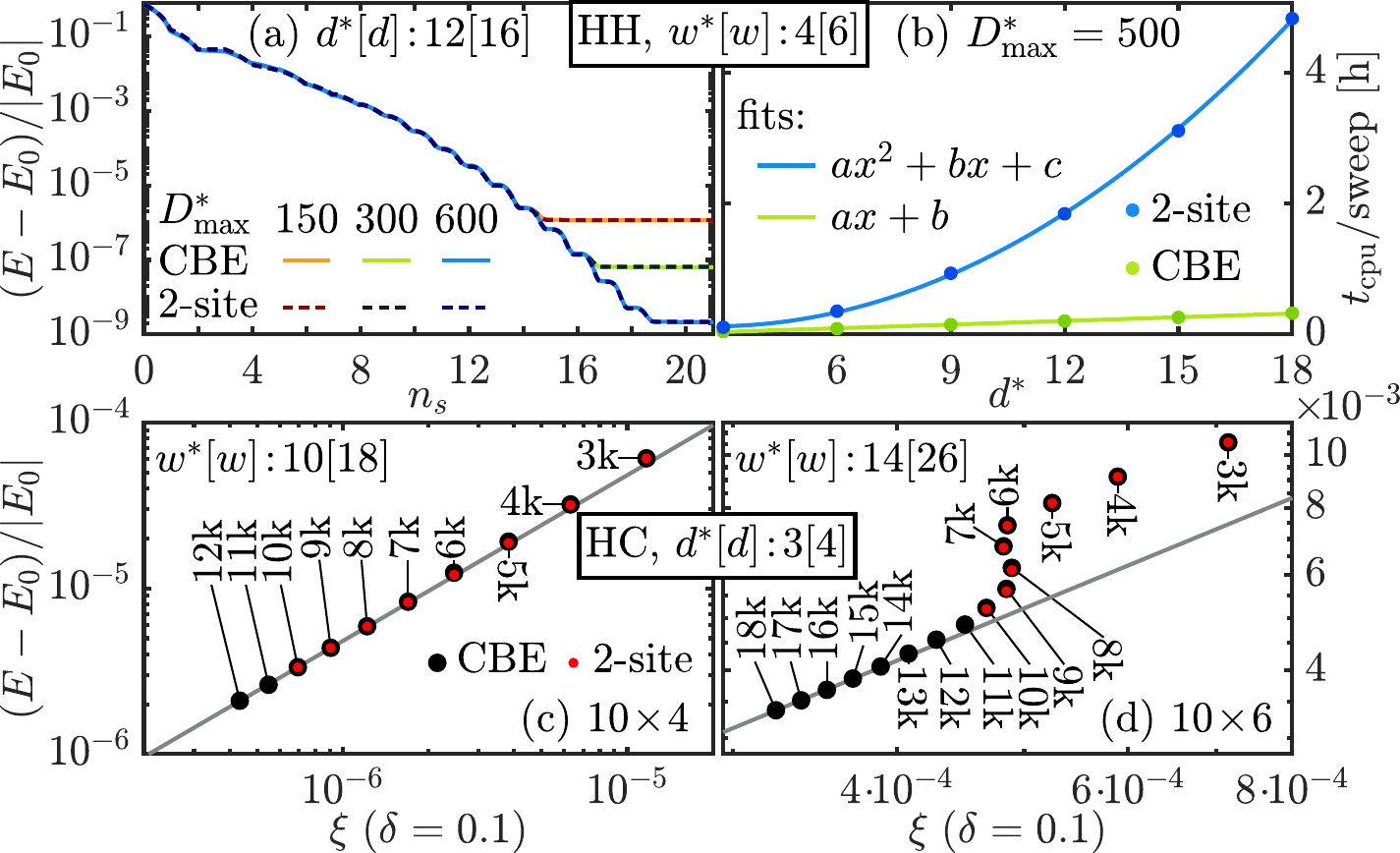} \vspace{-6mm}
 \caption{
 \label{fig:HubbardHolstein_HubbardCyl}
Hubbard-Holstein (HH) model:
(a) Convergence of the GS energy versus number of half-sweeps $n_s$
at fixed $\dast \!=\! 3(\Nphmax\!+\! 1)$.
$E_0$ was obtained by linear $\xi$-extrapolation of data from $\Dmax^{\ast}\in [1000,1200]$. (b) CPU time per sweep for various $\dast$ at fixed $\Dmax^{\ast}$,  showing  $\dast$ 
(CBE) vs.\ $d^{\ast 2}$ (2s) scaling.
Hubbard cylinders (HC):
Error in GS energy vs.\ $\xi$
for (c) $10\!\times\!4$ and (d) $10\!\times\!6$ HCs, 
obtained with CBE (black) and \twosite\ (red) DMRG,
 for various $\Dmaxast$ (legends). 
Since \twosite\ CPU times far exceed those of CBE,
\twosite\ data is only shown for $\Dmaxast \leq 10\mr{k}$.
 Reference energies $E_0 = -27.8816942$ ($10\!\times\!4$) and 
$-41.7474961$ ($10\!\times\!6$) are obtained by linear $\xi$-extrapolation of the four most accurate CBE results to $\xi\!=\!0$ (grey line). 
} \vspace{-4mm}
 \end{figure}
 Further benchmarks and comparison to DMRG3S are shown in 
 Ref.~\onlinecite{supplement},~Secs.~S-2,3. 
 We find that  CBE has similar run time per sweep but converges faster than DMRG3S \cite{Hubig2015}:
for given $\Dmaxast$, the energy converges in fewer sweeps 
and less run time,  and reaches a lower value.
\textit{Kondo-Heisenberg cylinders.---}
Finally, to include some new physics results in this paper, we
study the Kondo-Heisenberg~(KH) lattice model on a cylinder. 
The KH model is believed to describe the essential physics of heavy-fermion~(HF) materials~\cite{Coleman2007,Kirchner2020,Loehneysen2007,Stewart2001}, which feature many interesting phenomena. One of the most intriguing is the so-called Kondo breakdown~(KB) quantum critical point~(QCP)~\cite{Coleman2001,Coleman2002,Si2014},
where collective Kondo singlets~\cite{Si2014} formed at strong coupling break up, leading to
a FS reconstruction~\cite{Paschen2004,Shishido2005,Friedemann2010,Maksimovic2022}
at $T=0$.
 Strange metal behaviour is observed at finite temperatures with e.g. $\sim T$ resistivity~\cite{Martelli2019,Maksimovic2022,Prochaska2020,Trovarelli2000,Zhao2019} or $\sim T\log T$ specific heat~\cite{Trovarelli2000,Zhao2019,Loehneysen1996,Loehneysen1996a}.
Theoretical understanding of the KB--QCP is still incomplete, in part due to scarceness of numerical simulations.
Prior numerical studies used dynamical mean-field theory~\cite{DeLeo2008,DeLeo2008a,Tanaskovic2011,Si2001,Si2003} and Monte Carlo methods~\cite{Assaad1999,Capponi2001,Toldin2019,Danu2021}, but we are not aware of DMRG results on the KB--QCP.
Here, we take first steps in this direction by 
studying FS reconstruction on a KH cylinder: we show that at $T\!=\!0$, there are two distinct phases featuring different Fermi surfaces. 

We study a $\eLL_{\!x}\!\times\!\eLL_{\!y} \!=\! 40\!\times\!4$ KH cylinder, described by
\vspace{-1mm}
\begin{align*}
H_{\mr{KH}}  &= 
-\hspace{-2mm} \sum_{\langle \boldell, \boldellp \rangle,\sigma} 
\hspace{-2mm} 
\bigl(c^{\dagger}_{\boldell\sigma}c^{\phantom{\dagger}}_{\boldellp\sigma}\!\!  + \!  
\mathrm{h.c.}\bigr) 
\! + \!  J_{\mr{K}} \! \sum_{\boldell} \! \vec{S}_{\boldell} \! \cdot \!  \vec{s}_{\boldell} 
\!+\! 
\tfrac{1}{2} \!\! \sum_{\langle \boldell, \boldellp \rangle} \! \vec{S}_{\boldell}\! \cdot \! \vec{S}_{\boldellp}  . \vspace{-5mm}
\end{align*}
\vspace{-3mm}

\noindent 
Here, $\vec{s}_{\boldell} = \tfrac{1}{2} \sum_{\sigma \sigma'} c^\dag_{\boldell \sigma} \vec{\sigma}_{\! \sigma \sigma'} c^\pdag_{\boldell \sigma'}$
and $\vec{S}_{\boldell}$ are electron and local moment spin-$\tfrac{1}{2}$ operators 
at site $\boldell$. 
We search for the GS with total filling $N\!=\!1.25\eLL_{\!x}\eLL_{\!y}$ and spin $S\!=\!0$. 

For a $\eLL_{\!y}\!=\!4$ cylinder,  
the Brillouin zone consists of four lines, since 
$k_y \in \{0,\pm\tfrac{\pi}{2},\pi\}$ is discrete.
If such a line cuts the $\eLL_{\!y} \!\to\! \infty$ FS, that defines a 
``Fermi point'', with Fermi momentum $(k_{\mr{F}x}(k_y),k_y)$. We have 
extracted the corresponding $k_{\mr{F}x}(k_y)$ values
from CBE--DMRG results for the single-particle density matrix 
(see Ref.~\onlinecite{supplement}, Sec.~S-4~B  for details;
Fig.~S-13 shows controlled convergence of this quantity).
Figure~\ref{fig:KondoCyl_FS_main} shows the results 
for various values of $J_\mr{K}$. There are clearly two distinct phases with qualitatively different Fermi points $k_{\mr{F}x}(k_y)$. At small $J_{\mr{K}} \leq 2$, 
we find Fermi points at $(|k_{\mr{F}x}|,|k_y|) = (0.625\pi, \frac{\pi}{2})$
and $(0.256\pi, \pi)$, 
matching the free-electron values 
at $J_{\mr{K}} = 0$. By contrast, at large $J_{\mr{K}} \geq 2.8$,
we find Fermi points only at $(\tfrac{\pi}{2},0)$, 
suggesting a FS reconstruction at some $J_{\mr{Kc}}$ in between. Note also that $k_{\mr{F}x}(k_y)$ remains $J_\mr{K}$-independent
in each of the two
regimes. This is expected from Luttinger's 
sum rule~\cite{Luttinger1960,Oshikawa2000}, which links the effective number $n_{\mr{eff}}$ of mobile charge carriers 
(defined modulo 2, i.e.\ up to filled bands) to the
FS volume (see Ref.~\onlinecite{supplement}, Sec.~S-4~C for details).
 For small $J_{\mr{K}} \!\leq \! 0.75$, we find $n_{\mr{eff}} \!= \! 1.25$, consistent with $25\%$ electron doping. By contrast, at large $J_{\mr{K}} \! \geq \! 2.8$  we find $n_{\mr{eff}} = 0.25 =  2.25\,\mr{mod} \, 2$,
 consistent with the spins becoming mobile charge carriers by ``binding'' to the electrons~\cite{Si2014}. Pinpointing and studying a possible KB--QCP separating the two phases is left for future work.
 
 %
 \begin{figure}[t!]
 \includegraphics[width=\linewidth]{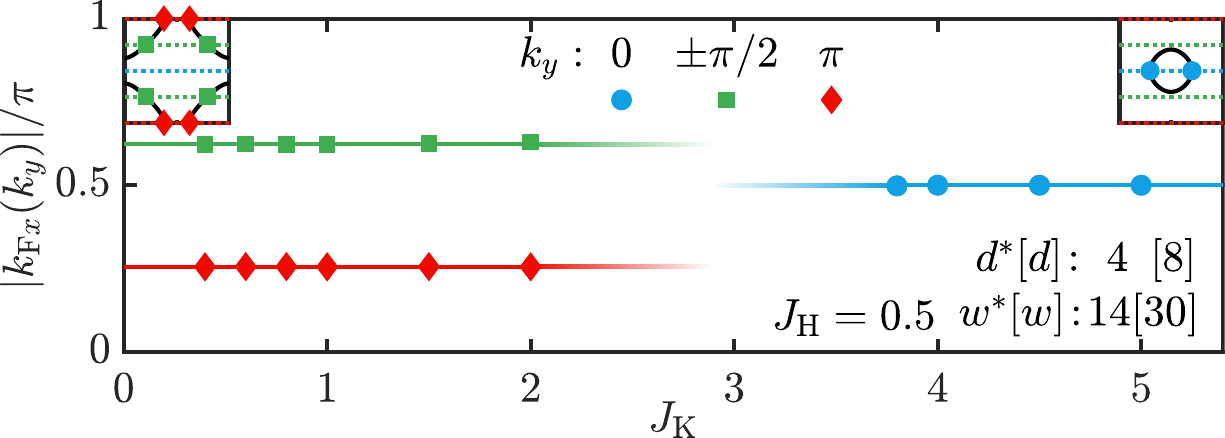}\vspace{-2mm}
 \caption{\label{fig:KondoCyl_FS_main}
Kondo-Heisenberg (KH) cylinder:
Fermi wavevectors $|k_{\mr{F}x}(k_y)|$ for a $40\times4$ KH cylinder for various values of $J_{\mr{K}}$.
Symbols are data points (error bars are below symbol size), lines are guides to the eye.
In the insets, black lines sketch the presumed FS  for $\eLL_y \! \to \! 
\infty$, dotted lines show the $k_y$ values allowed for $\eLL_y=4$.
 \vspace{-6mm}}
\end{figure}



%
\textit{Summary and outlook.---}
CBE expands bonds by adding subspaces on which  $\variance^\mtwosite$, the \twosite\ contribution to the energy variance, 
has significant weight, thus making these subspaces accessible to \onesite\ energy minimization.
CBE is fully variational 
and has 1s costs, since the variational space is
only slightly expanded relative to 1s DMRG.

By significantly saving costs,
CBE opens the door to studying challenging models of current interest at higher accuracy (larger $D$) than previously possible, or tackling more complex models, with $d$ or $w$ so large that they were hitherto out of reach. Examples are multi-band models with several different type of couplings, in particular in two-dimensional settings, models involving bosonic excitations, and quantum-chemical applications. 
We have made a first step in this direction by showing that the KH model on a width 4 cylinder features two phases with distinct FS volumes. Our study of the KH model opens the door to investigate this model in more depth; for example, follow-up work may aim to sort out 
the range of applicability of existing approximate approaches, e.g. parton mean-field theories~\cite{Senthil2003,Senthil2004} 
or DMFT based studies~\cite{DeLeo2008,DeLeo2008a,Tanaskovic2011,Si2001,Si2003}.
More generally, CBE can be used for any variational MPS optimization task.
Besides energy minimization, an example is approximating
a given $\Psi$ by a $\Psi'$ with smaller bond dimension through minimization
of $||\Psi' - \Psi||$. 
CBE can also be used to build Krylov spaces with \twosite\  accuracy at \onesite\ costs, 
relevant for 
all of the many 
MPS methods relying on Krylov methods. For example, in a follow-up paper \cite{Li2022} we focus on MPS time evolution using the time-dependent variational principle (TDVP), and use CBE to achieve dramatic improvements in performance. 
Finally, analogous statements hold for 
variational optimization or 
time evolution of MPOs. Thus, 
CBE will become a widely-used, indispensable tool in the MPS/MPO toolbox. 
\begin{acknowledgements}
We thank A.~Weichselbaum for inspiring discussions
and S.-S.B.~Lee, J.~Espinoza, M.~Lotem, J.~Shim and
A. Weichselbaum for comments on our manuscript.
This work was funded in part by the Deutsche Forschungsgemeinschaft under 
Germany's Excellence Strategy EXC-2111 (Project No.\ 390814868). It is part of the  Munich Quantum Valley, supported by the Bavarian state government with funds from the Hightech Agenda Bayern Plus. 
\end{acknowledgements}

\vspace{-4mm}
\bibliography{CBE-DMRG}

\begin{thebibliography}{76}%
\makeatletter
\providecommand \@ifxundefined [1]{%
 \@ifx{#1\undefined}
}%
\providecommand \@ifnum [1]{%
 \ifnum #1\expandafter \@firstoftwo
 \else \expandafter \@secondoftwo
 \fi
}%
\providecommand \@ifx [1]{%
 \ifx #1\expandafter \@firstoftwo
 \else \expandafter \@secondoftwo
 \fi
}%
\providecommand \natexlab [1]{#1}%
\providecommand \enquote  [1]{``#1''}%
\providecommand \bibnamefont  [1]{#1}%
\providecommand \bibfnamefont [1]{#1}%
\providecommand \citenamefont [1]{#1}%
\providecommand \href@noop [0]{\@secondoftwo}%
\providecommand \href [0]{\begingroup \@sanitize@url \@href}%
\providecommand \@href[1]{\@@startlink{#1}\@@href}%
\providecommand \@@href[1]{\endgroup#1\@@endlink}%
\providecommand \@sanitize@url [0]{\catcode `\\12\catcode `\$12\catcode
  `\&12\catcode `\#12\catcode `\^12\catcode `\_12\catcode `\%12\relax}%
\providecommand \@@startlink[1]{}%
\providecommand \@@endlink[0]{}%
\providecommand \url  [0]{\begingroup\@sanitize@url \@url }%
\providecommand \@url [1]{\endgroup\@href {#1}{\urlprefix }}%
\providecommand \urlprefix  [0]{URL }%
\providecommand \Eprint [0]{\href }%
\providecommand \doibase [0]{https://doi.org/}%
\providecommand \selectlanguage [0]{\@gobble}%
\providecommand \bibinfo  [0]{\@secondoftwo}%
\providecommand \bibfield  [0]{\@secondoftwo}%
\providecommand \translation [1]{[#1]}%
\providecommand \BibitemOpen [0]{}%
\providecommand \bibitemStop [0]{}%
\providecommand \bibitemNoStop [0]{.\EOS\space}%
\providecommand \EOS [0]{\spacefactor3000\relax}%
\providecommand \BibitemShut  [1]{\csname bibitem#1\endcsname}%
\let\auto@bib@innerbib\@empty
\bibitem [{\citenamefont {White}(1992)}]{White1992}%
  \BibitemOpen
  \bibfield  {author} {\bibinfo {author} {\bibfnamefont {S.~R.}\ \bibnamefont
  {White}},\ }\bibfield  {title} {\bibinfo {title} {Density matrix formulation
  for quantum renormalization groups},\ }\href
  {https://doi.org/10.1103/PhysRevLett.69.2863} {\bibfield  {journal} {\bibinfo
   {journal} {Phys. Rev. Lett.}\ }\textbf {\bibinfo {volume} {69}},\ \bibinfo
  {pages} {2863} (\bibinfo {year} {1992})}\BibitemShut {NoStop}%
\bibitem [{\citenamefont {White}(1993)}]{White1993}%
  \BibitemOpen
  \bibfield  {author} {\bibinfo {author} {\bibfnamefont {S.~R.}\ \bibnamefont
  {White}},\ }\bibfield  {title} {\bibinfo {title} {Density-matrix algorithms
  for quantum renormalization groups},\ }\href
  {https://doi.org/10.1103/PhysRevB.48.10345} {\bibfield  {journal} {\bibinfo
  {journal} {Phys. Rev. B}\ }\textbf {\bibinfo {volume} {48}},\ \bibinfo
  {pages} {10345} (\bibinfo {year} {1993})}\BibitemShut {NoStop}%
\bibitem [{\citenamefont {Verstraete}\ \emph {et~al.}(2004)\citenamefont
  {Verstraete}, \citenamefont {Porras},\ and\ \citenamefont
  {Cirac}}]{Verstraete2004}%
  \BibitemOpen
  \bibfield  {author} {\bibinfo {author} {\bibfnamefont {F.}~\bibnamefont
  {Verstraete}}, \bibinfo {author} {\bibfnamefont {D.}~\bibnamefont {Porras}},\
  and\ \bibinfo {author} {\bibfnamefont {J.~I.}\ \bibnamefont {Cirac}},\
  }\bibfield  {title} {\bibinfo {title} {Density matrix renormalization group
  and periodic boundary conditions: A quantum information perspective},\ }\href
  {https://doi.org/10.1103/PhysRevLett.93.227205} {\bibfield  {journal}
  {\bibinfo  {journal} {Phys. Rev. Lett.}\ }\textbf {\bibinfo {volume} {93}},\
  \bibinfo {pages} {227205} (\bibinfo {year} {2004})}\BibitemShut {NoStop}%
\bibitem [{\citenamefont {Schollw\"ock}(2005)}]{Schollwoeck2005}%
  \BibitemOpen
  \bibfield  {author} {\bibinfo {author} {\bibfnamefont {U.}~\bibnamefont
  {Schollw\"ock}},\ }\bibfield  {title} {\bibinfo {title} {The density-matrix
  renormalization group},\ }\href {https://doi.org/10.1103/RevModPhys.77.259}
  {\bibfield  {journal} {\bibinfo  {journal} {Rev. Mod. Phys.}\ }\textbf
  {\bibinfo {volume} {77}},\ \bibinfo {pages} {259} (\bibinfo {year}
  {2005})}\BibitemShut {NoStop}%
\bibitem [{\citenamefont {Schollw\"ock}(2011)}]{Schollwoeck2011}%
  \BibitemOpen
  \bibfield  {author} {\bibinfo {author} {\bibfnamefont {U.}~\bibnamefont
  {Schollw\"ock}},\ }\bibfield  {title} {\bibinfo {title} {The density-matrix
  renormalization group in the age of matrix product states},\ }\href
  {http://doi.org/10.1016/j.aop.2010.09.012} {\bibfield  {journal} {\bibinfo
  {journal} {Annals of Physics}\ }\textbf {\bibinfo {volume} {326}},\ \bibinfo
  {pages} {96} (\bibinfo {year} {2011})}\BibitemShut {NoStop}%
\bibitem [{\citenamefont {White}(1996)}]{White1996}%
  \BibitemOpen
  \bibfield  {author} {\bibinfo {author} {\bibfnamefont {S.~R.}\ \bibnamefont
  {White}},\ }\bibfield  {title} {\bibinfo {title} {Spin gaps in a frustrated
  {H}eisenberg model for cav${}_4$o${}_9$},\ }\href
  {https://doi.org/10.1103/PhysRevLett.77.3633} {\bibfield  {journal} {\bibinfo
   {journal} {Phys. Rev. Lett.}\ }\textbf {\bibinfo {volume} {77}},\ \bibinfo
  {pages} {3633} (\bibinfo {year} {1996})}\BibitemShut {NoStop}%
\bibitem [{\citenamefont {Stoudenmire}\ and\ \citenamefont
  {White}(2012)}]{Stoudenmire2012}%
  \BibitemOpen
  \bibfield  {author} {\bibinfo {author} {\bibfnamefont {E.}~\bibnamefont
  {Stoudenmire}}\ and\ \bibinfo {author} {\bibfnamefont {S.~R.}\ \bibnamefont
  {White}},\ }\bibfield  {title} {\bibinfo {title} {Studying two-dimensional
  systems with the density matrix renormalization group},\ }\href
  {https://doi.org/10.1146/annurev-conmatphys-020911-125018} {\bibfield
  {journal} {\bibinfo  {journal} {Ann. Rev. Cond. Mat. Phys.}\ }\textbf
  {\bibinfo {volume} {3}},\ \bibinfo {pages} {111} (\bibinfo {year}
  {2012})}\BibitemShut {NoStop}%
\bibitem [{\citenamefont {White}\ and\ \citenamefont
  {Scalapino}(1998)}]{White1998}%
  \BibitemOpen
  \bibfield  {author} {\bibinfo {author} {\bibfnamefont {S.~R.}\ \bibnamefont
  {White}}\ and\ \bibinfo {author} {\bibfnamefont {D.~J.}\ \bibnamefont
  {Scalapino}},\ }\bibfield  {title} {\bibinfo {title} {Density matrix
  renormalization group study of the striped phase in the 2d ${t-J}$ model},\
  }\href {https://doi.org/10.1103/PhysRevLett.80.1272} {\bibfield  {journal}
  {\bibinfo  {journal} {Phys. Rev. Lett.}\ }\textbf {\bibinfo {volume} {80}},\
  \bibinfo {pages} {1272} (\bibinfo {year} {1998})}\BibitemShut {NoStop}%
\bibitem [{\citenamefont {White}\ and\ \citenamefont
  {Scalapino}(2004)}]{White2004}%
  \BibitemOpen
  \bibfield  {author} {\bibinfo {author} {\bibfnamefont {S.~R.}\ \bibnamefont
  {White}}\ and\ \bibinfo {author} {\bibfnamefont {D.~J.}\ \bibnamefont
  {Scalapino}},\ }\bibfield  {title} {\bibinfo {title} {Checkerboard patterns
  in the {$t\text{\ensuremath{-}}J$} model},\ }\href
  {https://doi.org/10.1103/PhysRevB.70.220506} {\bibfield  {journal} {\bibinfo
  {journal} {Phys. Rev. B}\ }\textbf {\bibinfo {volume} {70}},\ \bibinfo
  {pages} {220506} (\bibinfo {year} {2004})}\BibitemShut {NoStop}%
\bibitem [{\citenamefont {White}\ and\ \citenamefont
  {Scalapino}(2009)}]{White2009a}%
  \BibitemOpen
  \bibfield  {author} {\bibinfo {author} {\bibfnamefont {S.~R.}\ \bibnamefont
  {White}}\ and\ \bibinfo {author} {\bibfnamefont {D.~J.}\ \bibnamefont
  {Scalapino}},\ }\bibfield  {title} {\bibinfo {title} {Pairing on striped
  $t$-$t'$-${J}$ lattices},\ }\href
  {https://doi.org/10.1103/PhysRevB.79.220504} {\bibfield  {journal} {\bibinfo
  {journal} {Phys. Rev. B}\ }\textbf {\bibinfo {volume} {79}},\ \bibinfo
  {pages} {220504} (\bibinfo {year} {2009})}\BibitemShut {NoStop}%
\bibitem [{\citenamefont {Jiang}\ \emph {et~al.}(2021)\citenamefont {Jiang},
  \citenamefont {Scalapino},\ and\ \citenamefont {White}}]{Jiang2021}%
  \BibitemOpen
  \bibfield  {author} {\bibinfo {author} {\bibfnamefont {S.}~\bibnamefont
  {Jiang}}, \bibinfo {author} {\bibfnamefont {D.~J.}\ \bibnamefont
  {Scalapino}},\ and\ \bibinfo {author} {\bibfnamefont {S.~R.}\ \bibnamefont
  {White}},\ }\bibfield  {title} {\bibinfo {title} {Ground-state phase diagram
  of the $t$-$t'$-${J}$ model},\ }\href
  {https://www.pnas.org/doi/abs/10.1073/pnas.2109978118} {\bibfield  {journal}
  {\bibinfo  {journal} {Proceedings of the National Academy of Sciences}\
  }\textbf {\bibinfo {volume} {118}},\ \bibinfo {pages} {e2109978118} (\bibinfo
  {year} {2021})}\BibitemShut {NoStop}%
\bibitem [{\citenamefont {LeBlanc}\ \emph {et~al.}(2015)\citenamefont
  {LeBlanc}, \citenamefont {Antipov}, \citenamefont {Becca}, \citenamefont
  {Bulik}, \citenamefont {Chan}, \citenamefont {Chung}, \citenamefont {Deng},
  \citenamefont {Ferrero}, \citenamefont {Henderson}, \citenamefont
  {Jim\'enez-Hoyos}, \citenamefont {Kozik}, \citenamefont {Liu}, \citenamefont
  {Millis}, \citenamefont {Prokof'ev}, \citenamefont {Qin}, \citenamefont
  {Scuseria}, \citenamefont {Shi}, \citenamefont {Svistunov}, \citenamefont
  {Tocchio}, \citenamefont {Tupitsyn}, \citenamefont {White}, \citenamefont
  {Zhang}, \citenamefont {Zheng}, \citenamefont {Zhu},\ and\ \citenamefont
  {Gull}}]{LeBlanc2015}%
  \BibitemOpen
  \bibfield  {author} {\bibinfo {author} {\bibfnamefont {J.~P.~F.}\
  \bibnamefont {LeBlanc}}, \bibinfo {author} {\bibfnamefont {A.~E.}\
  \bibnamefont {Antipov}}, \bibinfo {author} {\bibfnamefont {F.}~\bibnamefont
  {Becca}}, \bibinfo {author} {\bibfnamefont {I.~W.}\ \bibnamefont {Bulik}},
  \bibinfo {author} {\bibfnamefont {G.~K.-L.}\ \bibnamefont {Chan}}, \bibinfo
  {author} {\bibfnamefont {C.-M.}\ \bibnamefont {Chung}}, \bibinfo {author}
  {\bibfnamefont {Y.}~\bibnamefont {Deng}}, \bibinfo {author} {\bibfnamefont
  {M.}~\bibnamefont {Ferrero}}, \bibinfo {author} {\bibfnamefont {T.~M.}\
  \bibnamefont {Henderson}}, \bibinfo {author} {\bibfnamefont {C.~A.}\
  \bibnamefont {Jim\'enez-Hoyos}}, \bibinfo {author} {\bibfnamefont
  {E.}~\bibnamefont {Kozik}}, \bibinfo {author} {\bibfnamefont {X.-W.}\
  \bibnamefont {Liu}}, \bibinfo {author} {\bibfnamefont {A.~J.}\ \bibnamefont
  {Millis}}, \bibinfo {author} {\bibfnamefont {N.~V.}\ \bibnamefont
  {Prokof'ev}}, \bibinfo {author} {\bibfnamefont {M.}~\bibnamefont {Qin}},
  \bibinfo {author} {\bibfnamefont {G.~E.}\ \bibnamefont {Scuseria}}, \bibinfo
  {author} {\bibfnamefont {H.}~\bibnamefont {Shi}}, \bibinfo {author}
  {\bibfnamefont {B.~V.}\ \bibnamefont {Svistunov}}, \bibinfo {author}
  {\bibfnamefont {L.~F.}\ \bibnamefont {Tocchio}}, \bibinfo {author}
  {\bibfnamefont {I.~S.}\ \bibnamefont {Tupitsyn}}, \bibinfo {author}
  {\bibfnamefont {S.~R.}\ \bibnamefont {White}}, \bibinfo {author}
  {\bibfnamefont {S.}~\bibnamefont {Zhang}}, \bibinfo {author} {\bibfnamefont
  {B.-X.}\ \bibnamefont {Zheng}}, \bibinfo {author} {\bibfnamefont
  {Z.}~\bibnamefont {Zhu}},\ and\ \bibinfo {author} {\bibfnamefont
  {E.}~\bibnamefont {Gull}} (\bibinfo {collaboration} {Simons Collaboration on
  the Many-Electron Problem}),\ }\bibfield  {title} {\bibinfo {title}
  {Solutions of the two-dimensional {H}ubbard model: Benchmarks and results
  from a wide range of numerical algorithms},\ }\href
  {https://doi.org/10.1103/PhysRevX.5.041041} {\bibfield  {journal} {\bibinfo
  {journal} {Phys. Rev. X}\ }\textbf {\bibinfo {volume} {5}},\ \bibinfo {pages}
  {041041} (\bibinfo {year} {2015})}\BibitemShut {NoStop}%
\bibitem [{\citenamefont {Ehlers}\ \emph {et~al.}(2017)\citenamefont {Ehlers},
  \citenamefont {White},\ and\ \citenamefont {Noack}}]{Ehlers2017}%
  \BibitemOpen
  \bibfield  {author} {\bibinfo {author} {\bibfnamefont {G.}~\bibnamefont
  {Ehlers}}, \bibinfo {author} {\bibfnamefont {S.~R.}\ \bibnamefont {White}},\
  and\ \bibinfo {author} {\bibfnamefont {R.~M.}\ \bibnamefont {Noack}},\
  }\bibfield  {title} {\bibinfo {title} {Hybrid-space density matrix
  renormalization group study of the doped two-dimensional {H}ubbard model},\
  }\href {https://doi.org/10.1103/PhysRevB.95.125125} {\bibfield  {journal}
  {\bibinfo  {journal} {Phys. Rev. B}\ }\textbf {\bibinfo {volume} {95}},\
  \bibinfo {pages} {125125} (\bibinfo {year} {2017})}\BibitemShut {NoStop}%
\bibitem [{\citenamefont {Zheng}\ \emph {et~al.}(2017)\citenamefont {Zheng},
  \citenamefont {Chung}, \citenamefont {Corboz}, \citenamefont {Ehlers},
  \citenamefont {Qin}, \citenamefont {Noack}, \citenamefont {Shi},
  \citenamefont {White}, \citenamefont {Zhang},\ and\ \citenamefont
  {Chan}}]{Zheng2017}%
  \BibitemOpen
  \bibfield  {author} {\bibinfo {author} {\bibfnamefont {B.-X.}\ \bibnamefont
  {Zheng}}, \bibinfo {author} {\bibfnamefont {C.-M.}\ \bibnamefont {Chung}},
  \bibinfo {author} {\bibfnamefont {P.}~\bibnamefont {Corboz}}, \bibinfo
  {author} {\bibfnamefont {G.}~\bibnamefont {Ehlers}}, \bibinfo {author}
  {\bibfnamefont {M.-P.}\ \bibnamefont {Qin}}, \bibinfo {author} {\bibfnamefont
  {R.~M.}\ \bibnamefont {Noack}}, \bibinfo {author} {\bibfnamefont
  {H.}~\bibnamefont {Shi}}, \bibinfo {author} {\bibfnamefont {S.~R.}\
  \bibnamefont {White}}, \bibinfo {author} {\bibfnamefont {S.}~\bibnamefont
  {Zhang}},\ and\ \bibinfo {author} {\bibfnamefont {G.~K.-L.}\ \bibnamefont
  {Chan}},\ }\bibfield  {title} {\bibinfo {title} {Stripe order in the
  underdoped region of the two-dimensional {H}ubbard model},\ }\href
  {https://science.sciencemag.org/content/358/6367/1155} {\bibfield  {journal}
  {\bibinfo  {journal} {Science}\ }\textbf {\bibinfo {volume} {358}},\ \bibinfo
  {pages} {1155} (\bibinfo {year} {2017})}\BibitemShut {NoStop}%
\bibitem [{\citenamefont {Huang}\ \emph {et~al.}(2018)\citenamefont {Huang},
  \citenamefont {Mendl}, \citenamefont {Jiang}, \citenamefont {Moritz},\ and\
  \citenamefont {Devereaux}}]{Huang2018}%
  \BibitemOpen
  \bibfield  {author} {\bibinfo {author} {\bibfnamefont {E.~W.}\ \bibnamefont
  {Huang}}, \bibinfo {author} {\bibfnamefont {C.~B.}\ \bibnamefont {Mendl}},
  \bibinfo {author} {\bibfnamefont {H.-C.}\ \bibnamefont {Jiang}}, \bibinfo
  {author} {\bibfnamefont {B.}~\bibnamefont {Moritz}},\ and\ \bibinfo {author}
  {\bibfnamefont {T.~P.}\ \bibnamefont {Devereaux}},\ }\bibfield  {title}
  {\bibinfo {title} {Stripe order from the perspective of the {H}ubbard
  model},\ }\href {https://doi.org/10.1038/s41535-018-0097-0} {\bibfield
  {journal} {\bibinfo  {journal} {npj Quantum Materials}\ }\textbf {\bibinfo
  {volume} {3}},\ \bibinfo {pages} {22} (\bibinfo {year} {2018})}\BibitemShut
  {NoStop}%
\bibitem [{\citenamefont {Qin}\ \emph {et~al.}(2020)\citenamefont {Qin},
  \citenamefont {Chung}, \citenamefont {Shi}, \citenamefont {Vitali},
  \citenamefont {Hubig}, \citenamefont {Schollw\"ock}, \citenamefont {White},\
  and\ \citenamefont {Zhang}}]{Qin2020}%
  \BibitemOpen
  \bibfield  {author} {\bibinfo {author} {\bibfnamefont {M.}~\bibnamefont
  {Qin}}, \bibinfo {author} {\bibfnamefont {C.-M.}\ \bibnamefont {Chung}},
  \bibinfo {author} {\bibfnamefont {H.}~\bibnamefont {Shi}}, \bibinfo {author}
  {\bibfnamefont {E.}~\bibnamefont {Vitali}}, \bibinfo {author} {\bibfnamefont
  {C.}~\bibnamefont {Hubig}}, \bibinfo {author} {\bibfnamefont
  {U.}~\bibnamefont {Schollw\"ock}}, \bibinfo {author} {\bibfnamefont {S.~R.}\
  \bibnamefont {White}},\ and\ \bibinfo {author} {\bibfnamefont
  {S.}~\bibnamefont {Zhang}} (\bibinfo {collaboration} {Simons Collaboration on
  the Many-Electron Problem}),\ }\bibfield  {title} {\bibinfo {title} {Absence
  of superconductivity in the pure two-dimensional {H}ubbard model},\ }\href
  {https://doi.org/10.1103/PhysRevX.10.031016} {\bibfield  {journal} {\bibinfo
  {journal} {Phys. Rev. X}\ }\textbf {\bibinfo {volume} {10}},\ \bibinfo
  {pages} {031016} (\bibinfo {year} {2020})}\BibitemShut {NoStop}%
\bibitem [{\citenamefont {Jiang}\ \emph {et~al.}(2020)\citenamefont {Jiang},
  \citenamefont {Zaanen}, \citenamefont {Devereaux},\ and\ \citenamefont
  {Jiang}}]{Jiang2020}%
  \BibitemOpen
  \bibfield  {author} {\bibinfo {author} {\bibfnamefont {Y.-F.}\ \bibnamefont
  {Jiang}}, \bibinfo {author} {\bibfnamefont {J.}~\bibnamefont {Zaanen}},
  \bibinfo {author} {\bibfnamefont {T.~P.}\ \bibnamefont {Devereaux}},\ and\
  \bibinfo {author} {\bibfnamefont {H.-C.}\ \bibnamefont {Jiang}},\ }\bibfield
  {title} {\bibinfo {title} {Ground state phase diagram of the doped {H}ubbard
  model on the four-leg cylinder},\ }\href
  {https://doi.org/10.1103/PhysRevResearch.2.033073} {\bibfield  {journal}
  {\bibinfo  {journal} {Phys. Rev. Research}\ }\textbf {\bibinfo {volume}
  {2}},\ \bibinfo {pages} {033073} (\bibinfo {year} {2020})}\BibitemShut
  {NoStop}%
\bibitem [{\citenamefont {Jiang}\ and\ \citenamefont
  {Kivelson}(2022)}]{Jiang2022}%
  \BibitemOpen
  \bibfield  {author} {\bibinfo {author} {\bibfnamefont {H.-C.}\ \bibnamefont
  {Jiang}}\ and\ \bibinfo {author} {\bibfnamefont {S.~A.}\ \bibnamefont
  {Kivelson}},\ }\bibfield  {title} {\bibinfo {title} {Stripe order enhanced
  superconductivity in the {H}ubbard model},\ }\href
  {https://www.pnas.org/doi/abs/10.1073/pnas.2109406119} {\bibfield  {journal}
  {\bibinfo  {journal} {PNAS}\ }\textbf {\bibinfo {volume} {119}},\ \bibinfo
  {pages} {e2109406119} (\bibinfo {year} {2022})}\BibitemShut {NoStop}%
\bibitem [{\citenamefont {Yan}\ \emph {et~al.}(2011)\citenamefont {Yan},
  \citenamefont {Huse},\ and\ \citenamefont {White}}]{Yan2011}%
  \BibitemOpen
  \bibfield  {author} {\bibinfo {author} {\bibfnamefont {S.}~\bibnamefont
  {Yan}}, \bibinfo {author} {\bibfnamefont {D.~A.}\ \bibnamefont {Huse}},\ and\
  \bibinfo {author} {\bibfnamefont {S.~R.}\ \bibnamefont {White}},\ }\bibfield
  {title} {\bibinfo {title} {Spin-liquid ground state of the ${S = 1/2}$ kagome
  {H}eisenberg antiferromagnet},\ }\href
  {https://www.science.org/doi/abs/10.1126/science.1201080} {\bibfield
  {journal} {\bibinfo  {journal} {Science}\ }\textbf {\bibinfo {volume}
  {332}},\ \bibinfo {pages} {1173} (\bibinfo {year} {2011})}\BibitemShut
  {NoStop}%
\bibitem [{\citenamefont {Depenbrock}\ \emph {et~al.}(2012)\citenamefont
  {Depenbrock}, \citenamefont {McCulloch},\ and\ \citenamefont
  {Schollw\"ock}}]{Depenbrock2012}%
  \BibitemOpen
  \bibfield  {author} {\bibinfo {author} {\bibfnamefont {S.}~\bibnamefont
  {Depenbrock}}, \bibinfo {author} {\bibfnamefont {I.~P.}\ \bibnamefont
  {McCulloch}},\ and\ \bibinfo {author} {\bibfnamefont {U.}~\bibnamefont
  {Schollw\"ock}},\ }\bibfield  {title} {\bibinfo {title} {Nature of the
  spin-liquid ground state of the ${S}=1/2$ {H}eisenberg model on the kagome
  lattice},\ }\href {https://doi.org/10.1103/PhysRevLett.109.067201} {\bibfield
   {journal} {\bibinfo  {journal} {Phys. Rev. Lett.}\ }\textbf {\bibinfo
  {volume} {109}},\ \bibinfo {pages} {067201} (\bibinfo {year}
  {2012})}\BibitemShut {NoStop}%
\bibitem [{\citenamefont {Kolley}\ \emph {et~al.}(2015)\citenamefont {Kolley},
  \citenamefont {Depenbrock}, \citenamefont {McCulloch}, \citenamefont
  {Schollw\"ock},\ and\ \citenamefont {Alba}}]{Kolley2015}%
  \BibitemOpen
  \bibfield  {author} {\bibinfo {author} {\bibfnamefont {F.}~\bibnamefont
  {Kolley}}, \bibinfo {author} {\bibfnamefont {S.}~\bibnamefont {Depenbrock}},
  \bibinfo {author} {\bibfnamefont {I.~P.}\ \bibnamefont {McCulloch}}, \bibinfo
  {author} {\bibfnamefont {U.}~\bibnamefont {Schollw\"ock}},\ and\ \bibinfo
  {author} {\bibfnamefont {V.}~\bibnamefont {Alba}},\ }\bibfield  {title}
  {\bibinfo {title} {Phase diagram of the ${J}_{1}\text{-}{J}_{2}$ {H}eisenberg
  model on the kagome lattice},\ }\href
  {https://doi.org/10.1103/PhysRevB.91.104418} {\bibfield  {journal} {\bibinfo
  {journal} {Phys. Rev. B}\ }\textbf {\bibinfo {volume} {91}},\ \bibinfo
  {pages} {104418} (\bibinfo {year} {2015})}\BibitemShut {NoStop}%
\bibitem [{\citenamefont {He}\ \emph {et~al.}(2017)\citenamefont {He},
  \citenamefont {Zaletel}, \citenamefont {Oshikawa},\ and\ \citenamefont
  {Pollmann}}]{He2017a}%
  \BibitemOpen
  \bibfield  {author} {\bibinfo {author} {\bibfnamefont {Y.-C.}\ \bibnamefont
  {He}}, \bibinfo {author} {\bibfnamefont {M.~P.}\ \bibnamefont {Zaletel}},
  \bibinfo {author} {\bibfnamefont {M.}~\bibnamefont {Oshikawa}},\ and\
  \bibinfo {author} {\bibfnamefont {F.}~\bibnamefont {Pollmann}},\ }\bibfield
  {title} {\bibinfo {title} {Signatures of {Dirac} cones in a {DMRG} study of
  the kagome {H}eisenberg model},\ }\href
  {https://doi.org/10.1103/PhysRevX.7.031020} {\bibfield  {journal} {\bibinfo
  {journal} {Phys. Rev. X}\ }\textbf {\bibinfo {volume} {7}},\ \bibinfo {pages}
  {031020} (\bibinfo {year} {2017})}\BibitemShut {NoStop}%
\bibitem [{\citenamefont {\"Ostlund}\ and\ \citenamefont
  {Rommer}(1995)}]{Oestlund1995}%
  \BibitemOpen
  \bibfield  {author} {\bibinfo {author} {\bibfnamefont {S.}~\bibnamefont
  {\"Ostlund}}\ and\ \bibinfo {author} {\bibfnamefont {S.}~\bibnamefont
  {Rommer}},\ }\bibfield  {title} {\bibinfo {title} {Thermodynamic limit of
  density matrix renormalization},\ }\href
  {https://doi.org/10.1103/PhysRevLett.75.3537} {\bibfield  {journal} {\bibinfo
   {journal} {Phys. Rev. Lett.}\ }\textbf {\bibinfo {volume} {75}},\ \bibinfo
  {pages} {3537} (\bibinfo {year} {1995})}\BibitemShut {NoStop}%
\bibitem [{\citenamefont {Rommer}\ and\ \citenamefont
  {\"Ostlund}(1997)}]{Rommer1997}%
  \BibitemOpen
  \bibfield  {author} {\bibinfo {author} {\bibfnamefont {S.}~\bibnamefont
  {Rommer}}\ and\ \bibinfo {author} {\bibfnamefont {S.}~\bibnamefont
  {\"Ostlund}},\ }\bibfield  {title} {\bibinfo {title} {Class of ansatz wave
  functions for one-dimensional spin systems and their relation to the density
  matrix renormalization group},\ }\href
  {https://doi.org/10.1103/PhysRevB.55.2164} {\bibfield  {journal} {\bibinfo
  {journal} {Phys. Rev. B}\ }\textbf {\bibinfo {volume} {55}},\ \bibinfo
  {pages} {2164} (\bibinfo {year} {1997})}\BibitemShut {NoStop}%
\bibitem [{\citenamefont {White}(2005)}]{White2005}%
  \BibitemOpen
  \bibfield  {author} {\bibinfo {author} {\bibfnamefont {S.~R.}\ \bibnamefont
  {White}},\ }\bibfield  {title} {\bibinfo {title} {Density matrix
  renormalization group algorithms with a single center site},\ }\href
  {https://doi.org/10.1103/PhysRevB.72.180403} {\bibfield  {journal} {\bibinfo
  {journal} {Phys. Rev. B}\ }\textbf {\bibinfo {volume} {72}},\ \bibinfo
  {pages} {180403} (\bibinfo {year} {2005})}\BibitemShut {NoStop}%
\bibitem [{\citenamefont {Hubig}\ \emph {et~al.}(2015)\citenamefont {Hubig},
  \citenamefont {McCulloch}, \citenamefont {Schollw\"ock},\ and\ \citenamefont
  {Wolf}}]{Hubig2015}%
  \BibitemOpen
  \bibfield  {author} {\bibinfo {author} {\bibfnamefont {C.}~\bibnamefont
  {Hubig}}, \bibinfo {author} {\bibfnamefont {I.~P.}\ \bibnamefont
  {McCulloch}}, \bibinfo {author} {\bibfnamefont {U.}~\bibnamefont
  {Schollw\"ock}},\ and\ \bibinfo {author} {\bibfnamefont {F.~A.}\ \bibnamefont
  {Wolf}},\ }\bibfield  {title} {\bibinfo {title} {Strictly single-site {DMRG}
  algorithm with subspace expansion},\ }\href
  {https://doi.org/10.1103/PhysRevB.91.155115} {\bibfield  {journal} {\bibinfo
  {journal} {Phys. Rev. B}\ }\textbf {\bibinfo {volume} {91}},\ \bibinfo
  {pages} {155115} (\bibinfo {year} {2015})}\BibitemShut {NoStop}%
\bibitem [{\citenamefont {Gleis}\ \emph {et~al.}(2022)\citenamefont {Gleis},
  \citenamefont {Li},\ and\ \citenamefont {von Delft}}]{Gleis2022a}%
  \BibitemOpen
  \bibfield  {author} {\bibinfo {author} {\bibfnamefont {A.}~\bibnamefont
  {Gleis}}, \bibinfo {author} {\bibfnamefont {J.-W.}\ \bibnamefont {Li}},\ and\
  \bibinfo {author} {\bibfnamefont {J.}~\bibnamefont {von Delft}},\ }\bibfield
  {title} {\bibinfo {title} {Projector formalism for kept and discarded spaces
  of matrix product states},\ }\href
  {https://doi.org/10.1103/PhysRevB.106.195138} {\bibfield  {journal} {\bibinfo
   {journal} {Phys. Rev. B}\ }\textbf {\bibinfo {volume} {106}},\ \bibinfo
  {pages} {195138} (\bibinfo {year} {2022})}\BibitemShut {NoStop}%
\bibitem [{\citenamefont {Hubig}\ \emph {et~al.}(2018)\citenamefont {Hubig},
  \citenamefont {Haegeman},\ and\ \citenamefont {Schollw\"ock}}]{Hubig2018}%
  \BibitemOpen
  \bibfield  {author} {\bibinfo {author} {\bibfnamefont {C.}~\bibnamefont
  {Hubig}}, \bibinfo {author} {\bibfnamefont {J.}~\bibnamefont {Haegeman}},\
  and\ \bibinfo {author} {\bibfnamefont {U.}~\bibnamefont {Schollw\"ock}},\
  }\bibfield  {title} {\bibinfo {title} {Error estimates for extrapolations
  with matrix-product states},\ }\href
  {https://doi.org/10.1103/PhysRevB.97.045125} {\bibfield  {journal} {\bibinfo
  {journal} {Phys. Rev. B}\ }\textbf {\bibinfo {volume} {97}},\ \bibinfo
  {pages} {045125} (\bibinfo {year} {2018})}\BibitemShut {NoStop}%
\bibitem [{sup()}]{supplement}%
  \BibitemOpen
  \href@noop {} {\bibinfo  {journal} {See Supplemental Material at [url] for a
  detailed analysis of shrewd selection; a pseudocode for shrewd selection;
  additional simple benchmarks; a comparison to DMRG3S; and more details on the
  analysis of the Kondo-Heisenberg model on a 4-leg cylinder. The Supplemental
  Material includes
  Refs.~\cite{Jeckelmann1998,Tezuka2007,Fehske2008,Ejima2010,Reinhard2019,Coleman2007,Ye2022,Coleman2001,Luttinger1960,Senechal1999,Oshikawa2000,Si2014,Nishikawa2018}}\
  }\BibitemShut {NoStop}%
\bibitem [{Rel()}]{RelevantSubspace}%
  \BibitemOpen
\bibfield  {journal} {  }\href@noop {} {\bibinfo  {journal} {If \twosite\ DMRG
  has converged to an optimal MPS $\Psi_D$ with \textit{fixed} bond dimension
  $D$, the size of $\rDD$ is zero. Because $\Psi_D$ is already optimal~(at
  \textit{fixed} $D$), \textit{any} state in $\DD$ is less relevant than those
  already present in the kept space of $\Psi_D$. As a result, $\Psi_D$ cannot
  be further optimized unless $D$ is increased. Away from convergence, the size
  of $\rDD$ is usually still much smaller than the already somewhat optimized
  kept space, which in turn is much smaller than $\DD$. Expanding by
  $\rDD$~(CBE) instead of $\DD$~(\twosite) is similar in spirit to using an
  iterative eigensolver for Eq.~\eqref{subeq:SchroedingerMain} instead of full
  diagonalization}\ }\BibitemShut {NoStop}%
\bibitem [{Fir()}]{FirstTruncation}%
  \BibitemOpen
\bibfield  {journal} {  }\href@noop {} {\bibinfo  {journal} {We could achieve
  the desired reduction $\overline D \to \widetilde D$ already during
  preselection by choosing $D'= \widetilde D /w$ there, so that $\widehat D =
  \widetilde D$; however, that would neglect the information that in
  $H^{\mathrm{2s}} \psi^{\mathrm{2s}}$ the central MPO bond is closed. Final
  selection serves to include that information}\ }\BibitemShut {NoStop}%
\bibitem [{\citenamefont {Jeckelmann}\ and\ \citenamefont
  {White}(1998)}]{Jeckelmann1998}%
  \BibitemOpen
\bibfield  {journal} {  }\bibfield  {author} {\bibinfo {author} {\bibfnamefont
  {E.}~\bibnamefont {Jeckelmann}}\ and\ \bibinfo {author} {\bibfnamefont
  {S.~R.}\ \bibnamefont {White}},\ }\bibfield  {title} {\bibinfo {title}
  {Density-matrix renormalization-group study of the polaron problem in the
  {H}olstein model},\ }\href {https://doi.org/10.1103/PhysRevB.57.6376}
  {\bibfield  {journal} {\bibinfo  {journal} {Phys. Rev. B}\ }\textbf {\bibinfo
  {volume} {57}},\ \bibinfo {pages} {6376} (\bibinfo {year}
  {1998})}\BibitemShut {NoStop}%
\bibitem [{\citenamefont {Tezuka}\ \emph {et~al.}(2007)\citenamefont {Tezuka},
  \citenamefont {Arita},\ and\ \citenamefont {Aoki}}]{Tezuka2007}%
  \BibitemOpen
  \bibfield  {author} {\bibinfo {author} {\bibfnamefont {M.}~\bibnamefont
  {Tezuka}}, \bibinfo {author} {\bibfnamefont {R.}~\bibnamefont {Arita}},\ and\
  \bibinfo {author} {\bibfnamefont {H.}~\bibnamefont {Aoki}},\ }\bibfield
  {title} {\bibinfo {title} {Phase diagram for the one-dimensional
  {H}ubbard-{H}olstein model: A density-matrix renormalization group study},\
  }\href {https://doi.org/10.1103/PhysRevB.76.155114} {\bibfield  {journal}
  {\bibinfo  {journal} {Phys. Rev. B}\ }\textbf {\bibinfo {volume} {76}},\
  \bibinfo {pages} {155114} (\bibinfo {year} {2007})}\BibitemShut {NoStop}%
\bibitem [{\citenamefont {Fehske}\ \emph {et~al.}(2008)\citenamefont {Fehske},
  \citenamefont {Hager},\ and\ \citenamefont {Jeckelmann}}]{Fehske2008}%
  \BibitemOpen
  \bibfield  {author} {\bibinfo {author} {\bibfnamefont {H.}~\bibnamefont
  {Fehske}}, \bibinfo {author} {\bibfnamefont {G.}~\bibnamefont {Hager}},\ and\
  \bibinfo {author} {\bibfnamefont {E.}~\bibnamefont {Jeckelmann}},\ }\bibfield
   {title} {\bibinfo {title} {Metallicity in the half-filled
  {H}olstein-{H}ubbard model},\ }\href
  {https://iopscience.iop.org/article/10.1209/0295-5075/84/57001} {\bibfield
  {journal} {\bibinfo  {journal} {E. J. Phys.}\ }\textbf {\bibinfo {volume}
  {84}},\ \bibinfo {pages} {57001} (\bibinfo {year} {2008})}\BibitemShut
  {NoStop}%
\bibitem [{\citenamefont {Ejima}\ and\ \citenamefont
  {Fehske}(2010)}]{Ejima2010}%
  \BibitemOpen
  \bibfield  {author} {\bibinfo {author} {\bibfnamefont {S.}~\bibnamefont
  {Ejima}}\ and\ \bibinfo {author} {\bibfnamefont {H.}~\bibnamefont {Fehske}},\
  }\bibfield  {title} {\bibinfo {title} {{DMRG} analysis of the sdw-cdw
  crossover region in the 1d half-filled {H}ubbard-{H}olstein model},\ }\href
  {https://doi.org/10.1088/1742-6596/200/1/012031} {\bibfield  {journal}
  {\bibinfo  {journal} {J. Phys.: Conference Series}\ }\textbf {\bibinfo
  {volume} {200}},\ \bibinfo {pages} {012031} (\bibinfo {year}
  {2010})}\BibitemShut {NoStop}%
\bibitem [{\citenamefont {Reinhard}\ \emph {et~al.}(2019)\citenamefont
  {Reinhard}, \citenamefont {Mordovina}, \citenamefont {Hubig}, \citenamefont
  {Kretchmer}, \citenamefont {Schollw{\"o}ck}, \citenamefont {Appel},
  \citenamefont {Sentef},\ and\ \citenamefont {Rubio}}]{Reinhard2019}%
  \BibitemOpen
  \bibfield  {author} {\bibinfo {author} {\bibfnamefont {T.~E.}\ \bibnamefont
  {Reinhard}}, \bibinfo {author} {\bibfnamefont {U.}~\bibnamefont {Mordovina}},
  \bibinfo {author} {\bibfnamefont {C.}~\bibnamefont {Hubig}}, \bibinfo
  {author} {\bibfnamefont {J.~S.}\ \bibnamefont {Kretchmer}}, \bibinfo {author}
  {\bibfnamefont {U.}~\bibnamefont {Schollw{\"o}ck}}, \bibinfo {author}
  {\bibfnamefont {H.}~\bibnamefont {Appel}}, \bibinfo {author} {\bibfnamefont
  {M.~A.}\ \bibnamefont {Sentef}},\ and\ \bibinfo {author} {\bibfnamefont
  {A.}~\bibnamefont {Rubio}},\ }\bibfield  {title} {\bibinfo {title}
  {Density-matrix embedding theory study of the one-dimensional
  {H}ubbard--{H}olstein model},\ }\href
  {https://doi.org/10.1021/acs.jctc.8b01116} {\bibfield  {journal} {\bibinfo
  {journal} {J. Chem. Theory and Comp.}\ }\textbf {\bibinfo {volume} {15}},\
  \bibinfo {pages} {2221} (\bibinfo {year} {2019})}\BibitemShut {NoStop}%
\bibitem [{\citenamefont {Coleman}(2007)}]{Coleman2007}%
  \BibitemOpen
  \bibfield  {author} {\bibinfo {author} {\bibfnamefont {P.}~\bibnamefont
  {Coleman}},\ }\bibfield  {title} {\bibinfo {title} {Heavy fermions: Electrons
  at the edge of magnetism},\ }in\ \href
  {https://doi.org/10.1002/9780470022184.hmm105} {\emph {\bibinfo {booktitle}
  {Handbook of {Magnetism} and {Advanced} {Magnetic} {Materials}}}},\
  Vol.~\bibinfo {volume} {1},\ \bibinfo {editor} {edited by\ \bibinfo {editor}
  {\bibfnamefont {H.}~\bibnamefont {Kronm\"uller}}\ and\ \bibinfo {editor}
  {\bibfnamefont {S.}~\bibnamefont {Parkin}}}\ (\bibinfo  {publisher} {Wiley},\
  \bibinfo {year} {2007})\ pp.\ \bibinfo {pages} {95--148}\BibitemShut
  {NoStop}%
\bibitem [{\citenamefont {Ye}\ \emph {et~al.}(2022)\citenamefont {Ye},
  \citenamefont {Kung}, \citenamefont {Rosa}, \citenamefont {Bauer},
  \citenamefont {Haule},\ and\ \citenamefont {Blumberg}}]{Ye2022}%
  \BibitemOpen
  \bibfield  {author} {\bibinfo {author} {\bibfnamefont {M.}~\bibnamefont
  {Ye}}, \bibinfo {author} {\bibfnamefont {H.-H.}\ \bibnamefont {Kung}},
  \bibinfo {author} {\bibfnamefont {P.~F.~S.}\ \bibnamefont {Rosa}}, \bibinfo
  {author} {\bibfnamefont {E.~D.}\ \bibnamefont {Bauer}}, \bibinfo {author}
  {\bibfnamefont {K.}~\bibnamefont {Haule}},\ and\ \bibinfo {author}
  {\bibfnamefont {G.}~\bibnamefont {Blumberg}},\ }\bibfield  {title} {\bibinfo
  {title} {Anisotropy of {K}ondo-lattice coherence in momentum space for
  {$\mathrm{CeCoIn}{}_5$}},\ }\href {https://arxiv.org/abs/2202.09642}
  {\bibfield  {journal} {\bibinfo  {journal} {arXiv:2202.09642
  [cond-mat.str-el]}\ } (\bibinfo {year} {2022})}\BibitemShut {NoStop}%
\bibitem [{\citenamefont {Coleman}\ \emph {et~al.}(2001)\citenamefont
  {Coleman}, \citenamefont {P{\'e}pin}, \citenamefont {Si},\ and\ \citenamefont
  {Ramazashvili}}]{Coleman2001}%
  \BibitemOpen
  \bibfield  {author} {\bibinfo {author} {\bibfnamefont {P.}~\bibnamefont
  {Coleman}}, \bibinfo {author} {\bibfnamefont {C.}~\bibnamefont {P{\'e}pin}},
  \bibinfo {author} {\bibfnamefont {Q.}~\bibnamefont {Si}},\ and\ \bibinfo
  {author} {\bibfnamefont {R.}~\bibnamefont {Ramazashvili}},\ }\bibfield
  {title} {\bibinfo {title} {How do {Fermi} liquids get heavy and die?},\
  }\href {https://doi.org/10.1088/0953-8984/13/35/202} {\bibfield  {journal}
  {\bibinfo  {journal} {J. Phys. Cond. Mat.}\ }\textbf {\bibinfo {volume}
  {13}},\ \bibinfo {pages} {R723} (\bibinfo {year} {2001})}\BibitemShut
  {NoStop}%
\bibitem [{\citenamefont {Luttinger}(1960)}]{Luttinger1960}%
  \BibitemOpen
  \bibfield  {author} {\bibinfo {author} {\bibfnamefont {J.~M.}\ \bibnamefont
  {Luttinger}},\ }\bibfield  {title} {\bibinfo {title} {{Fermi} surface and
  some simple equilibrium properties of a system of interacting fermions},\
  }\href {https://doi.org/10.1103/PhysRev.119.1153} {\bibfield  {journal}
  {\bibinfo  {journal} {Phys. Rev.}\ }\textbf {\bibinfo {volume} {119}},\
  \bibinfo {pages} {1153} (\bibinfo {year} {1960})}\BibitemShut {NoStop}%
\bibitem [{\citenamefont {S\'en\'echal}(1999)}]{Senechal1999}%
  \BibitemOpen
  \bibfield  {author} {\bibinfo {author} {\bibfnamefont {D.}~\bibnamefont
  {S\'en\'echal}},\ }\bibfield  {title} {\bibinfo {title} {An introduction to
  bosonization}\ }\href {https://doi.org/10.48550/arxiv.cond-mat/9908262}
  {10.48550/arxiv.cond-mat/9908262} (\bibinfo {year} {1999})\BibitemShut
  {NoStop}%
\bibitem [{\citenamefont {Oshikawa}(2000)}]{Oshikawa2000}%
  \BibitemOpen
  \bibfield  {author} {\bibinfo {author} {\bibfnamefont {M.}~\bibnamefont
  {Oshikawa}},\ }\bibfield  {title} {\bibinfo {title} {Topological approach to
  {Luttinger}'s theorem and the {Fermi} surface of a {Kondo} lattice},\ }\href
  {https://doi.org/10.1103/PhysRevLett.84.3370} {\bibfield  {journal} {\bibinfo
   {journal} {Phys. Rev. Lett.}\ }\textbf {\bibinfo {volume} {84}},\ \bibinfo
  {pages} {3370} (\bibinfo {year} {2000})}\BibitemShut {NoStop}%
\bibitem [{\citenamefont {Si}\ \emph {et~al.}(2014)\citenamefont {Si},
  \citenamefont {Pixley}, \citenamefont {Nica}, \citenamefont {Yamamoto},
  \citenamefont {Goswami}, \citenamefont {Yu},\ and\ \citenamefont
  {Kirchner}}]{Si2014}%
  \BibitemOpen
  \bibfield  {author} {\bibinfo {author} {\bibfnamefont {Q.}~\bibnamefont
  {Si}}, \bibinfo {author} {\bibfnamefont {J.~H.}\ \bibnamefont {Pixley}},
  \bibinfo {author} {\bibfnamefont {E.}~\bibnamefont {Nica}}, \bibinfo {author}
  {\bibfnamefont {S.~J.}\ \bibnamefont {Yamamoto}}, \bibinfo {author}
  {\bibfnamefont {P.}~\bibnamefont {Goswami}}, \bibinfo {author} {\bibfnamefont
  {R.}~\bibnamefont {Yu}},\ and\ \bibinfo {author} {\bibfnamefont
  {S.}~\bibnamefont {Kirchner}},\ }\bibfield  {title} {\bibinfo {title}
  {{Kondo} destruction and quantum criticality in {Kondo} lattice systems},\
  }\href {https://doi.org/10.7566/JPSJ.83.061005} {\bibfield  {journal}
  {\bibinfo  {journal} {Journal of the Physical Society of Japan}\ }\textbf
  {\bibinfo {volume} {83}},\ \bibinfo {pages} {061005} (\bibinfo {year}
  {2014})}\BibitemShut {NoStop}%
\bibitem [{\citenamefont {Nishikawa}\ \emph {et~al.}(2018)\citenamefont
  {Nishikawa}, \citenamefont {Curtin}, \citenamefont {Hewson},\ and\
  \citenamefont {Crow}}]{Nishikawa2018}%
  \BibitemOpen
  \bibfield  {author} {\bibinfo {author} {\bibfnamefont {Y.}~\bibnamefont
  {Nishikawa}}, \bibinfo {author} {\bibfnamefont {O.~J.}\ \bibnamefont
  {Curtin}}, \bibinfo {author} {\bibfnamefont {A.~C.}\ \bibnamefont {Hewson}},\
  and\ \bibinfo {author} {\bibfnamefont {D.~J.~G.}\ \bibnamefont {Crow}},\
  }\bibfield  {title} {\bibinfo {title} {Magnetic field induced quantum
  criticality and the {Luttinger} sum rule},\ }\href
  {https://doi.org/10.1103/PhysRevB.98.104419} {\bibfield  {journal} {\bibinfo
  {journal} {Phys. Rev. B}\ }\textbf {\bibinfo {volume} {98}},\ \bibinfo
  {pages} {104419} (\bibinfo {year} {2018})}\BibitemShut {NoStop}%
\bibitem [{\citenamefont {Stoudenmire}\ and\ \citenamefont
  {White}(2013)}]{Stoudenmire2013}%
  \BibitemOpen
  \bibfield  {author} {\bibinfo {author} {\bibfnamefont {E.~M.}\ \bibnamefont
  {Stoudenmire}}\ and\ \bibinfo {author} {\bibfnamefont {S.~R.}\ \bibnamefont
  {White}},\ }\bibfield  {title} {\bibinfo {title} {Real-space parallel density
  matrix renormalization group},\ }\href
  {https://doi.org/10.1103/PhysRevB.87.155137} {\bibfield  {journal} {\bibinfo
  {journal} {Phys. Rev. B}\ }\textbf {\bibinfo {volume} {87}},\ \bibinfo
  {pages} {155137} (\bibinfo {year} {2013})}\BibitemShut {NoStop}%
\bibitem [{\citenamefont {Zauner-Stauber}\ \emph {et~al.}(2018)\citenamefont
  {Zauner-Stauber}, \citenamefont {Vanderstraeten}, \citenamefont {Fishman},
  \citenamefont {Verstraete},\ and\ \citenamefont
  {Haegeman}}]{ZaunerStauber2018}%
  \BibitemOpen
  \bibfield  {author} {\bibinfo {author} {\bibfnamefont {V.}~\bibnamefont
  {Zauner-Stauber}}, \bibinfo {author} {\bibfnamefont {L.}~\bibnamefont
  {Vanderstraeten}}, \bibinfo {author} {\bibfnamefont {M.~T.}\ \bibnamefont
  {Fishman}}, \bibinfo {author} {\bibfnamefont {F.}~\bibnamefont
  {Verstraete}},\ and\ \bibinfo {author} {\bibfnamefont {J.}~\bibnamefont
  {Haegeman}},\ }\bibfield  {title} {\bibinfo {title} {Variational optimization
  algorithms for uniform matrix product states},\ }\href
  {https://doi.org/10.1103/PhysRevB.97.045145} {\bibfield  {journal} {\bibinfo
  {journal} {Phys. Rev. B}\ }\textbf {\bibinfo {volume} {97}},\ \bibinfo
  {pages} {045145} (\bibinfo {year} {2018})}\BibitemShut {NoStop}%
\bibitem [{\citenamefont {Vanderstraeten}\ \emph {et~al.}(2019)\citenamefont
  {Vanderstraeten}, \citenamefont {Haegeman},\ and\ \citenamefont
  {Verstraete}}]{Vanderstraeten2019}%
  \BibitemOpen
  \bibfield  {author} {\bibinfo {author} {\bibfnamefont {L.}~\bibnamefont
  {Vanderstraeten}}, \bibinfo {author} {\bibfnamefont {J.}~\bibnamefont
  {Haegeman}},\ and\ \bibinfo {author} {\bibfnamefont {F.}~\bibnamefont
  {Verstraete}},\ }\bibfield  {title} {\bibinfo {title} {Tangent-space methods
  for uniform matrix product states},\ }\href
  {https://scipost.org/10.21468/SciPostPhysLectNotes.7} {\bibfield  {journal}
  {\bibinfo  {journal} {SciPost Phys. Lect. Notes}\ }\textbf {\bibinfo {volume}
  {7}} (\bibinfo {year} {2019})}\BibitemShut {NoStop}%
\bibitem [{\citenamefont {Weichselbaum}(2012)}]{Weichselbaum2012}%
  \BibitemOpen
  \bibfield  {author} {\bibinfo {author} {\bibfnamefont {A.}~\bibnamefont
  {Weichselbaum}},\ }\bibfield  {title} {\bibinfo {title} {Non-abelian
  symmetries in tensor networks: {A} quantum symmetry space approach},\ }\href
  {https://doi.org/10.1016/j.aop.2012.07.009} {\bibfield  {journal} {\bibinfo
  {journal} {Ann. of Phys.}\ }\textbf {\bibinfo {volume} {327}},\ \bibinfo
  {pages} {2972} (\bibinfo {year} {2012})}\BibitemShut {NoStop}%
\bibitem [{\citenamefont {Weichselbaum}(2020)}]{Weichselbaum2020}%
  \BibitemOpen
  \bibfield  {author} {\bibinfo {author} {\bibfnamefont {A.}~\bibnamefont
  {Weichselbaum}},\ }\bibfield  {title} {\bibinfo {title} {X-symbols for
  non-abelian symmetries in tensor networks},\ }\href
  {https://doi.org/10.1103/PhysRevResearch.2.023385} {\bibfield  {journal}
  {\bibinfo  {journal} {Phys. Rev. Research}\ }\textbf {\bibinfo {volume}
  {2}},\ \bibinfo {pages} {023385} (\bibinfo {year} {2020})}\BibitemShut
  {NoStop}%
\bibitem [{\citenamefont {Motruk}\ \emph {et~al.}(2016)\citenamefont {Motruk},
  \citenamefont {Zaletel}, \citenamefont {Mong},\ and\ \citenamefont
  {Pollmann}}]{Motruk2016}%
  \BibitemOpen
  \bibfield  {author} {\bibinfo {author} {\bibfnamefont {J.}~\bibnamefont
  {Motruk}}, \bibinfo {author} {\bibfnamefont {M.~P.}\ \bibnamefont {Zaletel}},
  \bibinfo {author} {\bibfnamefont {R.~S.~K.}\ \bibnamefont {Mong}},\ and\
  \bibinfo {author} {\bibfnamefont {F.}~\bibnamefont {Pollmann}},\ }\bibfield
  {title} {\bibinfo {title} {Density matrix renormalization group on a cylinder
  in mixed real and momentum space},\ }\href
  {https://doi.org/10.1103/PhysRevB.93.155139} {\bibfield  {journal} {\bibinfo
  {journal} {Phys. Rev. B}\ }\textbf {\bibinfo {volume} {93}},\ \bibinfo
  {pages} {155139} (\bibinfo {year} {2016})}\BibitemShut {NoStop}%
\bibitem [{\citenamefont {Kirchner}\ \emph {et~al.}(2020)\citenamefont
  {Kirchner}, \citenamefont {Paschen}, \citenamefont {Chen}, \citenamefont
  {Wirth}, \citenamefont {Feng}, \citenamefont {Thompson},\ and\ \citenamefont
  {Si}}]{Kirchner2020}%
  \BibitemOpen
  \bibfield  {author} {\bibinfo {author} {\bibfnamefont {S.}~\bibnamefont
  {Kirchner}}, \bibinfo {author} {\bibfnamefont {S.}~\bibnamefont {Paschen}},
  \bibinfo {author} {\bibfnamefont {Q.}~\bibnamefont {Chen}}, \bibinfo {author}
  {\bibfnamefont {S.}~\bibnamefont {Wirth}}, \bibinfo {author} {\bibfnamefont
  {D.}~\bibnamefont {Feng}}, \bibinfo {author} {\bibfnamefont {J.~D.}\
  \bibnamefont {Thompson}},\ and\ \bibinfo {author} {\bibfnamefont
  {Q.}~\bibnamefont {Si}},\ }\bibfield  {title} {\bibinfo {title} {Colloquium:
  Heavy-electron quantum criticality and single-particle spectroscopy},\ }\href
  {https://doi.org/10.1103/RevModPhys.92.011002} {\bibfield  {journal}
  {\bibinfo  {journal} {Rev. Mod. Phys.}\ }\textbf {\bibinfo {volume} {92}},\
  \bibinfo {pages} {011002} (\bibinfo {year} {2020})}\BibitemShut {NoStop}%
\bibitem [{\citenamefont {L{\"o}hneysen}\ \emph {et~al.}(2007)\citenamefont
  {L{\"o}hneysen}, \citenamefont {Rosch}, \citenamefont {Vojta},\ and\
  \citenamefont {W{\"o}lfle}}]{Loehneysen2007}%
  \BibitemOpen
  \bibfield  {author} {\bibinfo {author} {\bibfnamefont {H.~v.}\ \bibnamefont
  {L{\"o}hneysen}}, \bibinfo {author} {\bibfnamefont {A.}~\bibnamefont
  {Rosch}}, \bibinfo {author} {\bibfnamefont {M.}~\bibnamefont {Vojta}},\ and\
  \bibinfo {author} {\bibfnamefont {P.}~\bibnamefont {W{\"o}lfle}},\ }\bibfield
   {title} {\bibinfo {title} {Fermi-liquid instabilities at magnetic quantum
  phase transitions},\ }\href {https://doi.org/10.1103/RevModPhys.79.1015}
  {\bibfield  {journal} {\bibinfo  {journal} {Rev. Mod. Phys.}\ }\textbf
  {\bibinfo {volume} {79}},\ \bibinfo {pages} {1015} (\bibinfo {year}
  {2007})}\BibitemShut {NoStop}%
\bibitem [{\citenamefont {Stewart}(2001)}]{Stewart2001}%
  \BibitemOpen
  \bibfield  {author} {\bibinfo {author} {\bibfnamefont {G.~R.}\ \bibnamefont
  {Stewart}},\ }\bibfield  {title} {\bibinfo {title} {Non-{Fermi}-liquid
  behavior in $d$- and $f$-electron metals},\ }\href
  {https://doi.org/10.1103/RevModPhys.73.797} {\bibfield  {journal} {\bibinfo
  {journal} {Rev. Mod. Phys.}\ }\textbf {\bibinfo {volume} {73}},\ \bibinfo
  {pages} {797} (\bibinfo {year} {2001})}\BibitemShut {NoStop}%
\bibitem [{\citenamefont {Coleman}\ and\ \citenamefont
  {P\'epin}(2002)}]{Coleman2002}%
  \BibitemOpen
  \bibfield  {author} {\bibinfo {author} {\bibfnamefont {P.}~\bibnamefont
  {Coleman}}\ and\ \bibinfo {author} {\bibfnamefont {C.}~\bibnamefont
  {P\'epin}},\ }\bibfield  {title} {\bibinfo {title} {What is the fate of the
  heavy electron at a quantum critical point?},\ }\href
  {https://doi.org/https://doi.org/10.1016/S0921-4526(01)01342-4} {\bibfield
  {journal} {\bibinfo  {journal} {Physica B: Condensed Matter}\ }\textbf
  {\bibinfo {volume} {312-313}},\ \bibinfo {pages} {383} (\bibinfo {year}
  {2002})},\ \bibinfo {note} {the International Conference on Strongly
  Correlated Electron Systems}\BibitemShut {NoStop}%
\bibitem [{\citenamefont {Paschen}\ \emph {et~al.}(2004)\citenamefont
  {Paschen}, \citenamefont {L{\"u}hmann}, \citenamefont {Wirth}, \citenamefont
  {Gegenwart}, \citenamefont {Trovarelli}, \citenamefont {Geibel},
  \citenamefont {Steglich}, \citenamefont {Coleman},\ and\ \citenamefont
  {Si}}]{Paschen2004}%
  \BibitemOpen
  \bibfield  {author} {\bibinfo {author} {\bibfnamefont {S.}~\bibnamefont
  {Paschen}}, \bibinfo {author} {\bibfnamefont {T.}~\bibnamefont
  {L{\"u}hmann}}, \bibinfo {author} {\bibfnamefont {S.}~\bibnamefont {Wirth}},
  \bibinfo {author} {\bibfnamefont {P.}~\bibnamefont {Gegenwart}}, \bibinfo
  {author} {\bibfnamefont {O.}~\bibnamefont {Trovarelli}}, \bibinfo {author}
  {\bibfnamefont {C.}~\bibnamefont {Geibel}}, \bibinfo {author} {\bibfnamefont
  {F.}~\bibnamefont {Steglich}}, \bibinfo {author} {\bibfnamefont
  {P.}~\bibnamefont {Coleman}},\ and\ \bibinfo {author} {\bibfnamefont
  {Q.}~\bibnamefont {Si}},\ }\bibfield  {title} {\bibinfo {title} {Hall-effect
  evolution across a heavy-fermion quantum critical point},\ }\href
  {https://doi.org/10.1038/nature03129} {\bibfield  {journal} {\bibinfo
  {journal} {Nature}\ }\textbf {\bibinfo {volume} {432}},\ \bibinfo {pages}
  {881} (\bibinfo {year} {2004})}\BibitemShut {NoStop}%
\bibitem [{\citenamefont {Shishido}\ \emph {et~al.}(2005)\citenamefont
  {Shishido}, \citenamefont {Settai}, \citenamefont {Harima},\ and\
  \citenamefont {\={O}nuki}}]{Shishido2005}%
  \BibitemOpen
  \bibfield  {author} {\bibinfo {author} {\bibfnamefont {H.}~\bibnamefont
  {Shishido}}, \bibinfo {author} {\bibfnamefont {R.}~\bibnamefont {Settai}},
  \bibinfo {author} {\bibfnamefont {H.}~\bibnamefont {Harima}},\ and\ \bibinfo
  {author} {\bibfnamefont {Y.}~\bibnamefont {\={O}nuki}},\ }\bibfield  {title}
  {\bibinfo {title} {A drastic change of the {Fermi} surface at a critical
  pressure in {$\mathrm{CeRhIn}_5$}: dhva study under pressure},\ }\href
  {https://doi.org/10.1143/JPSJ.74.1103} {\bibfield  {journal} {\bibinfo
  {journal} {Journal of the Physical Society of Japan}\ }\textbf {\bibinfo
  {volume} {74}},\ \bibinfo {pages} {1103} (\bibinfo {year} {2005})},\ \Eprint
  {https://arxiv.org/abs/https://doi.org/10.1143/JPSJ.74.1103}
  {https://doi.org/10.1143/JPSJ.74.1103} \BibitemShut {NoStop}%
\bibitem [{\citenamefont {Friedemann}\ \emph {et~al.}(2010)\citenamefont
  {Friedemann}, \citenamefont {Oeschler}, \citenamefont {Wirth}, \citenamefont
  {Krellner}, \citenamefont {Geibel}, \citenamefont {Steglich}, \citenamefont
  {Paschen}, \citenamefont {Kirchner},\ and\ \citenamefont
  {Si}}]{Friedemann2010}%
  \BibitemOpen
  \bibfield  {author} {\bibinfo {author} {\bibfnamefont {S.}~\bibnamefont
  {Friedemann}}, \bibinfo {author} {\bibfnamefont {N.}~\bibnamefont
  {Oeschler}}, \bibinfo {author} {\bibfnamefont {S.}~\bibnamefont {Wirth}},
  \bibinfo {author} {\bibfnamefont {C.}~\bibnamefont {Krellner}}, \bibinfo
  {author} {\bibfnamefont {C.}~\bibnamefont {Geibel}}, \bibinfo {author}
  {\bibfnamefont {F.}~\bibnamefont {Steglich}}, \bibinfo {author}
  {\bibfnamefont {S.}~\bibnamefont {Paschen}}, \bibinfo {author} {\bibfnamefont
  {S.}~\bibnamefont {Kirchner}},\ and\ \bibinfo {author} {\bibfnamefont
  {Q.}~\bibnamefont {Si}},\ }\bibfield  {title} {\bibinfo {title}
  {Fermi-surface collapse and dynamical scaling near a quantum-critical
  point},\ }\href {https://doi.org/10.1073/pnas.1009202107} {\bibfield
  {journal} {\bibinfo  {journal} {Proc. Natl. Acad. Sci.}\ }\textbf {\bibinfo
  {volume} {107}},\ \bibinfo {pages} {14547} (\bibinfo {year}
  {2010})}\BibitemShut {NoStop}%
\bibitem [{\citenamefont {Maksimovic}\ \emph {et~al.}(2022)\citenamefont
  {Maksimovic}, \citenamefont {Eilbott}, \citenamefont {Cookmeyer},
  \citenamefont {Wan}, \citenamefont {Rusz}, \citenamefont {Nagarajan},
  \citenamefont {Haley}, \citenamefont {Maniv}, \citenamefont {Gong},
  \citenamefont {Faubel}, \citenamefont {Hayes}, \citenamefont {Bangura},
  \citenamefont {Singleton}, \citenamefont {Palmstrom}, \citenamefont {Winter},
  \citenamefont {McDonald}, \citenamefont {Jang}, \citenamefont {Ai},
  \citenamefont {Lin}, \citenamefont {Ciocys}, \citenamefont {Gobbo},
  \citenamefont {Werman}, \citenamefont {Oppeneer}, \citenamefont {Altman},
  \citenamefont {Lanzara},\ and\ \citenamefont {Analytis}}]{Maksimovic2022}%
  \BibitemOpen
  \bibfield  {author} {\bibinfo {author} {\bibfnamefont {N.}~\bibnamefont
  {Maksimovic}}, \bibinfo {author} {\bibfnamefont {D.~H.}\ \bibnamefont
  {Eilbott}}, \bibinfo {author} {\bibfnamefont {T.}~\bibnamefont {Cookmeyer}},
  \bibinfo {author} {\bibfnamefont {F.}~\bibnamefont {Wan}}, \bibinfo {author}
  {\bibfnamefont {J.}~\bibnamefont {Rusz}}, \bibinfo {author} {\bibfnamefont
  {V.}~\bibnamefont {Nagarajan}}, \bibinfo {author} {\bibfnamefont {S.~C.}\
  \bibnamefont {Haley}}, \bibinfo {author} {\bibfnamefont {E.}~\bibnamefont
  {Maniv}}, \bibinfo {author} {\bibfnamefont {A.}~\bibnamefont {Gong}},
  \bibinfo {author} {\bibfnamefont {S.}~\bibnamefont {Faubel}}, \bibinfo
  {author} {\bibfnamefont {I.~M.}\ \bibnamefont {Hayes}}, \bibinfo {author}
  {\bibfnamefont {A.}~\bibnamefont {Bangura}}, \bibinfo {author} {\bibfnamefont
  {J.}~\bibnamefont {Singleton}}, \bibinfo {author} {\bibfnamefont {J.~C.}\
  \bibnamefont {Palmstrom}}, \bibinfo {author} {\bibfnamefont {L.}~\bibnamefont
  {Winter}}, \bibinfo {author} {\bibfnamefont {R.}~\bibnamefont {McDonald}},
  \bibinfo {author} {\bibfnamefont {S.}~\bibnamefont {Jang}}, \bibinfo {author}
  {\bibfnamefont {P.}~\bibnamefont {Ai}}, \bibinfo {author} {\bibfnamefont
  {Y.}~\bibnamefont {Lin}}, \bibinfo {author} {\bibfnamefont {S.}~\bibnamefont
  {Ciocys}}, \bibinfo {author} {\bibfnamefont {J.}~\bibnamefont {Gobbo}},
  \bibinfo {author} {\bibfnamefont {Y.}~\bibnamefont {Werman}}, \bibinfo
  {author} {\bibfnamefont {P.~M.}\ \bibnamefont {Oppeneer}}, \bibinfo {author}
  {\bibfnamefont {E.}~\bibnamefont {Altman}}, \bibinfo {author} {\bibfnamefont
  {A.}~\bibnamefont {Lanzara}},\ and\ \bibinfo {author} {\bibfnamefont {J.~G.}\
  \bibnamefont {Analytis}},\ }\bibfield  {title} {\bibinfo {title} {Evidence
  for a delocalization quantum phase transition without symmetry breaking in
  {$\mathrm{CeCoIn}_5$}},\ }\href {https://doi.org/10.1126/science.aaz4566}
  {\bibfield  {journal} {\bibinfo  {journal} {Science}\ }\textbf {\bibinfo
  {volume} {375}},\ \bibinfo {pages} {76} (\bibinfo {year} {2022})}\BibitemShut
  {NoStop}%
\bibitem [{\citenamefont {Martelli}\ \emph {et~al.}(2019)\citenamefont
  {Martelli}, \citenamefont {Cai}, \citenamefont {Nica}, \citenamefont
  {Taupin}, \citenamefont {Prokofiev}, \citenamefont {Liu}, \citenamefont
  {Lai}, \citenamefont {Yu}, \citenamefont {Ingersent}, \citenamefont
  {K\"uchler}, \citenamefont {Strydom}, \citenamefont {Geiger}, \citenamefont
  {Haenel}, \citenamefont {Larrea}, \citenamefont {Si},\ and\ \citenamefont
  {Paschen}}]{Martelli2019}%
  \BibitemOpen
  \bibfield  {author} {\bibinfo {author} {\bibfnamefont {V.}~\bibnamefont
  {Martelli}}, \bibinfo {author} {\bibfnamefont {A.}~\bibnamefont {Cai}},
  \bibinfo {author} {\bibfnamefont {E.~M.}\ \bibnamefont {Nica}}, \bibinfo
  {author} {\bibfnamefont {M.}~\bibnamefont {Taupin}}, \bibinfo {author}
  {\bibfnamefont {A.}~\bibnamefont {Prokofiev}}, \bibinfo {author}
  {\bibfnamefont {C.-C.}\ \bibnamefont {Liu}}, \bibinfo {author} {\bibfnamefont
  {H.-H.}\ \bibnamefont {Lai}}, \bibinfo {author} {\bibfnamefont
  {R.}~\bibnamefont {Yu}}, \bibinfo {author} {\bibfnamefont {K.}~\bibnamefont
  {Ingersent}}, \bibinfo {author} {\bibfnamefont {R.}~\bibnamefont
  {K\"uchler}}, \bibinfo {author} {\bibfnamefont {A.~M.}\ \bibnamefont
  {Strydom}}, \bibinfo {author} {\bibfnamefont {D.}~\bibnamefont {Geiger}},
  \bibinfo {author} {\bibfnamefont {J.}~\bibnamefont {Haenel}}, \bibinfo
  {author} {\bibfnamefont {J.}~\bibnamefont {Larrea}}, \bibinfo {author}
  {\bibfnamefont {Q.}~\bibnamefont {Si}},\ and\ \bibinfo {author}
  {\bibfnamefont {S.}~\bibnamefont {Paschen}},\ }\bibfield  {title} {\bibinfo
  {title} {Sequential localization of a complex electron fluid},\ }\href
  {https://doi.org/10.1073/pnas.1908101116} {\bibfield  {journal} {\bibinfo
  {journal} {Proceedings of the National Academy of Sciences}\ }\textbf
  {\bibinfo {volume} {116}},\ \bibinfo {pages} {17701} (\bibinfo {year}
  {2019})}\BibitemShut {NoStop}%
\bibitem [{\citenamefont {Prochaska}\ \emph {et~al.}(2020)\citenamefont
  {Prochaska}, \citenamefont {Li}, \citenamefont {MacFarland}, \citenamefont
  {Andrews}, \citenamefont {Bonta}, \citenamefont {Bianco}, \citenamefont
  {Yazdi}, \citenamefont {Schrenk}, \citenamefont {Detz}, \citenamefont
  {Limbeck}, \citenamefont {Si}, \citenamefont {Ringe}, \citenamefont
  {Strasser}, \citenamefont {Kono},\ and\ \citenamefont
  {Paschen}}]{Prochaska2020}%
  \BibitemOpen
  \bibfield  {author} {\bibinfo {author} {\bibfnamefont {L.}~\bibnamefont
  {Prochaska}}, \bibinfo {author} {\bibfnamefont {X.}~\bibnamefont {Li}},
  \bibinfo {author} {\bibfnamefont {D.~C.}\ \bibnamefont {MacFarland}},
  \bibinfo {author} {\bibfnamefont {A.~M.}\ \bibnamefont {Andrews}}, \bibinfo
  {author} {\bibfnamefont {M.}~\bibnamefont {Bonta}}, \bibinfo {author}
  {\bibfnamefont {E.~F.}\ \bibnamefont {Bianco}}, \bibinfo {author}
  {\bibfnamefont {S.}~\bibnamefont {Yazdi}}, \bibinfo {author} {\bibfnamefont
  {W.}~\bibnamefont {Schrenk}}, \bibinfo {author} {\bibfnamefont
  {H.}~\bibnamefont {Detz}}, \bibinfo {author} {\bibfnamefont {A.}~\bibnamefont
  {Limbeck}}, \bibinfo {author} {\bibfnamefont {Q.}~\bibnamefont {Si}},
  \bibinfo {author} {\bibfnamefont {E.}~\bibnamefont {Ringe}}, \bibinfo
  {author} {\bibfnamefont {G.}~\bibnamefont {Strasser}}, \bibinfo {author}
  {\bibfnamefont {J.}~\bibnamefont {Kono}},\ and\ \bibinfo {author}
  {\bibfnamefont {S.}~\bibnamefont {Paschen}},\ }\bibfield  {title} {\bibinfo
  {title} {Singular charge fluctuations at a magnetic quantum critical point},\
  }\href {https://doi.org/10.1126/science.aag1595} {\bibfield  {journal}
  {\bibinfo  {journal} {Science}\ }\textbf {\bibinfo {volume} {367}},\ \bibinfo
  {pages} {285} (\bibinfo {year} {2020})}\BibitemShut {NoStop}%
\bibitem [{\citenamefont {Trovarelli}\ \emph {et~al.}(2000)\citenamefont
  {Trovarelli}, \citenamefont {Geibel}, \citenamefont {Mederle}, \citenamefont
  {Langhammer}, \citenamefont {Grosche}, \citenamefont {Gegenwart},
  \citenamefont {Lang}, \citenamefont {Sparn},\ and\ \citenamefont
  {Steglich}}]{Trovarelli2000}%
  \BibitemOpen
  \bibfield  {author} {\bibinfo {author} {\bibfnamefont {O.}~\bibnamefont
  {Trovarelli}}, \bibinfo {author} {\bibfnamefont {C.}~\bibnamefont {Geibel}},
  \bibinfo {author} {\bibfnamefont {S.}~\bibnamefont {Mederle}}, \bibinfo
  {author} {\bibfnamefont {C.}~\bibnamefont {Langhammer}}, \bibinfo {author}
  {\bibfnamefont {F.~M.}\ \bibnamefont {Grosche}}, \bibinfo {author}
  {\bibfnamefont {P.}~\bibnamefont {Gegenwart}}, \bibinfo {author}
  {\bibfnamefont {M.}~\bibnamefont {Lang}}, \bibinfo {author} {\bibfnamefont
  {G.}~\bibnamefont {Sparn}},\ and\ \bibinfo {author} {\bibfnamefont
  {F.}~\bibnamefont {Steglich}},\ }\bibfield  {title} {\bibinfo {title}
  {$\mathrm{YbRh}_2\mathrm{Si}_2$: Pronounced non-{Fermi}-liquid effects above
  a low-lying magnetic phase transition},\ }\href
  {https://doi.org/10.1103/PhysRevLett.85.626} {\bibfield  {journal} {\bibinfo
  {journal} {Phys. Rev. Lett.}\ }\textbf {\bibinfo {volume} {85}},\ \bibinfo
  {pages} {626} (\bibinfo {year} {2000})}\BibitemShut {NoStop}%
\bibitem [{\citenamefont {Zhao}\ \emph {et~al.}(2019)\citenamefont {Zhao},
  \citenamefont {Zhang}, \citenamefont {Lyu}, \citenamefont {Bachus},
  \citenamefont {Tokiwa}, \citenamefont {Gegenwart}, \citenamefont {Zhang},
  \citenamefont {Cheng}, \citenamefont {Yang}, \citenamefont {Chen},
  \citenamefont {Isikawa}, \citenamefont {Si}, \citenamefont {Steglich},\ and\
  \citenamefont {Sun}}]{Zhao2019}%
  \BibitemOpen
  \bibfield  {author} {\bibinfo {author} {\bibfnamefont {H.}~\bibnamefont
  {Zhao}}, \bibinfo {author} {\bibfnamefont {J.}~\bibnamefont {Zhang}},
  \bibinfo {author} {\bibfnamefont {M.}~\bibnamefont {Lyu}}, \bibinfo {author}
  {\bibfnamefont {S.}~\bibnamefont {Bachus}}, \bibinfo {author} {\bibfnamefont
  {Y.}~\bibnamefont {Tokiwa}}, \bibinfo {author} {\bibfnamefont
  {P.}~\bibnamefont {Gegenwart}}, \bibinfo {author} {\bibfnamefont
  {S.}~\bibnamefont {Zhang}}, \bibinfo {author} {\bibfnamefont
  {J.}~\bibnamefont {Cheng}}, \bibinfo {author} {\bibfnamefont {Y.-f.}\
  \bibnamefont {Yang}}, \bibinfo {author} {\bibfnamefont {G.}~\bibnamefont
  {Chen}}, \bibinfo {author} {\bibfnamefont {Y.}~\bibnamefont {Isikawa}},
  \bibinfo {author} {\bibfnamefont {Q.}~\bibnamefont {Si}}, \bibinfo {author}
  {\bibfnamefont {F.}~\bibnamefont {Steglich}},\ and\ \bibinfo {author}
  {\bibfnamefont {P.}~\bibnamefont {Sun}},\ }\bibfield  {title} {\bibinfo
  {title} {Quantum-critical phase from frustrated magnetism in a strongly
  correlated metal},\ }\href {https://doi.org/10.1038/s41567-019-0666-6}
  {\bibfield  {journal} {\bibinfo  {journal} {Nature Phys.}\ }\textbf {\bibinfo
  {volume} {15}},\ \bibinfo {pages} {1261} (\bibinfo {year}
  {2019})}\BibitemShut {NoStop}%
\bibitem [{\citenamefont {L\"ohneysen}\ \emph {et~al.}(1996)\citenamefont
  {L\"ohneysen}, \citenamefont {Sieck}, \citenamefont {Stockert},\ and\
  \citenamefont {Waffenschmidt}}]{Loehneysen1996}%
  \BibitemOpen
  \bibfield  {author} {\bibinfo {author} {\bibfnamefont {H.}~\bibnamefont
  {L\"ohneysen}}, \bibinfo {author} {\bibfnamefont {M.}~\bibnamefont {Sieck}},
  \bibinfo {author} {\bibfnamefont {O.}~\bibnamefont {Stockert}},\ and\
  \bibinfo {author} {\bibfnamefont {M.}~\bibnamefont {Waffenschmidt}},\
  }\bibfield  {title} {\bibinfo {title} {Investigation of non-fermi-liquid
  behavior in {CeCu${}_{6-x}$Au${}_x$}},\ }\href
  {https://doi.org/https://doi.org/10.1016/0921-4526(96)00151-2} {\bibfield
  {journal} {\bibinfo  {journal} {Physica B: Condensed Matter}\ }\textbf
  {\bibinfo {volume} {223-224}},\ \bibinfo {pages} {471} (\bibinfo {year}
  {1996})},\ \bibinfo {note} {proceedings of the International Conference on
  Strongly Correlated Electron Systems}\BibitemShut {NoStop}%
\bibitem [{\citenamefont {von L\"ohneysen}(1996)}]{Loehneysen1996a}%
  \BibitemOpen
  \bibfield  {author} {\bibinfo {author} {\bibfnamefont {H.}~\bibnamefont {von
  L\"ohneysen}},\ }\bibfield  {title} {\bibinfo {title} {Non-fermi-liquid
  behaviour in the heavy-fermion system},\ }\href
  {https://doi.org/10.1088/0953-8984/8/48/003} {\bibfield  {journal} {\bibinfo
  {journal} {Journal of Physics: Condensed Matter}\ }\textbf {\bibinfo {volume}
  {8}},\ \bibinfo {pages} {9689} (\bibinfo {year} {1996})}\BibitemShut
  {NoStop}%
\bibitem [{\citenamefont {De~Leo}\ \emph
  {et~al.}(2008{\natexlab{a}})\citenamefont {De~Leo}, \citenamefont {Civelli},\
  and\ \citenamefont {Kotliar}}]{DeLeo2008}%
  \BibitemOpen
  \bibfield  {author} {\bibinfo {author} {\bibfnamefont {L.}~\bibnamefont
  {De~Leo}}, \bibinfo {author} {\bibfnamefont {M.}~\bibnamefont {Civelli}},\
  and\ \bibinfo {author} {\bibfnamefont {G.}~\bibnamefont {Kotliar}},\
  }\bibfield  {title} {\bibinfo {title} {Cellular dynamical mean-field theory
  of the periodic {Anderson} model},\ }\href
  {https://doi.org/10.1103/PhysRevB.77.075107} {\bibfield  {journal} {\bibinfo
  {journal} {Phys. Rev. B}\ }\textbf {\bibinfo {volume} {77}},\ \bibinfo
  {pages} {075107} (\bibinfo {year} {2008}{\natexlab{a}})}\BibitemShut
  {NoStop}%
\bibitem [{\citenamefont {De~Leo}\ \emph
  {et~al.}(2008{\natexlab{b}})\citenamefont {De~Leo}, \citenamefont {Civelli},\
  and\ \citenamefont {Kotliar}}]{DeLeo2008a}%
  \BibitemOpen
  \bibfield  {author} {\bibinfo {author} {\bibfnamefont {L.}~\bibnamefont
  {De~Leo}}, \bibinfo {author} {\bibfnamefont {M.}~\bibnamefont {Civelli}},\
  and\ \bibinfo {author} {\bibfnamefont {G.}~\bibnamefont {Kotliar}},\
  }\bibfield  {title} {\bibinfo {title} {{$T =0$} heavy-fermion quantum
  critical point as an orbital-selective {Mott} transition},\ }\href
  {https://doi.org/10.1103/PhysRevLett.101.256404} {\bibfield  {journal}
  {\bibinfo  {journal} {Phys. Rev. Lett.}\ }\textbf {\bibinfo {volume} {101}},\
  \bibinfo {pages} {256404} (\bibinfo {year} {2008}{\natexlab{b}})}\BibitemShut
  {NoStop}%
\bibitem [{\citenamefont {Tanaskovi{\'c}}\ \emph {et~al.}(2011)\citenamefont
  {Tanaskovi{\'c}}, \citenamefont {Haule}, \citenamefont {Kotliar},\ and\
  \citenamefont {Dobrosavljevi{\'c}}}]{Tanaskovic2011}%
  \BibitemOpen
  \bibfield  {author} {\bibinfo {author} {\bibfnamefont {D.}~\bibnamefont
  {Tanaskovi{\'c}}}, \bibinfo {author} {\bibfnamefont {K.}~\bibnamefont
  {Haule}}, \bibinfo {author} {\bibfnamefont {G.}~\bibnamefont {Kotliar}},\
  and\ \bibinfo {author} {\bibfnamefont {V.}~\bibnamefont
  {Dobrosavljevi{\'c}}},\ }\bibfield  {title} {\bibinfo {title} {Phase diagram,
  energy scales, and nonlocal correlations in the {Anderson} lattice model},\
  }\href {https://doi.org/10.1103/PhysRevB.84.115105} {\bibfield  {journal}
  {\bibinfo  {journal} {Phys. Rev. B}\ }\textbf {\bibinfo {volume} {84}},\
  \bibinfo {pages} {115105} (\bibinfo {year} {2011})}\BibitemShut {NoStop}%
\bibitem [{\citenamefont {Si}\ \emph {et~al.}(2001)\citenamefont {Si},
  \citenamefont {Rabello}, \citenamefont {Ingersent},\ and\ \citenamefont
  {Smith}}]{Si2001}%
  \BibitemOpen
  \bibfield  {author} {\bibinfo {author} {\bibfnamefont {Q.}~\bibnamefont
  {Si}}, \bibinfo {author} {\bibfnamefont {S.}~\bibnamefont {Rabello}},
  \bibinfo {author} {\bibfnamefont {K.}~\bibnamefont {Ingersent}},\ and\
  \bibinfo {author} {\bibfnamefont {J.~L.}\ \bibnamefont {Smith}},\ }\bibfield
  {title} {\bibinfo {title} {Locally critical quantum phase transitions in
  strongly correlated metals},\ }\href {https://doi.org/10.1038/35101507}
  {\bibfield  {journal} {\bibinfo  {journal} {Nature}\ }\textbf {\bibinfo
  {volume} {413}},\ \bibinfo {pages} {804} (\bibinfo {year}
  {2001})}\BibitemShut {NoStop}%
\bibitem [{\citenamefont {Si}\ \emph {et~al.}(2003)\citenamefont {Si},
  \citenamefont {Rabello}, \citenamefont {Ingersent},\ and\ \citenamefont
  {Smith}}]{Si2003}%
  \BibitemOpen
  \bibfield  {author} {\bibinfo {author} {\bibfnamefont {Q.}~\bibnamefont
  {Si}}, \bibinfo {author} {\bibfnamefont {S.}~\bibnamefont {Rabello}},
  \bibinfo {author} {\bibfnamefont {K.}~\bibnamefont {Ingersent}},\ and\
  \bibinfo {author} {\bibfnamefont {J.~L.}\ \bibnamefont {Smith}},\ }\bibfield
  {title} {\bibinfo {title} {Local fluctuations in quantum critical metals},\
  }\href {https://doi.org/10.1103/PhysRevB.68.115103} {\bibfield  {journal}
  {\bibinfo  {journal} {Phys. Rev. B}\ }\textbf {\bibinfo {volume} {68}},\
  \bibinfo {pages} {115103} (\bibinfo {year} {2003})}\BibitemShut {NoStop}%
\bibitem [{\citenamefont {Assaad}(1999)}]{Assaad1999}%
  \BibitemOpen
  \bibfield  {author} {\bibinfo {author} {\bibfnamefont {F.~F.}\ \bibnamefont
  {Assaad}},\ }\bibfield  {title} {\bibinfo {title} {Quantum monte carlo
  simulations of the half-filled two-dimensional {Kondo} lattice model},\
  }\href {https://doi.org/10.1103/PhysRevLett.83.796} {\bibfield  {journal}
  {\bibinfo  {journal} {Phys. Rev. Lett.}\ }\textbf {\bibinfo {volume} {83}},\
  \bibinfo {pages} {796} (\bibinfo {year} {1999})}\BibitemShut {NoStop}%
\bibitem [{\citenamefont {Capponi}\ and\ \citenamefont
  {Assaad}(2001)}]{Capponi2001}%
  \BibitemOpen
  \bibfield  {author} {\bibinfo {author} {\bibfnamefont {S.}~\bibnamefont
  {Capponi}}\ and\ \bibinfo {author} {\bibfnamefont {F.~F.}\ \bibnamefont
  {Assaad}},\ }\bibfield  {title} {\bibinfo {title} {Spin and charge dynamics
  of the ferromagnetic and antiferromagnetic two-dimensional half-filled
  {Kondo} lattice model},\ }\href {https://doi.org/10.1103/PhysRevB.63.155114}
  {\bibfield  {journal} {\bibinfo  {journal} {Phys. Rev. B}\ }\textbf {\bibinfo
  {volume} {63}},\ \bibinfo {pages} {155114} (\bibinfo {year}
  {2001})}\BibitemShut {NoStop}%
\bibitem [{\citenamefont {Parisen~Toldin}\ \emph {et~al.}(2019)\citenamefont
  {Parisen~Toldin}, \citenamefont {Sato},\ and\ \citenamefont
  {Assaad}}]{Toldin2019}%
  \BibitemOpen
  \bibfield  {author} {\bibinfo {author} {\bibfnamefont {F.}~\bibnamefont
  {Parisen~Toldin}}, \bibinfo {author} {\bibfnamefont {T.}~\bibnamefont
  {Sato}},\ and\ \bibinfo {author} {\bibfnamefont {F.~F.}\ \bibnamefont
  {Assaad}},\ }\bibfield  {title} {\bibinfo {title} {Mutual information in
  heavy-fermion systems},\ }\href {https://doi.org/10.1103/PhysRevB.99.155158}
  {\bibfield  {journal} {\bibinfo  {journal} {Phys. Rev. B}\ }\textbf {\bibinfo
  {volume} {99}},\ \bibinfo {pages} {155158} (\bibinfo {year}
  {2019})}\BibitemShut {NoStop}%
\bibitem [{\citenamefont {Danu}\ \emph {et~al.}(2021)\citenamefont {Danu},
  \citenamefont {Liu}, \citenamefont {Assaad},\ and\ \citenamefont
  {Raczkowski}}]{Danu2021}%
  \BibitemOpen
  \bibfield  {author} {\bibinfo {author} {\bibfnamefont {B.}~\bibnamefont
  {Danu}}, \bibinfo {author} {\bibfnamefont {Z.}~\bibnamefont {Liu}}, \bibinfo
  {author} {\bibfnamefont {F.~F.}\ \bibnamefont {Assaad}},\ and\ \bibinfo
  {author} {\bibfnamefont {M.}~\bibnamefont {Raczkowski}},\ }\bibfield  {title}
  {\bibinfo {title} {Zooming in on heavy fermions in {Kondo} lattice models},\
  }\href {https://doi.org/10.1103/PhysRevB.104.155128} {\bibfield  {journal}
  {\bibinfo  {journal} {Phys. Rev. B}\ }\textbf {\bibinfo {volume} {104}},\
  \bibinfo {pages} {155128} (\bibinfo {year} {2021})}\BibitemShut {NoStop}%
\bibitem [{\citenamefont {Senthil}\ \emph {et~al.}(2003)\citenamefont
  {Senthil}, \citenamefont {Sachdev},\ and\ \citenamefont
  {Vojta}}]{Senthil2003}%
  \BibitemOpen
  \bibfield  {author} {\bibinfo {author} {\bibfnamefont {T.}~\bibnamefont
  {Senthil}}, \bibinfo {author} {\bibfnamefont {S.}~\bibnamefont {Sachdev}},\
  and\ \bibinfo {author} {\bibfnamefont {M.}~\bibnamefont {Vojta}},\ }\bibfield
   {title} {\bibinfo {title} {Fractionalized fermi liquids},\ }\href
  {https://doi.org/10.1103/PhysRevLett.90.216403} {\bibfield  {journal}
  {\bibinfo  {journal} {Phys. Rev. Lett.}\ }\textbf {\bibinfo {volume} {90}},\
  \bibinfo {pages} {216403} (\bibinfo {year} {2003})}\BibitemShut {NoStop}%
\bibitem [{\citenamefont {Senthil}\ \emph {et~al.}(2004)\citenamefont
  {Senthil}, \citenamefont {Vojta},\ and\ \citenamefont
  {Sachdev}}]{Senthil2004}%
  \BibitemOpen
  \bibfield  {author} {\bibinfo {author} {\bibfnamefont {T.}~\bibnamefont
  {Senthil}}, \bibinfo {author} {\bibfnamefont {M.}~\bibnamefont {Vojta}},\
  and\ \bibinfo {author} {\bibfnamefont {S.}~\bibnamefont {Sachdev}},\
  }\bibfield  {title} {\bibinfo {title} {Weak magnetism and non-fermi liquids
  near heavy-fermion critical points},\ }\href
  {https://doi.org/10.1103/PhysRevB.69.035111} {\bibfield  {journal} {\bibinfo
  {journal} {Phys. Rev. B}\ }\textbf {\bibinfo {volume} {69}},\ \bibinfo
  {pages} {035111} (\bibinfo {year} {2004})}\BibitemShut {NoStop}%
\bibitem [{\citenamefont {Li}\ \emph {et~al.}(2022)\citenamefont {Li},
  \citenamefont {Gleis},\ and\ \citenamefont {von Delft}}]{Li2022}%
  \BibitemOpen
  \bibfield  {author} {\bibinfo {author} {\bibfnamefont {J.-W.}\ \bibnamefont
  {Li}}, \bibinfo {author} {\bibfnamefont {A.}~\bibnamefont {Gleis}},\ and\
  \bibinfo {author} {\bibfnamefont {J.}~\bibnamefont {von Delft}},\ }\bibfield
  {title} {\bibinfo {title} {Time-dependent variational principle with
  controlled bond expansion for matrix product states},\ }\href
  {https://doi.org/10.48550/arXiv.2208.10972} {\bibfield  {journal} {\bibinfo
  {journal} {arXiv:2208.10972 [cond-mat.str-el]}\ } (\bibinfo {year}
  {2022})}\BibitemShut {NoStop}%
\end{thebibliography}%
\clearpage
\title{Supplemental material: \\ Controlled bond expansion for DMRG ground state search at single-site costs}

\date{\today}
\maketitle

\setcounter{secnumdepth}{2} 
\renewcommand{\thefigure}{S-\arabic{figure}}
\setcounter{figure}{0}
\setcounter{section}{0}
\setcounter{equation}{0}
\renewcommand{\thesection}{S-\arabic{section}}
\renewcommand{\theequation}{S\arabic{equation}}

\begin{figure}[!b] 
\includegraphics[width=\linewidth]{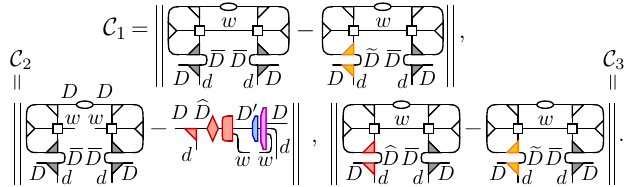}
\caption{
Shrewd selection (concept). 
During a right-to-left CBE sweep, bond $\ell$
is expan\-ded from $A_\ell (\TriangleWhiteA)$ to $A^\pdag_\ell \oplus \Atrunc_\ell (\TriangleWhiteA \oplus \TriangleOrangeA)$, 
 where $\Atrunc_\ell (\TriangleOrangeA)$, with image dimen\-sion $\Dt$, is a truncation of $\Ab_\ell (\TriangleGreyA)$, with image dimension $\Db\! = \! D(d\! - \! 1)$. This expansion will reduce $\variance^\mtwosite$ significantly if
$\Atrunc_\ell \oplus \Bb_\ellplusone^\pdag (\TriangleOrangeA \otimes \TriangleGreyB)$ targets $\rDD$, a $\Dt \Db$-dimensional subspace of
the $\Db{}^2\!$-dimensional space $\D\D$ on which $H^\mtwosite_\ell \psi^\mtwosite_\ell$ has significant weight.
As explained in the main text, ideally, $\Atrunc_\ell (\TriangleOrangeA)$ should minimize the cost function $\Cc_1$. 
To achieve this at \onesite\ costs, we instead find $\Atrunc_\ell (\TriangleOrangeA)$ using shrewd selection, involving two separate
truncations. The first truncation (\textit{preselection}) truncates the central MPS bond from 
$D \!\to \! D'$ in the presence of its environment by minimizing $\Cc_2$; this replaces the full complement 
by a preselected complement, $\Ab_\ell 
\TriangleGreyA \!\to \!\! \Apruned_\ell \TriangleRedA$, 
with reduced image dimension,
$\Db \! \to \! \Dh \!=\! D' w$ \cite{FirstTruncation}. The second truncation
(\textit{final selection}) minimizes $\Cc_3$ 
with central MPO bond closed as appropriate for $H^\mtwosite_\ell\psi^\mtwosite_\ell$: it further truncates  $\Apruned_\ell$ to yield the final truncated complement, $\Atrunc_\ell$, $\TriangleRedA \!\to \!\!
\TriangleOrangeA$, 
$\Dh \! \to \! \Dt < \! D$. To ensure \onesite\ costs for final selection we need $\Dh \!=\! D$, and thus choose $D' \!=\! D/w$ for preselection. The truncations underlying preselection and final selection are explained in detail in Fig.~\ref{fig:pruningexplained_supp}.}
\vspace{-4.5mm}
\label{fig:cost-function_supp} 
\end{figure}

\begin{figure}[b!] 
\raisebox{-7.3mm}{
 \includegraphics[width=0.99\linewidth]{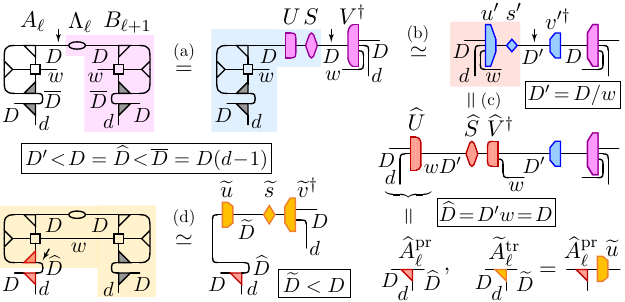}}
 \caption{\Shrewd\ selection (details). Computation of (a-c) the preselected complement $\Apruned_\ell \, (\TriangleRedA)$
 to minimize $\Cc_2$, and (d) the final truncated complement $\Atrunc_\ell \, (\TriangleOrangeA)$
to minimize $\Cc_3$, using four SVDs, all with
at most \onesite\ costs.
For each, an arrow indicates a bond being opened before
doing the SVD, shading and  symbols in matching colors indicate the SVD input and output, and the latter is written as $USV^\dagger$ or $usv^\dagger$ when involving no or some truncation, respectively. 
Importantly, we express 
$\Ab^\pdag_\ell \Ab_\ell^\dag$
and $\Bb_\ellplusone^\dag \Bb^\pdag_\ellplusone$ (grey)
as $\mathbbm{1}^\ps_\ell -\! A^\pdag_\ell A_\ell^\dag$ and $\mathbbm{1}^\ps_\ell \!-\! B^\dag_\ellplusone B_\ellplusone^\pdag$
(\Eq{eq:CompletenessMain}), 
avoiding the 
computation of $\Ab_\ell$ and $\Bb_\ellplusone$.
(a) The first SVD canonicalizes the right side of the diagram,
assigning its weights to the central MPS bond.  
(b) The second SVD and truncation reduces the dimension of this bond, 
$D \!\to \! D'\!=\! D/w$. (c) The third SVD regroups indices to combine 
the truncated MPS bond and the MPO bond into a composite bond of dimension
$\Dh \! = \! D'w \!=\! D$, 
yielding the preselected complement $\Apruned_\ell \!=\! \widehat U \, (\TriangleRedA)$. Nominally, step (c) would require no truncation if exact arithmetic were used, but in practice 
(numerically) zero singular values, of order $\mathcal{O}(10^{-16})$, may arise; these  must be discarded to ensure $A^\dag_\ell \Apruned_\ell \!=\! 0$.
(d) The fourth SVD and truncation yields the final truncated  complement $\Atrunc_\ell \!=\! \Apruned_\ell \widetilde u \, (\TriangleOrangeA)$, with  bond reduction 
$\Dh \!\to \!\Dt \!< \!D$. 
Table~\ref{tab:getAtr} gives a pseudocode for shrewd selection.
} \vspace{-4mm}
\label{fig:pruningexplained_supp}
\end{figure}

This supplement offers additional material on five issues:
in Sec.~\ref{sec:preselection}, details on the implementation of shrewd selection, including pseudocode, and a detailed analysis
of  preselection and final selection; 
in Sec.~\ref{sec:simple_benchmark}, a simple additional benchmark of CBE--DMRG on free Fermions;
in Sec.~\ref{sec:DMRG3S}, a comparison to DMRG3S;
and in Sec.~\ref{sec:cylinder}, more details on the analysis of the Kondo-Heisenberg model on a 4-leg cylinder.


%
\section{\Shrewd\ selection}
\label{sec:preselection}

Figures~\Figureone\ and~2  in the main text introduce
a novel scheme needed for CBE, called \textit{\shrewd\ selection}. 
In this section, we discuss it in detail. Section~\ref{sec:ShrewdSelectionDetails} provides algorithmic details; Sec.~\ref{sec:MotivationPreselection} discusses various options for choosing
the parameters involved in perselection and final selection; Secs.~\ref{sec:SingularValues} and \ref{sec:SingularVectors} discusses the properties of the singular values and singular vectors obtained; 
and Sec.~\ref{sec:ConvergenceRates} discusses the convergence rate per sweep. 

 \subsection{Algorithmic details}
\label{sec:ShrewdSelectionDetails}

For convenience, Fig.~\Figureone\ of the main text is shown again in Fig.~\ref{fig:cost-function_supp}, with a caption summarizing 
the main ideas underlying shrewd selection. Its two ingredients, preselection and final selection, are explained in detail in  Fig.~\ref{fig:pruningexplained_supp} using tensor network diagrams.
Table~\ref{tab:getAtr} provides pseudocode for the tensor network diagrams in Fig.~\ref{fig:pruningexplained_supp}.

In the remainder of this section we discuss preselection and final selection in more detail, and illustrate their effects on the properties of various singular value spectra and singular vectors. We here write bond dimensions with $\ast$, indicating numbers of multiplets (not states), since these determine computational complexities and truncation thresholds and are the quantities shown in the figures. Relations such as
$\Dh = D' w$, exact for Abelian symmetries where all symmetry multiplets have dimension 1, become approximate, $\Dhast \simeq \Dpast \wast$, when
written for non-Abelian symmetries. 

\begin{table}[t!]
\begin{minipage}{\linewidth}
\begin{algorithm}[H]
\caption{Computation of truncated complement using \shrewd\ selection}
\begin{algorithmic}[1]
  \Require \twosite\ Hamiltonian $H^\mtwosite_{\ell} = L_{\ell-1}W_{\ell}W_{\ell+1}R_{\ell+2}$, \twosite\ wavefunction $\psi^\mtwosite = A_\ell\Lambda_{\ell}B_{\ell+1}$ in bond-canonical form, preselection bond dimension $D'$, truncated complement dimension $\Dt$
  \Ensure truncated complement $\Atrunc_{\ell}$  (\scalebox{1}{\TriangleOrangeA})
  \Function{getRorth}{$R_{\ell+2}$,$W_{\ell+1}$,$B_{\ell+1}$,$\Lambda_{\ell}$}
    \State Compute $R^{\mr{tmp}}_{\ell+1} = \Lambda_{\ell}B_{\ell+1}W_{\ell+1}R_{\ell+2}$
    \State Compute $R^{\mr{orth}}_{\ell+1} = R^{\mr{tmp}}_{\ell+1} - R^{\mr{tmp}}_{\ell+1}B^{\dag}_{\ell+1}B^{\pdag}_{\ell+1}$
    \State \Return $R^{\mr{orth}}_{\ell+1}$
  \EndFunction
  \State (Fig.~\ref{fig:pruningexplained_supp}(a)): SVD $\ell$-bond of $R^{\mr{orth}}_{\ell+1} = U S V^\dag$ 
 \Function{getLorth}{$L_{\ell-1}$,$W_{\ell}$,$A_{\ell}$,$U$,$S$}
    \State Compute $L^{\mr{tmp}}_{\ell} = L_{\ell-1}W_{\ell} U S$
    \State Compute $L^{\mr{orth}}_{\ell} = L^{\mr{tmp}}_{\ell} - A_{\ell}^{\pdag}A_{\ell}^{\dag}L^{\mr{tmp}}_{\ell}$
    \State \Return $L^{\mr{orth}}_{\ell}$
  \EndFunction
  \State (Fig.~\ref{fig:pruningexplained_supp}(b)): SVD $L^{\mr{orth}}_{\ell} \!=\! U' S' V^{\prime\dag}$ and truncate all except the largest $D'$ singular values in $S'$: $U' S' V^{\prime\dag} \overset{\mr{trunc}}{\to} u' s' v^{\prime\dag}$
  \State (Fig.~\ref{fig:pruningexplained_supp}(c)): Redirect the MPO-leg of $u' s'$ and perform an SVD on its combined MPO- and $\ell$-bond, $u' s' = \Uh \Sh \Vh{}^\dag$. Truncate all singular values in $\Sh$ which are numerically zero to ensure $A^\dag_{\ell}\Uh = 0$. \Comment{{\bf warning:} $A^\dag_{\ell}\Uh = 0$ is crucial and \textit{must} be ensured!}
  \State (Optional): safety orthogonalization of $\Uh$ by SVD on $\Uh - A^\pdag_{\ell}A^\dag_{\ell}\Uh$ plus truncation of small singular values. 
  \State Assign $\Ah^\pruned_{\ell} = \Uh$ (\scalebox{1}{\TriangleRedA})
 \Function{getCorth}{$L_{\ell-1}$,$W_{\ell}$,$W_{\ell+1}$,$R_{\ell+2}$,$A_{\ell}$,$\Lambda_{\ell}$,$\B_{\ell+1}$,$\Ah^\pruned_{\ell}$}
    \State Compute $L^{\mr{pr}}_{\ell} = (\Ah^\pruned_{\ell})^{\dag}L_{\ell-1}W_{\ell}A_{\ell}$
    \State Compute $C^{\mr{tmp}}_{\ell+1} = L^{\mr{pr}}_{\ell}\Lambda_{\ell}B_{\ell+1} W_{\ell+1}R_{\ell+2}$
    \State Compute $C^{\mr{orth}}_{\ell+1} = C^{\mr{tmp}}_{\ell+1} - C^{\mr{tmp}}_{\ell+1}B^{\dag}_{\ell+1}B^{\pdag}_{\ell+1}$
    \State \Return $C^{\mr{orth}}_{\ell+1}$
  \EndFunction
 \State (Fig.~\ref{fig:pruningexplained_supp}(d)): SVD $C^{\mr{orth}}_{\ell+1} = \Ut \, \St \, \Vt{}^\dag$ and truncate all except the largest $\Dt$ singular values: $\Ut \, \St \, \Vt{}^\dag \overset{\mr{trunc}}{\to} \ut \, \st \, \vt{}^\dag$
 \State Compute $\Atrunc_{\ell} = \Ah^\pruned_{\ell}\ut$ (\scalebox{1}{\TriangleOrangeA})
\end{algorithmic}
\end{algorithm}

\end{minipage}
\caption{Pseudocode for computing the truncated complement $\Atrunc_{\ell}$ using shrewd selection.}
\label{tab:getAtr}
\end{table}
\subsection{Options for preselection and final selection}
\label{sec:MotivationPreselection}
The key idea of CBE is to expand the isometry $A_\ell (\TriangleWhiteA)$, 
whose image (the kept space) initially has dimension $\Diast$, through a direct sum with a so-called 
truncated complement, an isometry
with image dimension $\Dtast$ ($<\! \Diast$). The latter is obtained through 
a suitable truncation of the full complement, $\Ab_\ell (\TriangleGreyA)$, 
whose image (the discarded space)
initially has dimension $\Dbast \!\simeq\! \Diast (\dast \!-\!1)$.
Figure~\Figureone\ defines three cost functions, $\Cc_1$, $\Cc_2$ and $\Cc_3$, relevant
for constructing the truncated complement. 
The optimal choice for the truncated complement, 
to be denoted $\Abtrunc_\ell (\TriangleYellowA)$ here,
is obtained by exact minimization of $\Cc_1$, but that requires  
\twosite\ costs. Therefore, the main text proposes 
an alternative two-step strategy, requiring
only \onesite\ costs. First perform preselection:
obtain a preselected complement $\Apruned_\ell$ (\TriangleRedA), with 
image dimension $\Dhast \! \simeq \! \Dpast \wast$, 
through minimization of $\Cc_2$ (Fig.~\ref{fig:pruningexplained_supp}, steps (a-c)). Then
perform final selection: obtain the desired truncated complement, denoted $\Atrunc_\ell$ (\TriangleOrangeA),  through minimization of $\Cc_3$ (Fig.~\ref{fig:pruningexplained_supp}, step (d)). 
The minimization of the cost functions $\Cc_1$ and $\Cc_3$ defined in 
Fig.~\Figureone\ involves performing SVDs and truncations of the following two tensors, 
respectively: 
\begin{subequations}
\label{subeq:SVDs}
\begin{align}
\label{eq:compareSingularvalues-grey-grey}
\Mb{}^\full   & = 	\! \! \raisebox{-9.0mm}{ \includegraphics[width=0.66\linewidth]{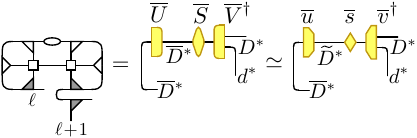}} \, , \hspace{0cm} 
\\ 
\label{eq:compareSingularvalues-red-grey}
\Mh{}^\pruned
& = \! \! \raisebox{-9.0mm}{ \includegraphics[width=0.66\linewidth]{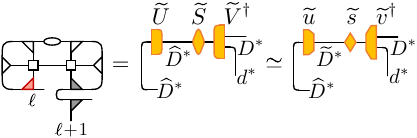}} \hspace{0cm} \, .
\end{align}
\end{subequations}
They differ only in one ingredient, $\Ab^\dag_\ell (\TriangleGreyAdagger)$ 
vs.\ $\Apruned_\ell{}^\dag (\TriangleRedAdagger)$,
but since these have vastly different open leg dimensions, 
$\Dbast$ vs.\ $\Dhast$, the SVD costs differ vastly too, \twosite\ vs.\ \onesite. 
The isometries $\ubar (\SlopingRectangleYellowU)$ or $\ut (\SlopingRectangleOrangeU)$ obtained from the above SVDs and truncations,
both with image dimension $\Dtast$, can then be used to construct 
$\Ab^\trunc_\ell (\TriangleYellowA)$ or  $\Atrunc_\ell (\TriangleOrangeA)$
as follows: 
\begin{subequations} 
\label{eq:three-A-truncs}
\begin{align}
\raisebox{-0.5\height}{ \includegraphics[width=0.333\linewidth]{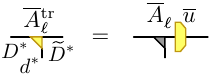}} \hspace{0cm} \, , 
\\
\raisebox{-0.5\height}{ \includegraphics[width=0.333\linewidth]{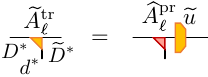}} \hspace{0cm} \, .
\intertext{Both $\Ab^\trunc_\ell (\TriangleYellowA)$ and 
$\Atrunc_\ell (\TriangleOrangeA)$ have image dimension $\Dtast$;
the former serves as reference
(equivalent to using no preselection, 
$\Dpast = \Dast$, the latter is an approximation to the former. An even cruder approximation is obtained if one performs
preselection without final selection: for that, truncate $\Uh \simeq \uh$  
in step (c) of Fig.~\ref{fig:pruningexplained_supp}\ using 
$\Dhast \!=\! \Dtast$ (not $\Dpast \wast$), 
and use the resulting isometry, $\Ahtrunc_\ell (\TrianglePinkA) = \uh$,  as approximation for $\Ab^\trunc_\ell (\TriangleYellowA)$, omitting step (d) altogether:} 
\raisebox{-0.5\height}{ \includegraphics[width=0.333\linewidth]{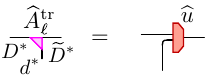}} \hspace{0cm} \, .
\end{align}
\end{subequations}

To illustrate the effects of preselection, we will compare
four settings: (I) the reference, $\Abtrunc (\TriangleYellowA)$; or three versions of preselection with  
$\Dpast \!=\! \Dfast/\wast$, $0.1\Dfast/\wast$ or $1$, to be called 
(II) \textit{moderate}, (III) \textit{severe} or (IV) \textit{extreme} preselection, respectively,
all followed by final selection, yielding three versions of 
 $\Atrunc (\TriangleOrangeA)$.
Here, $\Dfast$ is the final bond dimension after an update,
obtained by expanding the bond from dimension $\Diast$ to 
$\Diast \!+\! \Dtast \!=\! \Dfast(1 \!+\! \delta)$, then 
trimming it back to $\Dfast$. To illustrate the
importance of final selection we also consider a fifth setting:
(V) moderate preselection and $\Uh \! \simeq \! \uh$ truncation, without final selection, yielding 
$\Ahtrunc_\ell (\TrianglePinkA)$.

In the main text, we recommended performing CBE updates 
using moderate preselection followed by final selection.
We showed (Fig.~\Figurefour(a)) that this yields equally fast convergence per sweep 
for the GS energy as  \twosite\ update. 
Below, we elucidate why moderate preselection works so well. To this end, we analyze various singular value spectra (Sec.~\ref{sec:SingularValues}) and left singular 
vectors (Sec.~\ref{sec:SingularVectors}), with $\Dfast = \Dmaxast$ fixed.
We also show that severe and even extreme preselection likewise yield full convergence, albeit at slower rates, by comparing various convergence rates per sweep while increasing $\Dfast$ (Sec.~\ref{sec:ConvergenceRates}). 

\subsection{Singular values}
\label{sec:SingularValues}
We start by comparing  the singular values  
of the tensors $\Mb{}^\full $ and $\Mh{}^\pruned$, 
i.e.\ the diagonal elements of the diagonal matrices $\Sb (\DiamondYellowS)$ and $\St (\DiamondOrangeS)$ 
in \Eqs{subeq:SVDs}, denoted $\Scb_{i}$ ($i= 1, \dots, \Dbast$) and $\Sct_i$ ($i=1, \dots , \Dhast$), respectively. They differ strongly in 
number, but if the largest $\Sct_i$ values roughly mimic
the largest $\Scb_i$ values, serving as reference, then preselection is ``efficient'',
in that it yields essentially optimal results
for the dominant singular values.
%

%
 \begin{figure}[t!]
 \includegraphics[width=\linewidth]{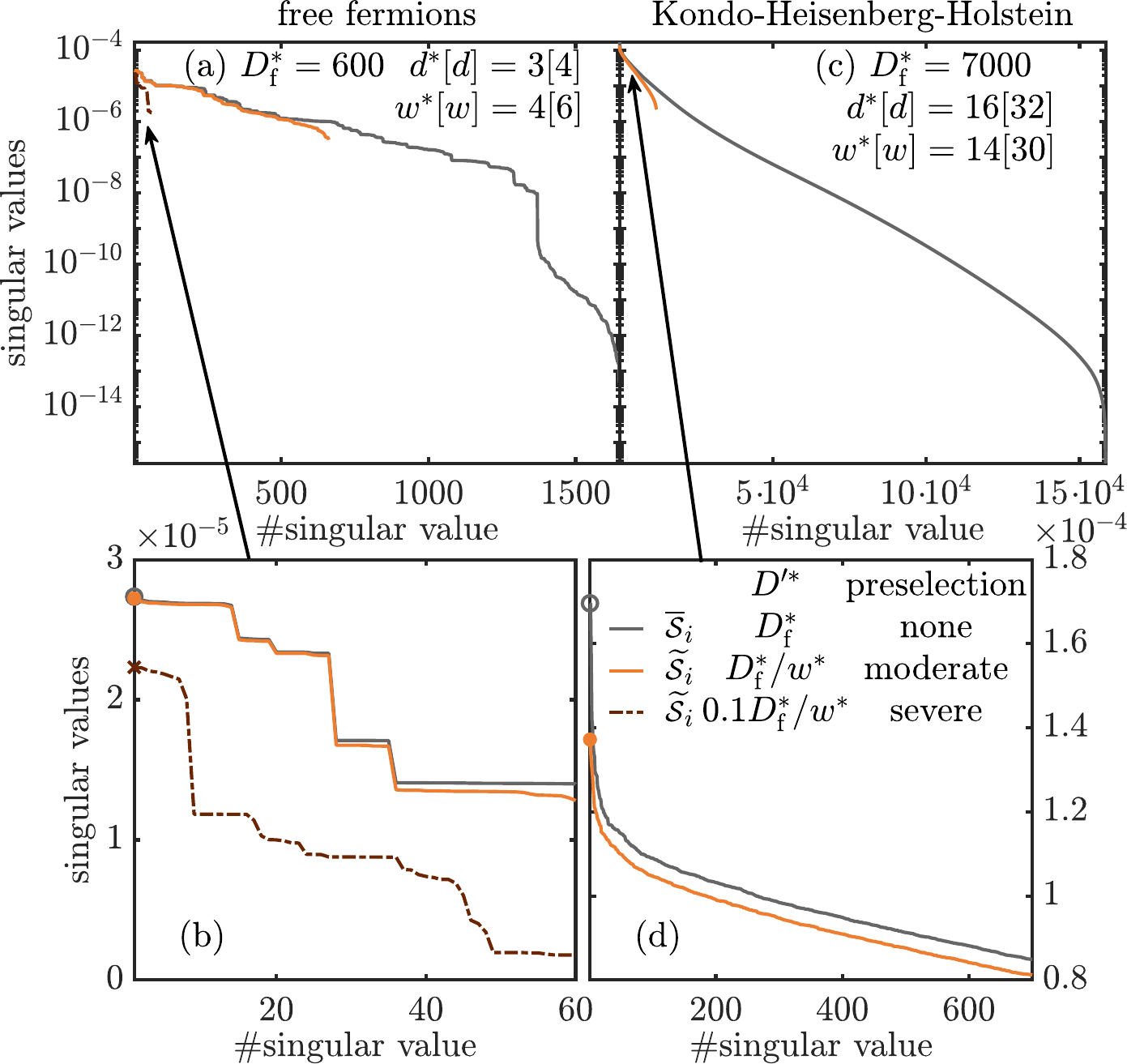} 
 \caption{
 \label{fig:preselection_recommended} 
Comparison of singular values for three truncation settings (I-III) defined in Sec.~\ref{sec:MotivationPreselection}:
the singular values $\protect\Scb_i$ of the tensor $\Mb{}^\full$,
obtained (I) without preselection (reference, grey); and the singular values $\protect\Sct_i$ of the tensor $\Mh{}^\pruned$,
obtained using (II) moderate 
preselection ($\Dpast \!=\! \Dfast /w^\ast$, orange) and (III) severe preselection 
($\Dpast \!=\! 0.1 \Dfast/\wast$, brown), all
followed by final selection with $\Dtast \!=\! 0.1\Dfast$. 
They are all computed for 
bond $\ell \!=\! \eLL/2$ of (a,b) the free fermion chain of Fig.~\Figurethree, 
and (c,d) the KHH cylinder of Fig.~\ref{fig:KondoCyl}(d).
(b,d) Subsets of the data from (a,c), shown on 
linear scales, focusing on the range of the
largest $\Dtast \!=\! \Dfast \delta$ 
 singular values $\protect\Scb_i$ and  $\protect\Sct_i$ (with $\delta \!=\! 0.1$). This range contains all singular vectors
comprising the truncated complement
$\Atrunc (\TriangleOrangeA)$ obtained after final selection and used for bond expansion.
The singular values found with moderate (orange) or no (grey) preselection agree  rather well, but those from severe preselection (brown) 
 differ significantly from these.}
 \end{figure}
 
 \begin{figure}[t!]
  \includegraphics[width=\linewidth]{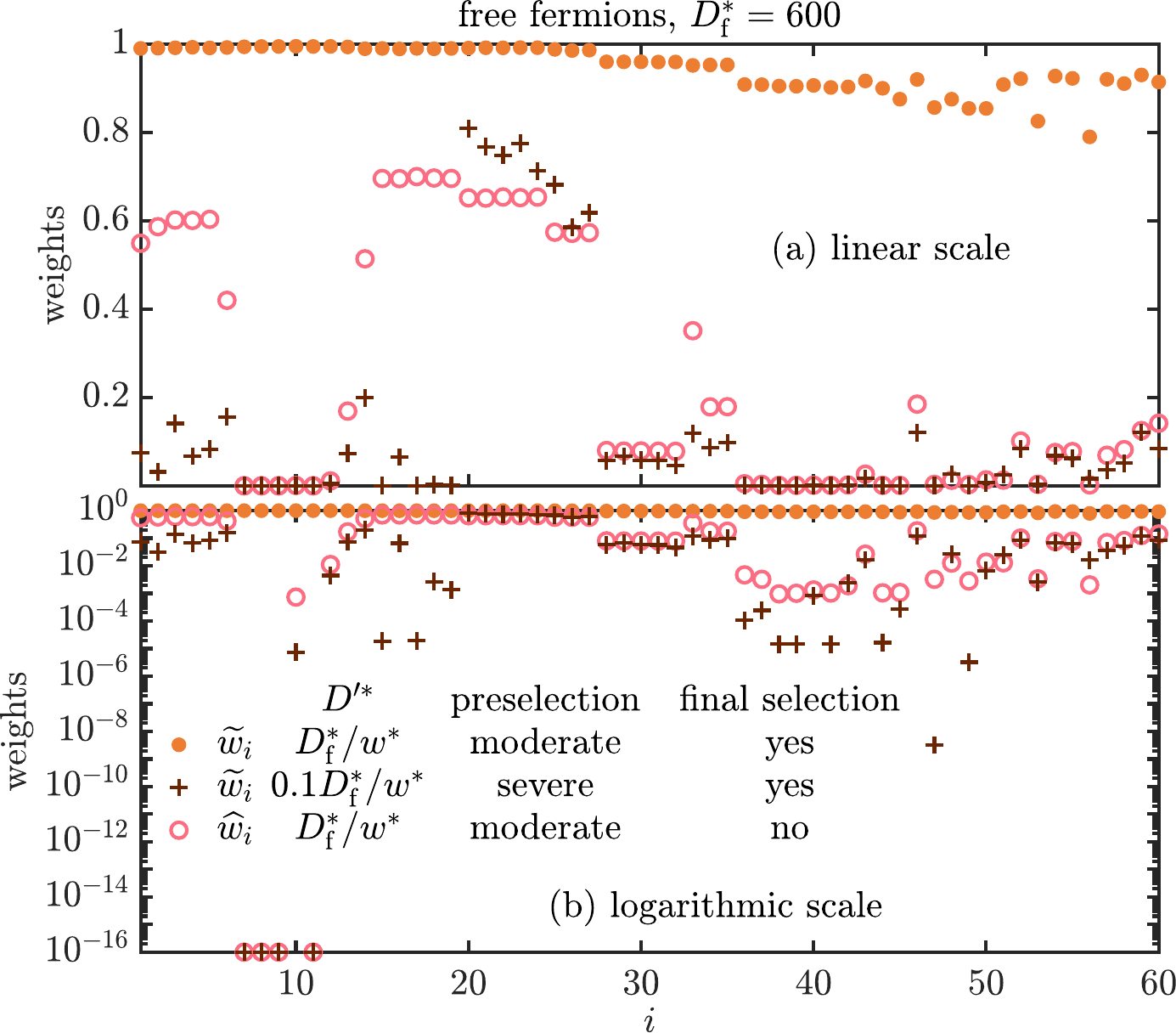} 
 \caption{
 \label{fig:preselection_vectors} 
Comparison of weights \eqref{subeq:weights} with which reference singular vectors 
$\protect\Scb_i$ from $\Abtrunc_\ell (\TriangleYellowA)$
are supported in truncated spaces obtained
with three truncation settings (II-IV) defined in Sec.~\ref{sec:MotivationPreselection}:
$\wt_i$ gives the weight of $\ket{\protect\Scb_i}$ in 
$\mathrm{span}\{\ket{\protect\Sct_j}\}$, 
the image of $\Atrunc_\ell (\TriangleOrangeA)$,
computed through 
\shrewd\ selection, using either (II) moderate (orange dots) or (III) severe (brown crosses) preselection;
and $\wh_i$ gives the weight of $\ket{\protect\Scb_i}$ in 
$\mathrm{span}\{\ket{\protect\Sch_j}\}$, 
the image of $\Apruned_\ell$ (\TriangleRedA), computed using (IV) moderate preselection without final selection (pink circles). Both panels show the same data, 
on (a) a linear and (b) a log scale.
}
 \end{figure}

Figure~\ref{fig:preselection_recommended} compares $\Scb_i$ (grey) and $\Sct_i$ (orange:
moderate or brown: severe preselection) for bond $\ell = \eLL/2$ of both the least and most challenging models considered in this work:  
(a,b) the free fermion chain of Fig.~\Figurethree,
and (c,d) the KHH cylinder of  Fig.~\ref{fig:KondoCyl}.
Here, we consider the case that  $\Dfast$ has reached $\Dmaxast$ and is not grown further,
and hence choose $\Dtast \!=\!  \Dfast \delta $ (with $\delta = 0.1$),
so that $\Diast=\Dfast$.
For (II) moderate preselection ($\Dpast \!=\! \Dfast/\wast$)
the $\Sct_i$ (orange) and $\Scb_i$ (grey) values coincide
quite well in the range where they are largest, and eventually drift apart 
as they get smaller. Especially for the largest $\Dtast \!=\! \Dfast \delta$ ($\delta \!=\! 0.1$) singular values, i.e.\ the ones that survive final selection and are used for bond expansion,  the agreement is rather good  (Figs.~\ref{fig:preselection_recommended}~(b,d)). This is a very important finding---it indicates that moderate preselection 
is efficient.
By contrast, (III) severe preselection ($\Dpast \!=\! 0.1 \Dfast/\wast$), shown
only in Fig.~\ref{fig:preselection_recommended}~(a,b), yields $\Scb_i$ (brown) values
that differ substantially from their $\Sct_i$ (grey) counterparts, even in the range of largest values. Therefore, in this case preselection is too severe to be very efficient. 
(We note in passing that when using severe preselection, the corresponding final selection involves
almost no further truncation, since $\Dhast$ (given by $\simeq \Dpast \wast
= 0.1 \Dast$) is almost equal to $\Dtast$ (given by $\Dast \delta$).
For the present example, we have $\Dhast \!=\! 63$ and $\Dtast \!=\! 60$.) 
In Figs.~\ref{fig:preselection_recommended}~(a,b), the length 
of the grey vs.\ orange lines visually illustrates the main rationale for our
CBE strategy: the number of $\Scb_i$ values is generally very much larger than needed 
for successful bond expansion, $\Db^\ast \gg \Dtast$. Thus, the \twosite\ full complement subspace (obtained by excluding the \onesite\ variational 
space from the \twosite\ variational space), is likewise much 
larger than needed for energy minimization---only a small subspace thereof 
really matters. CBE aims to identify parts of that small subspace; \shrewd\ selection offers a cheap way of doing so, yielding a notable speedup when computing the truncated complement.

\subsection{Singular vectors}
\label{sec:SingularVectors}

We next turn to a comparison of singular vectors to further quantify the benefits of using (II) moderate rather than (III) severe preselection, and of using final selection. 

For the latter purpose, we consider a truncation scheme (V)
involving moderate preselection but no final selection: 
after the minimization of the cost function
$\Cc_2$ (see Fig.~\ref{fig:pruningexplained_supp}(c)), 
we directly truncate  $\Uh \, \Sh \, \Vh{}^\dag \simeq \uh  \, \sh \, \vh{}^\dag$
from $\Dhast$ to $\Dtast$, and define the truncated complement as $\Ahtrunc_{\ell} = \uh$~(\TrianglePinkA), with singular vectors $\ket{\Sch_i}$.

To compare singular vectors we compute the weights 
\begin{subequations}
\label{subeq:weights}
\begin{align}
\widetilde{w}_i &= \sum_{j=1}^{\Dtast} |\langle \Sct_j | \Scb_i \rangle|^2  =
\! \! \raisebox{-5.5mm}{ \includegraphics[width=0.1\linewidth]{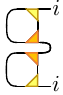}}
\, , 
\\
\widehat{w}_i &= \sum_{j=1}^{\Dtast} |\langle \Sch_j | \Scb_i \rangle|^2
=\! \! \raisebox{-5.5mm}{ \includegraphics[width=0.1\linewidth]{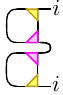}} \, .
\end{align}
\end{subequations}
Here, $\wt_i$ is the weight with which a singular vector 
$\ket{\Scb_i}$ (ordered by size of corresponding singular value) from the image of $\Ab{}^\trunc_{\ell}$ (\TriangleYellowA)
is supported in the subspace $\textrm{span}\{\ket{\Sct_j}\}$,
the image of $\Atrunc_{\ell}$ (\TriangleOrangeA); 
and $\wh_i$ gives its weight in 
$\textrm{span}\{\ket{\Sch_j}\}$, the image of $\Ah{}^\trunc_{\ell}$ (\TrianglePinkA).
In less technical terms, the weights characterize how well  reference singular vectors can be represented in these truncated spaces. 

These weights are shown in Fig.~\ref{fig:preselection_vectors} for the free fermion data corresponding to Fig.~\ref{fig:preselection_recommended}(a). For 
(II) moderate preselection plus final selection ($\widetilde{w}_i$, orange dots), 
all weights are close to one. Thus, this truncation scheme almost
perfectly captures that part of the \twosite\ subspace most relevant for 
minimizing the GS energy. By contrast, for both (III) severe preselection plus final selection ($\wt_i$, brown crosses) and (V) moderate preselection without final selection ($\wh_i$, pinc circles), most weights are significantly smaller than 1; four are numerically zero. Thus, both these schemes discard a significant part of the space relevant for minimizing the GS energy. 

The above analysis illustrates that final selection 
includes valuable additional information for the
$\Dhast \to \Dtast$ 
truncation, which is not available when 
truncating $\Sh$ from $\Dhast \to \Dtast$ directly after preselection. This is 
because the central MPO bond, open during preselection,
is \textit{closed} during final selection (compare
their cost functions, $\Cc_2$ and $\Cc_3$ in Fig.~\Figureone).
Closing the central MPO bond, as appropriate for $H^\mtwosite_\ell\psi^\mtwosite_\ell$,
brings in additional information.
The SVD in step (d) of Fig.~\ref{fig:pruningexplained_supp}\ involves an additional rotation (encoded in $\ut$) before the $\Dhast \to \Dtast$ truncation, incorporating this additional information.

\subsection{Convergence rate per sweep}
\label{sec:ConvergenceRates}

\begin{figure}[t!]
 \includegraphics[width=\linewidth]{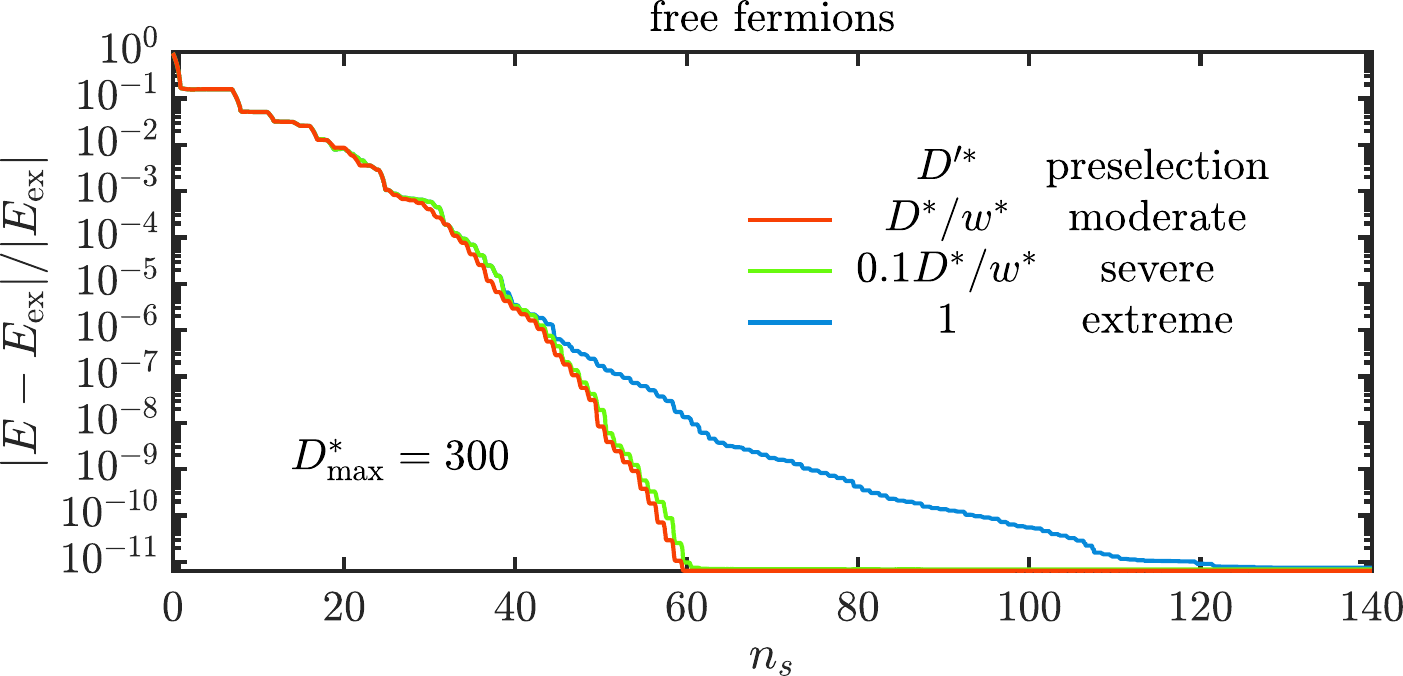} 
 \caption{
 \label{fig:extreme_preselection}
Influence of preselection on CBE--DMRG convergence rate, for a half-filled free-fermion chain ($\eLL = N = 20$).
The GS energy is plotted as a function of the number of half-sweeps, $n_s$, for three values of $\Dpast$, used for preselection. We start from a $\Dast_i = 1$ valence bond state, set $\delta=0.1$, 
increase $\Dast$ using $\alpha=1.1$ until $D^\ast = 300$ is reached, and continue
sweeping with $\alpha=1$ thereafter. }
 \end{figure}

The weights obtained for severe preselection ($\Dpast \!= \delta D^\ast/w^\ast$) in Fig.~\ref{fig:preselection_vectors} pose the question whether $\Dpast$ can be too small to give converged results.
In this case, preselection would not only be inefficient, but actually unsuccessful.
To explore this, Fig.~\ref{fig:extreme_preselection} compares the CBE--DMRG convergence rate for several choices of $\Dpast$, corresponding to  (II) moderate (red), (III) severe (green), and (IV) extreme (blue) preselection.

As expected, convergence slows down with smaller $\Dpast$. Remarkably, however, once convergence has been reached, the converged results agree (even for $\Dpast=1$, a truly
extreme choice!). In this sense, the preselection strategy is robust---converged
results don't depend on $\Dpast$. Note, though, that the computation time 
does not depend significantly on $\Dpast$ (provided it is clearly smaller than 
$\Dast$). On the other hand, it obviously does depend on the number of sweeps,
and the time per sweep can be very large for expensive models.  Therefore, $\Dpast$
should not be chosen too small, to avoid a time-costly increase in 
the number of sweeps. 
To summarize: a bond expansion is 
\textit{efficient}, yielding a significant reduction in GS energy and therefore quick convergence, if $\Dpast$ is large enough that the ``most important''  states $\ket{\Scb_i}$, i.e.\ those with the largest singular values $\Scb_i$, are well represented in the expanded space, i.e.\ have weights $\wt_i \simeq 1$. 

However, even if $\Dpast$ is so small that most 
of the important states $\ket{\Scb_i}$ are represented with small weights, 
a bond expansion can nevertheless be \textit{successful}, in the sense of adding some relevant new states,  provided that these weights are non-zero, $\wt_i\neq0$.
The reason is that the states $\ket{\Sct_i}$ added to $A_\ell (\TriangleWhiteA)$ contain information about the optimal states $\ket{\Scb_i}$ with finite $\wt_i$, i.e.\ those $\ket{\Scb_i}$ are not orthogonal to the expanded kept space.
As long as this information is available, subsequent \onesite\ updates will optimize the kept sector accordingly; the states $\ket{\Sct_i}$ just offer a somewhat less optimal starting point for that than the $\ket{\Scb_i}$. 

Note that it is of utmost importance for successful bond expansion that information on the \textit{most important} $\ket{\Scb_i}$ is included. Since only a small set of states is in the end used for expansion, the most important states must be prioritized; otherwise, inferior information is included in the kept space, rendering the bond expansion unsuccessful:
Subsequent \onesite\ updates may then optimize towards a suboptimal kept sector, as the optimal one may not be available to the \onesite\ update, e.g. due to symmetry constraints. The energy will still decrease due to the unsuccessful bond expansion plus \onesite\ update, but not as much as if the correct information on the most important $\ket{\Scb_i}$ is correctly included. The result will be a suboptimal final state at the desired finite bond dimension $\Dmax^\ast$, i.e.\ we have wasted resources.

Fig.~\ref{fig:extreme_preselection} shows that CBE--DMRG correctly includes information on the most optimal states when expanding the bond, independent of $\Dpast$. Even with extreme preselection ($\Dpast=1$), it does not get stuck with some sub-optimal state at $\Dmax^\ast = 300$, but eventually converges (albeit slowly) to the 
same GS as found with larger choices of $\Dmax$.
\section{Simple Benchmark: Free Fermions\label{sec:simple_benchmark}}

In this section, we benchmark CBE--DMRG for  free fermions in one dimension (1D).
The main purpose is to evaluate the validity of the CBE discarded weight as an error measure
usable for extrapolation on an exactly solvable model and compare it to other established error measures.
All CPU time measurements were done on a single core of an Intel Core i7-9750H processor.

Consider a chain of spinful free fermions, exactly solvable but non-trivial for DMRG, with Hamiltonian 
$H_{\mathrm{FF}} = -\sum_{i=1}^{\sseLL-1} \sum_{\sigma} 
\bigl(c^{\dagger}_{i\sigma} c^{\phantom{\dagger}}_{i+1\sigma} + \mathrm{h.c.}\bigr)$
and $\eLL=100$ sites.
We exploit $\mathrm{U}(1)_{\mathrm{ch}}\otimes \mathrm{SU}(2)_{\mathrm{sp}}$ charge and spin symmetry, with local dimension $d^\ast [d] \!=\! 3[4]$.
The MPO dimension is $w^\ast [w] \!=\! 4[6]$.
We seek the GS in the sector with total spin $S\!=\!0$, at half-filling,
with particle number $N\!=\!\eLL$.
%

%
 \begin{figure}
 \includegraphics[width=\linewidth]{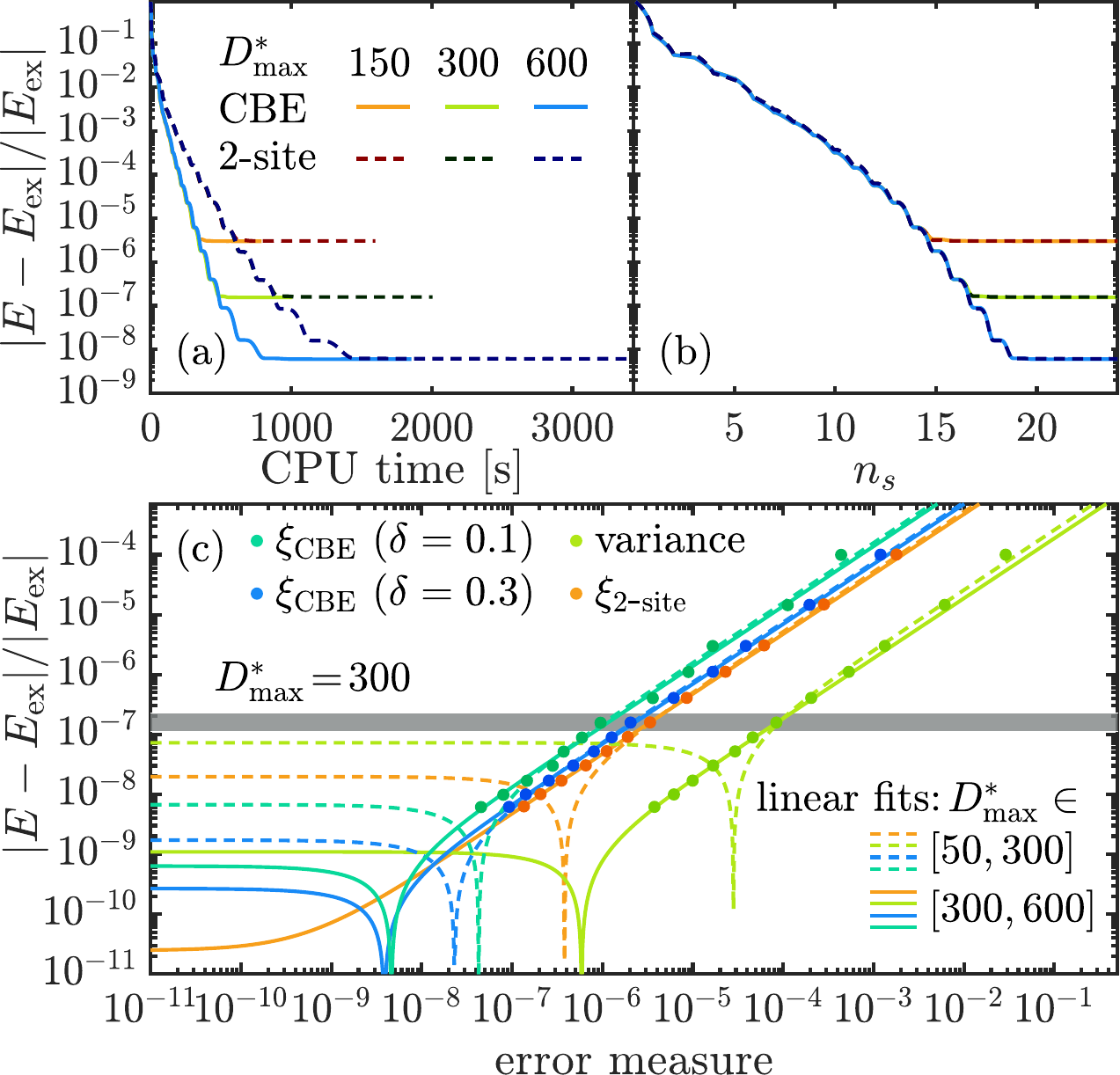} \vspace{-6mm}
 \caption{
 \label{fig:FreeFermions}
Benchmark results for free fermions. Relative error in GS energy vs.\ (a) CPU time xt
and (b) number of half-sweeps $n_s$, for CBE and \twosite\  DMRG.
$E_{\mr{ex}}$ is the exact GS energy. 
(c)
Quality of linear extrapolation of the GS energy using various error measures. 
Dashed (solid) lines show linear fits to
data points lying on or above (on or below) the grey
bar, computed using
$\Dmaxast \le 300$ ($\ge 300$), 
representing intermediate (high) accuracy calculations;
when these lines touch zero, the extrapolated error changes sign.}
\vspace{-4mm}
 \end{figure}

Figure~\ref{fig:FreeFermions}(a) plots the relative error in energy 
vs.\  
CPU time for different $D^\ast_\maximum$ for both CBE and \twosite\ schemes;  Fig.~\ref{fig:FreeFermions}(b)
plots it vs.\  
the number of half-sweeps $n_s$. While convergence with $n_s$ is comparable for CBE and \twosite,
CBE requires less CPU time than \twosite\
by a factor of $\simeq 2$. 
(This speedup factor is less than $\dast = 3$, since $\dast$ is quite small and steps not involving the iterative eigensolver have the same numerical cost for both CBE and \twosite\ schemes.)

Figure~\ref{fig:FreeFermions}(c) shows 
linear-fit extrapolations of the energy in terms of the  discarded weight $\xi$ and the \twosite\ variance (the latter computed
following Ref.~\cite{Hubig2018}).
The quality of the extrapolations 
is comparable for all considered methods: they all reduce the error in energy by roughly one order compared to the most accurate data point considered, as expected~\cite{White2005,Hubig2018}. The 
error is smaller for  $\delta = 0.3$ than for $\delta = 0.1$, 
and its dependence on discarded weight is slightly less noisy (though this hardly affects the extrapolation). 
%

\section{Comparison of CBE to DMRG3S}
\label{sec:DMRG3S}
In this section, we provide a comparison between DMRG3S and CBE--DMRG. First, we formulate DMRG3S in terms of the kept-discarded~($\K\D$) space language developed by us in Ref.~\onlinecite{Gleis2022a}
and also used in this paper.
Based on that, we then discuss to what extent the bond-expansion term in DMRG3S is different to that occurring in CBE-DMRG. We then compare the performance of DMRG3S and CBE--DMRG based on two models. 
\subsection{DMRG3S in $\K\D$ language}
In case of a right-to-left sweep, DMRG3S expands and truncates the right isometry as follows:
\begin{align}
\hspace{-0.5cm}
\raisebox{-0.1\height}{
\includegraphics[width=\linewidth]{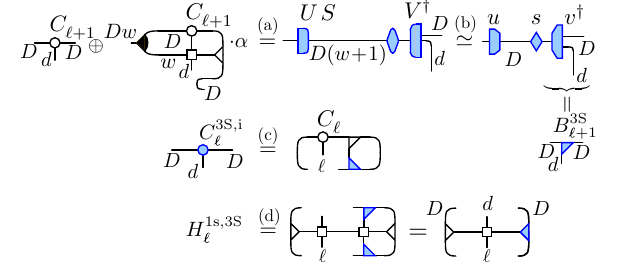}}\hspace{-1cm} \, ,
\label{eq:DMRG3S-mixing}
\end{align}
where \QuaterCircleBlackThreeS represents a unitary, in analogy to Eq.~(6) in the main text.
Here, $C_{\ell+1}$ is direct-summed with the expansion term multiplied by a mixing parameter $\alpha$, then (a) singular value decomposed and (b) truncated to bond dimension $D$, yielding the new isometry $B^{\mr{3S}}_{\ell+1}$. Finally in steps (c) and (d), $C_{\ell}$ and $H^{\onesite}_{\ell}$ are ``expanded'',
respectively, similar to CBE 
(DMRG3S first updates $C_{\ellplusone}$ and then uses the mixing expansion of Eq.~\eqref{eq:DMRG3S-mixing} to expand $B_{\ellplusone}$;
by contrast, CBE first expands $A_{\ell}$ via Eq.~\eqref{eq:expand-bond}, then updates $C_{\ellplusone}$). Note that in step (c),
$C_{\ell}^{\mr{3S,i}}$ needs to be normalized explicitly because
\begin{align}
\raisebox{-3.0mm}{\includegraphics[width=0.21\linewidth]{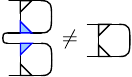}} \,\, ,
\end{align}
i.e.\ the kept space spanned by the old isometry $B_{\ell+1}$ is not fully contained in the new one, $B^{\mr{3S}}_{\ell+1}$, since part of the kept space has been truncated.
Finally, $C_{\ell}^{\mr{3S}}$ is updated with the GS of $H^{\monesite,\mr{3S}}_{\ell}$ obtained with an iterative eigensolver (Lanczos in our case), 
initialized with $C_{\ell}^{\mr{3S,i}}$.
Our CBE strategy differs from DMRG3S in the following ways:
\noindent (i)
When constructing the expansion term, CBE considers $H^{\mtwosite}_{\ell}\psi^{\mtwosite}_{\ell}$, 
i.e.\ the action of the full \twosite\ Hamiltonian on the \twosite\ wavefunction.
By contrast, DMRG3S only considers part of $H^{\mtwosite}_{\ell}\psi^{\mtwosite}_{\ell}$ (the right ``half'' in the right-to-left sweep discussed here).
We found however that considering $H^{\twosite}_{\ell} \psi^{\twosite}_{\ell}$ fully is crucial to not experience convergence issues.
Note that the expansion term in DMRG3S is more heuristic than that in CBE and does not have the interpretation of an effective Hamiltonian acting on a wavefunction.

\noindent (ii)
CBE projects $H^{\mtwosite}_{\ell}\psi^{\mtwosite}_{\ell}$ fully to the $\D\D$ sector, i.e.\ the image of the orthogonal complements $\Ab_\ell \otimes \Bb_\ellplusone \, (\TriangleGreyA \otimes \TriangleGreyB)$. This ensures that the kept space is not truncated during the bond expansion and crucially, the energy of the variational wavefunction 
remains the same. By contrast, DMRG3S does not involve any $\D\D$ or $\D$ projections. Thus, part of the $\K$ sector is usually truncated during the DMRG3S
bond expansion, raising the energy of the variational wavefunction.
Thus, CBE--DMRG is fully variational (bond expansion does not lead to a less optimal wavefunction) while DMRG3S is not (see also the discussion of Fig.~1 of Ref.~\onlinecite{Hubig2015}).

\noindent (iii)
Because DMRG3S changes the variational wavefunction by truncating part of the $\K$ sector, the weight of the expansion term in DMRG3S
has to be controlled by a heuristic mixing factor $\alpha$ to ensure the variational energy is not 
raised too much. This mixing factor has to be carefully adapted during the calculation to ensure reliable convergence 
and is model dependent (see Ref.~\onlinecite{Hubig2015} Sec.~VI). By contrast, there is no such mixing parameter in CBE. 
In CBE, there is a parameter $\delta$ which controls the amount of bond expansion.
We found however that CBE--DMRG is not at all sensitive to the value of $\delta$
and most important, $\delta$ is not model dependent. Indeed, we have 
set $\delta=0.1$ in our CBE calculations independent of the model.
Further, $\delta$ remains constant during the calculation.

Note that if 3S would include projections to the $\D$ sector and would not truncate part of the kept space during expansion, it would be similar to CBE without preselection and final selection. However, leaving out preselection is expensive while leaving out final selection is inefficient (see Sec.~\ref{sec:preselection}).
\subsection{Results}
We now benchmark the accuracy and speed of DMRG3S against that of CBE--DMRG. 
For that, we use three models: a 1D Hubbard-Holstein model, spinful free fermions on a short 4-leg cylinder
and a free fermion chain with only next-nearest neighbor hopping.
All CPU time measurements were done on a single core of an Intel Core i7-9750H processor. 
%

 \begin{figure}[t!]
 \includegraphics[width=\linewidth]{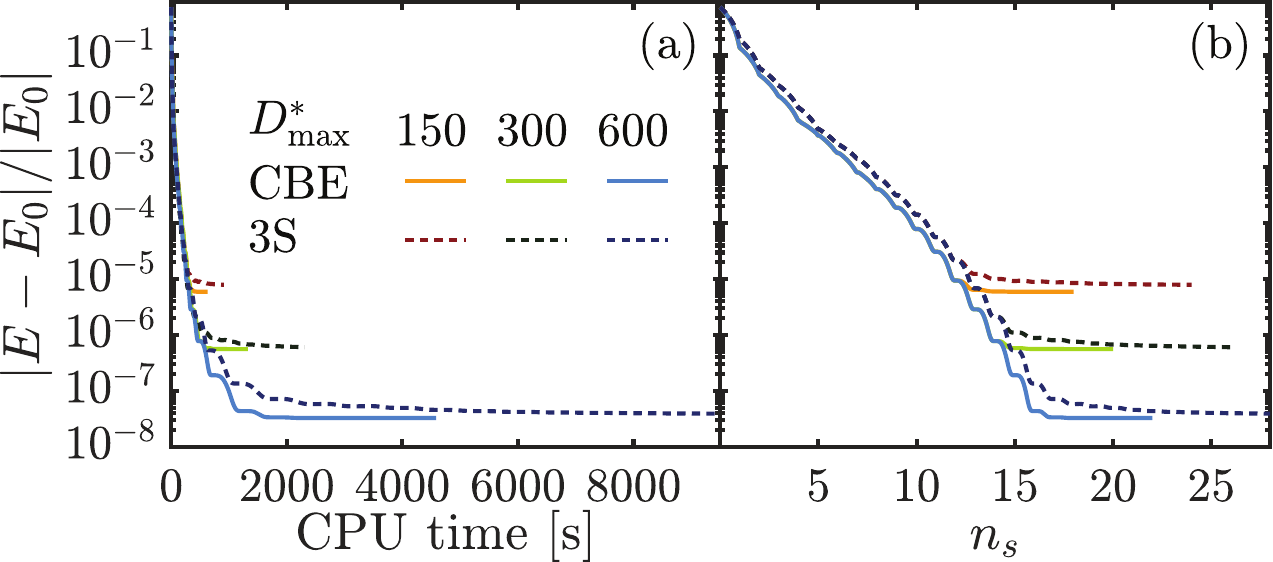} 
 \caption{
 \label{fig:HubbardHolstein_CBE_vs_3S} 
Error in energy for the Hubbard-Holstein (HH) model versus (a) CPU time and (b) number of half-sweeps $n_s$,
computed using CBE--DMRG (solid) or DMRG3S (dashed).
$E_{0}$ is obtained via $\xi$-extrapolation of calculations done at $\Dast \geq 1000$.}
 \end{figure}
\textit{Hubbard-Holstein model.---}
We start our comparison with the 1D Hubbard-Holstein model~\cite{Jeckelmann1998,Tezuka2007,Fehske2008,Ejima2010,Reinhard2019}, 
with  Hamiltonian 
\begin{align}
H_{\mr{HH}} 
&= -\sum_{i\sigma} \bigl(c^{\dagger}_{i\sigma}c^{\phantom{\dagger}}_{i+1\sigma} + \textrm{h.c.}\bigr) 
+ U \sum_{i} n_{i\uparrow}n_{i\downarrow} \\  \nonumber 
&+\omega_\mr{ph} \sum_{i} b^{\dagger}_i b^{\phantom{\dagger}}_i
+ g\sum_{i} \bigl(n_{i\uparrow}+n_{i\downarrow} - 1\bigr)\bigl(b^{\dagger}_i + b^{\phantom{\dagger}}_i\bigr)
\, .
\end{align}
We chose  $U=0.8$, $g=\sqrt{0.2}$, $\omega_\mr{ph}=0.5$, $\eLL = N = 100$, total spin $S\!=\!0$, and restrict the maximum local number of excited phonons to $\Nphmax=3$.
Both CBE--DMRG and DMRG3S are initialized with the same $D^\ast=1$ MPS with uniform charge distribution and the bond dimension is grown by a factor of $\sqrt{2}$ every half sweep, i.e.\ it is doubled every sweep. 
The DMRG3S mixing parameter is adapted 
according to the prescription described in Ref.~\onlinecite{Hubig2015}, Sec.~VI.
Figures~\ref{fig:HubbardHolstein_CBE_vs_3S} (a) and (b) show a comparison between the error in energy versus CPU time and number of sweeps, respectively, for different bond dimensions. As a function of CPU time, the error in energy of DMRG3S initially converges at the same rate as CBE--DMRG.
Subsequently, however,
the convergence of 3S slows down compared to CBE,
ultimately requiring
significantly more CPU time to reach the final converged result. Further, the final converged 3S result is not as accurate as the CBE result,
though this is more severe at small $D^\ast$ than at large $D^\ast$. 
At $\Dast = 150$,
the relative error from 3S is about 1.3 times that of CBE.
%

 \begin{figure}[t!]
 \includegraphics[width=\linewidth]{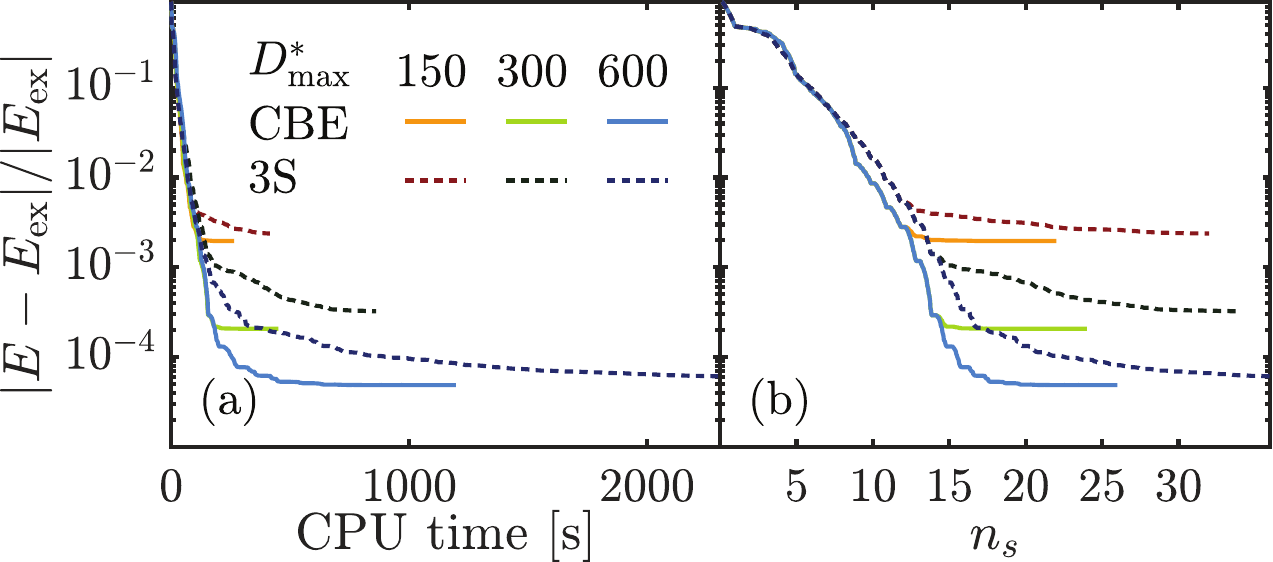} 
 \caption{
 \label{fig:FFCyl_10x4_CBE_vs_3S} 
Error in energy for spinful free fermions on a $10\times4$ cylinder versus (a) CPU time and (b) number of half-sweeps $n_s$.
$E_{\mr{ex}}$ is the exact ground-state energy.}
 \end{figure}
\textit{Spinful free fermion cylinder.---}
For our next benchmark, we use free fermions on a $\eLL_{\!x}\times\eLL_{\!y} \! = \!10 \times 4$ cylinder,
described by $H_{\mr{cyl}} = -\sum_{\langle \boldell, \boldellp \rangle,\sigma} 
\bigl(c^{\dagger}_{\boldell\sigma}c^{\phantom{\dagger}}_{\boldellp\sigma}\!   +\!  
\mathrm{h.c.}\bigr) $.
We search for the GS with $N = \eLL_{\!x}\cdot\eLL_{\!y}$ and $S=0$. Again, we start with a $D^\ast=1$ state with uniform charge distribution and
increase the bond dimension by a factor of $\sqrt{2}$ every half-sweep.
Figures~\ref{fig:FFCyl_10x4_CBE_vs_3S} (a) and (b) show a comparison of the error in energy versus CPU time and the number of sweeps obtained with both CBE and 3S, respectively. 
Again, CBE and 3S initially converge at the same rate w.r.t. CPU time, but DMRG3S eventually slows down and takes longer to reach final convergence compared to CBE.
Further, for all considered bond dimensions, 3S converges now to a noticeably larger error, about 1.2 to $>\!1.5$ times that of CBE.
%
 
 \begin{figure}[b!]
 \includegraphics[width=\linewidth]{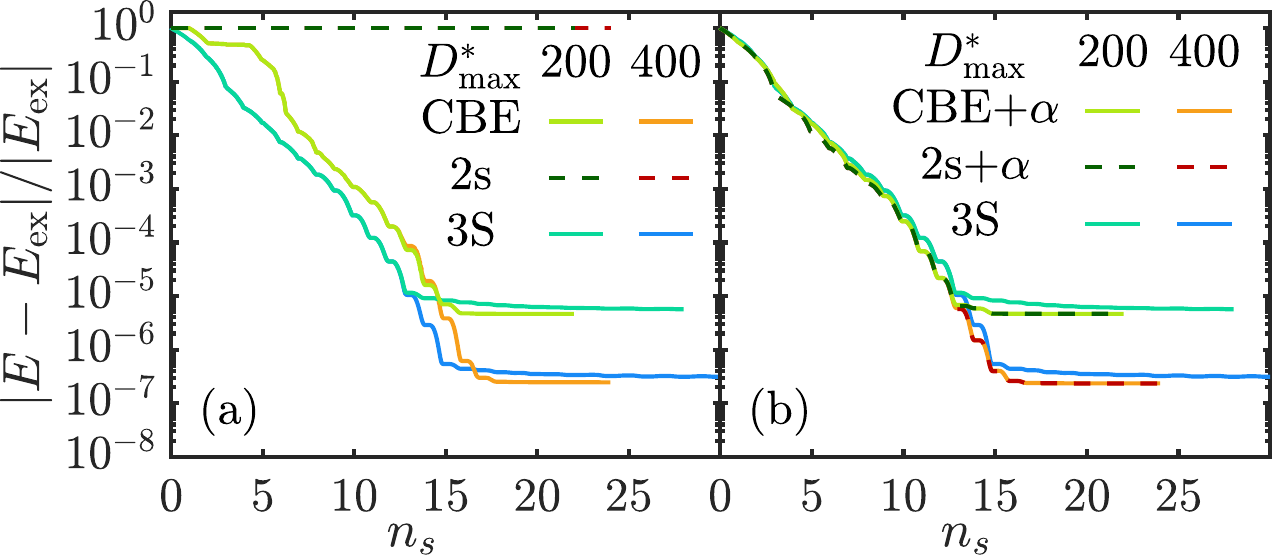} 
 \caption{
 \label{fig:FFLadder_CBE_vs_3S_vs_2S} 
Error in energy for the next-nearest neighbor free fermion chain, 
computed using CBE and \twosite, (a) without mixing,
and (b) with mixing ($\alpha=0.1$), during the initial 14 half-sweeps. DMRG3S results 
in (a) and (b) are the same data. $E_{\mr{ex}}$ is the exact ground-state energy.}
 \end{figure}

 \textit{Next-nearest neighbor free fermion chain.---}
 As a last model for our comparison, we choose free fermions on a chain with only next-nearest neighbor hopping, described
 by $H_{\mr{nnn}} = -\sum_{\ell=1}^{\scripteLL-2} (c^\dag_{\ell} c^\pdag_{\ell+2}  + \mr{h.c.})$. 
 Choosing $\eLL=100$ and exploiting $\mathrm{U}(1)_{\mathrm{ch}}$ symmetry, 
 we initialize DMRG with a 
 half-filled product state consisting of a succession of two occupied sites followed by two empty sites.
 
As shown in Fig.~\ref{fig:FFLadder_CBE_vs_3S_vs_2S}(a), this rather simple model initialized with the product state described above poses a serious challenge to \twosite\ DMRG, which does not converge. 
The reason for the failure of \twosite\ DMRG is that 
the initial state has $\variance^{1\perp}=0$ and $\variance^{2\perp}=0$ (c.f. Eq.~(8)),
implying that $H^{\mtwosite}_{\ell}\psi^{\mtwosite}_{\ell}$ is
parallel to $\psi^{\mtwosite}_{\ell}$. From the perspective of \twosite\ DMRG, the initial state is therefore an eigenstate.

By contrast, both DMRG3S and CBE--DMRG do converge, with CBE--DMRG again reaching convergence faster in terms of number of sweeps and converging to a sightly lower energy than DMRG3S. During the initial few sweeps,  CBE--DMRG lowers the energy somewhat more slowly than DMRG3S, reflecting the close relation between CBE--DMRG and \twosite\ DMRG. 
In contrast to the latter, however, CBE--DMRG eventually does converge. 
The reason is that CBE expands the MPS bond from $D$ to $D+\Dt$ even if the projection of $H^{\mtwosite}_{\ell}\psi^{\mtwosite}_{\ell}$ to $\D\D$ yields zero --- indeed, 
final selection (Fig.~2(d)) generates $\Dt$ additional states even if some or all of the 
associated singular values (from $\st$ in Fig.~2(d)) are numerically zero. This enlarges the kept space from $D$ to $D+\Dt$, such that eventually $\variance^{1\perp}$ becomes nonzero and the energy can be lowered during the CBE \onesite\ update.

As suggested in Ref.~\onlinecite{Stoudenmire2012}, Section~3.1, adding noise terms 
in the spirit of DMRG3S or density matrix perturbation of Ref.~\onlinecite{White2005} during the initial few sweeps can help \twosite\ DMRG to converge.
The same is true for CBE--DMRG, which also struggles during the initial sweeps in the present case, as mentioned above. To demonstrate this, we therefore performed \twosite\ and CBE calculations combined with
DMRG3S mixing, dubbed \twosite$+\alpha$ and CBE$+\alpha$, respectively. 
(For  CBE$+\alpha$, first the CBE expansion of $A_{\ell}$ according to \Eq{eq:expand-bond} is used, then $C_{\ellplusone}$ is updated,
and finally a mixing expansion of $B_{\ellplusone}$ according to \Eq{eq:DMRG3S-mixing} is used.) We choose $\alpha=0.1$ during the initial 7 sweeps (i.e.\ 14 half-sweeps) and then continue without mixing. 
Note that we do not need to fine-tune $\alpha$, in contrast to DMRG3S. The results of this strategy are displayed in Fig.~\ref{fig:FFLadder_CBE_vs_3S_vs_2S}(b), which shows that 
both \twosite$+\alpha$ and CBE$+\alpha$ converge similarly w.r.t.\ the number of sweeps. 

\textit{Summary of CBE to DMRG3S comparison.---} CBE generically converges significantly faster w.r.t. number of sweeps than 3S but takes about the same CPU time per sweep. This
leads to overall significantly faster convergence of CBE compared to 3S. Further, the accuracy of CBE is generically better than that of 3S at the same
bond dimension, meaning that CBE uses variational resources more efficiently than 3S. This seems to be especially the case for more challenging models
where single-site methods provide the most benefit due to reduced computational demands.
An exception are situations where \twosite\ DMRG fails entirely. 
In such cases, the convergence during the initial few sweeps is significantly slower for CBE than DMRG3S (though CBE eventually catches up, utimately reaching a lower final energy than DMRG3S). 
 The initial CBE convergence can be sped up, if desired, by including some mixing during the first few sweeps, using a  mixing parameter that need not be fine-tuned.
This strategy is the one we would recommend as standard practice when dealing with challenging models.

\section{Kondo-Heisenberg cylinders\label{sec:cylinder}}
In this section, 
we provide supplementary information on the two most challenging models considered in this work, both defined on a 4-leg cylinder: the Kondo-Heisenberg~(KH) model discussed in the main text, where we presented evidence for Fermi surface (FS) reconstruction; and the  Kondo-Heisenberg-Holstein (KHH) model, included here to
demonstrate the feasibility of using CBE for tackling truly complex models.
The KH model is relevant for heavy-fermion materials, which consist of itinerant conduction electrons, hybridizing with localized $f$ orbitals~\cite{Coleman2007}. At low energies, only the spin degree of freedom of the $f$ electrons remain, describable by a KH model,
\begin{align}
H_{\mr{KH}} 
&= 
-\sum_{\sigma=\uparrow,\downarrow}\sum_{\langle \boldell, \boldellp \rangle} 
\bigl(c^{\dagger}_{\boldell\sigma}c^{\phantom{\dagger}}_{\boldellp\sigma}  + h.c.\bigr) \nopagebreak \\ \nonumber 
&+ J_{\mr{K}}\sum_{\boldell} \vec{S}_{\boldell} \cdot \vec{s}_{\boldell} 
+ J_{\mr{H}} \sum_{\langle \boldell, \boldellp \rangle} \vec{S}_{\boldell}\cdot \vec{S}_{\boldellp} \,. 
\end{align}
Here, $c^{\dag}_{\boldell\sigma}$ is a fermionic creation operator at site $\boldell = (x,y)$ with spin $\sigma$,
$\vec{s}_{\boldell} = \tfrac{1}{2}\sum_{ss'} c^{\dag}_{\boldell s}\vec{\sigma}_{ss'} c^{\pdag}_{\boldell s'}$ is the corresponding electron spin operator and  $\vec{S}_{\ell}$ the spin operator of a spin-$\tfrac{1}{2}$ local moment, all for site $\boldell$.
The KHH model is obtained from the KH model by additionally including Holstein phonons, motivated by experimental data suggesting that phonons
may play a role in heavy-fermion physics~\cite{Ye2022}:
\begin{align}
H_{\mr{KHH}} 
 = H_{\mr{KH}} &+ \omega_{\mr{ph}} \sum_{\boldell} b^{\dag}_{\boldell} b^{\pdag}_{\boldell} 
+ g
\sum_{\boldell \sigma} (n_{\boldell \sigma} - \tfrac{1}{2})
\bigl(b^{\dagger}_{\boldell} + b^{\phantom{\dagger}}_{\boldell}\bigr) \, .
\end{align}%
Here, $b_{\boldell}^{\dag}$ is phonon creation operator for site $\boldell$. To deal with the infinite local phonon Hilbert space, we restrict the maximum number of local phonon excitations to $\Nphmax$ (specified below) in our DMRG calculations.
In Sec.~\ref{sec:supp_KHH} we first show stable convergence of CBE--DMRG for the KHH model on a $10\times4$ cylinder. 
Then, in Sec.~\ref{sec:cylinderFS}, we describe how to extract information on the FS in $40\times4$ 
KH cylinders from ground states computed with CBE--DMRG.
Finally, in Sec.~\ref{sec:LuttingersTheorem}, we show that our
KH cylinder results are consistent with 
Luttinger's sum rule, relating the electron density to the FS volume. 
\vspace{0.5cm}
\subsection{Kondo-Heisenberg-Holstein cylinders: convergence}\label{sec:supp_KHH}

\begin{figure}
 \includegraphics[width=\linewidth]{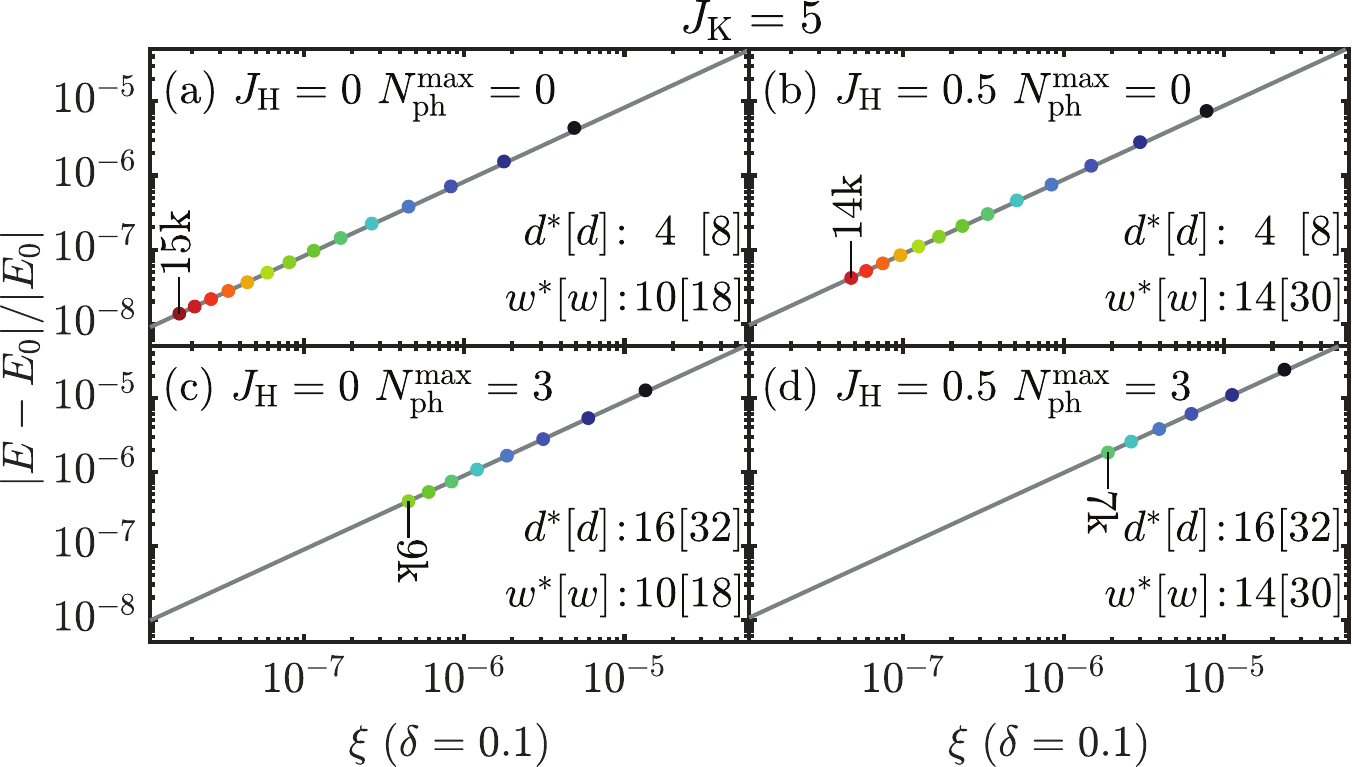} \vspace{-6mm}
 \caption{
 \label{fig:KondoCyl}
Error in GS energy versus discarded weight for the 
Kondo-Heisenberg-Holstein (KHH) model on a $10\times 4$ cylinder, with (a) only Kondo coupling, (b) Kondo and Heisenberg coupling, (c) Kondo and Holstein coupling and (c) Kondo, Heisenberg and Holstein coupling. Legends state our choices for $J_{\mathrm{H}} $ and $\Nphmax$, and corresponding values of $d^\ast [d]$ and 
$w^\ast [w]$. For each panel, 
$E_0$ was obtained by linear $\xi$-extrapolation to $\xi \!=\! 0$ (grey line) using
the four largest $D^\ast$ values. The very largest $D^\ast$  is 
shown next to its data point; 
$D^{\ast}$ changes by 1k between adjacent data points.
 } \vspace{-4mm}
 \end{figure}

 Our intention is to show that CBE--DMRG is stable for 
the KHH model, which is
at the edge of what is possible with current DMRG techniques.
To check the applicability of
CBE--DMRG 
to the KHH model on a $10\times4$ cylinder,
we use Kondo coupling $J_{\mathrm{K}}\! = \!5$, Holstein coupling $g\!=\!0.5$ to the phonons and optical phonon frequency $\omega_\mr{ph} \!=\! 0.5$.
We considered two different values for $\Nphmax \in \{0,3\}$ and the Heisenberg coupling $J_{\mr{H}} \in \{0,0.5\}$.
We performed GS searches for $N\!=\!\eLL(1\!+\!\tfrac{1}{4}) \!=\! 50$ and $S\!=\!0$, i.e.\ at 25\% electron doping. Figure~\ref{fig:KondoCyl} shows the energy error vs.\ $\xi$ for four 
parameter combinations (see legends). 
The linear $\xi$-dependence of $E$ demonstrates proper convergence of CBE--DMRG.  
Very large $D^{\ast}$ values are achievable despite the rather huge values of $d$ and $w$. This is remarkable especially for $J_{\mathrm{H}} \!=\! 0.5$ and $\Nphmax \!=\!3$ (Fig.~\ref{fig:KondoCyl}(d)), where $d^\ast[d] = 16[32]$ and $w^\ast[w] = 14[30]$ are large, so that \twosite\ schemes become excessively costly.
These results encouragingly illustrate the potential of CBE for handling very complex models.
\subsection{Kondo-Heisenberg cylinders: Fermi surface\label{sec:cylinderFS}}
%

 %

Having established stable convergence of CBE--DMRG for the KHH model on a $10\times4$ cylinder, we turn to the Kondo-Heisenberg~(KH) model on longer $40\times4$ cylinders.
In this section, we provide some supplementary information on our discussion of the Fermi surface~(FS) reconstruction in the KH model.
Heavy-fermion materials feature many interesting phenomena. One that is not so well understood is the so-called Kondo breakdown~(KB) quantum critical point~(QCP). When the system is tuned across this
KB--QCP, the FS volume abruptly changes~\cite{Coleman2001}, leading to a violation of Luttinger's sum rule
\cite{Luttinger1960} and strange metal behavior at finite temperatures.
In Fig.~6 of the main text, we have shown strong evidence for the existence of two distinct phases with different FS volumes in the KH model on a 4-leg cylinder.
This in turn strongly suggests the existence of a KB-QCP in the KH model on 4-leg cylinders, which can be studied in a non-perturbative, controlled and unbiased way using our newly developed CBE--DMRG method.
Here, our goal is 
to explain in detail how we extracted the Fermi points from our CBE--DMRG data
on the $40\times4$ KH cylinder,
thereby
establishing the two distinct phases reported in the main text.
We leave the
study and
discussion of 
a possible KB--QCP and its
 rich physics to future work.

 \begin{figure}[t!]
 \includegraphics[width=\linewidth]{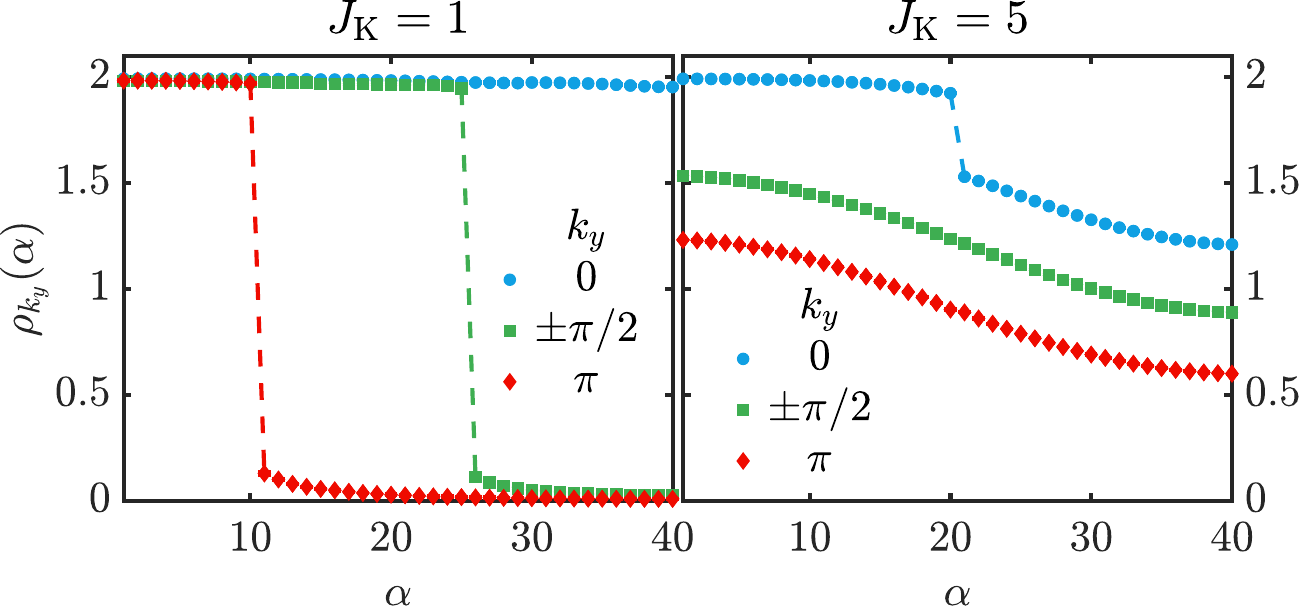} \vspace{-6mm}
 \caption{
 \label{fig:KondoCyl_rhok_ev}
$k_y$-resolved eigenvalues, $\rho_{k_y}(\alpha)$ of the single-particle density matrix of the Kondo-Heisenberg (KH) model on a $40\times4$ cylinder at $25\%$ electron doping, $J_{\mr{H}}=0.5$ and (a) $J_{\mr{K}} = 0.5$ and (b) $J_{\mr{K}} = 2.5$. Eigenvalues are extrapolated to truncation error $\xi \to 0$, error bars are below symbol sizes.
Dashed lines highlight jumps in the spectra.
 } \vspace{-4mm}
 \end{figure}
To illustrate our Fermi point extraction strategy, we here focus on $J_{\mr{K}} = 1$ and $J_{\mr{K}} = 5$, representative for the two phases with different Fermi surfaces at small and large $J_{\mr{K}}$, respectively.
We extract the Fermi points from the single-particle density matrix,
\begin{align}
\rho_{k_y}(x,x') = \sum_{\sigma} \langle c^{\dagger}_{x k_y \sigma} c^{\pdag}_{x' k_y \sigma} \rangle \, ,
\end{align}
where $c_{x k_y \sigma} = \tfrac{1}{2} \sum_{y=1}^{4} \mr{e}^{\mr{i} k_y y} c_{x y \sigma}$
, with $k_y \in \{0,\pm\tfrac{\pi}{2},\pi\}$,
is the $y$-Fourier transform of the fermionic annihilation operator $c_{x y \sigma} = c_{\boldell\sigma}$. 

Figure~\ref{fig:KondoCyl_rhok_ev} shows the eigenvalues of $\rho_{k_y}(x,x')$, dubbed $\rho_{k_y}(\alpha)$, for given $k_y$ (extrapolated to zero discarded weight $\xi$).
The structure of the eigenvalue spectra for $J_{\mr{K}} = 1$ and $J_{\mr{K}} = 5$ differ qualitatively:
For $J_{\mr{K}} = 1$, they show a jump for $k_y = \pm\tfrac{\pi}{2}$ and $k_y = \pi$, but not for $k_y = 0$,
while for $J_{\mr{K}} = 5$ it is the other way around.
%

 \begin{figure}[b!]
 \includegraphics[width=\linewidth]{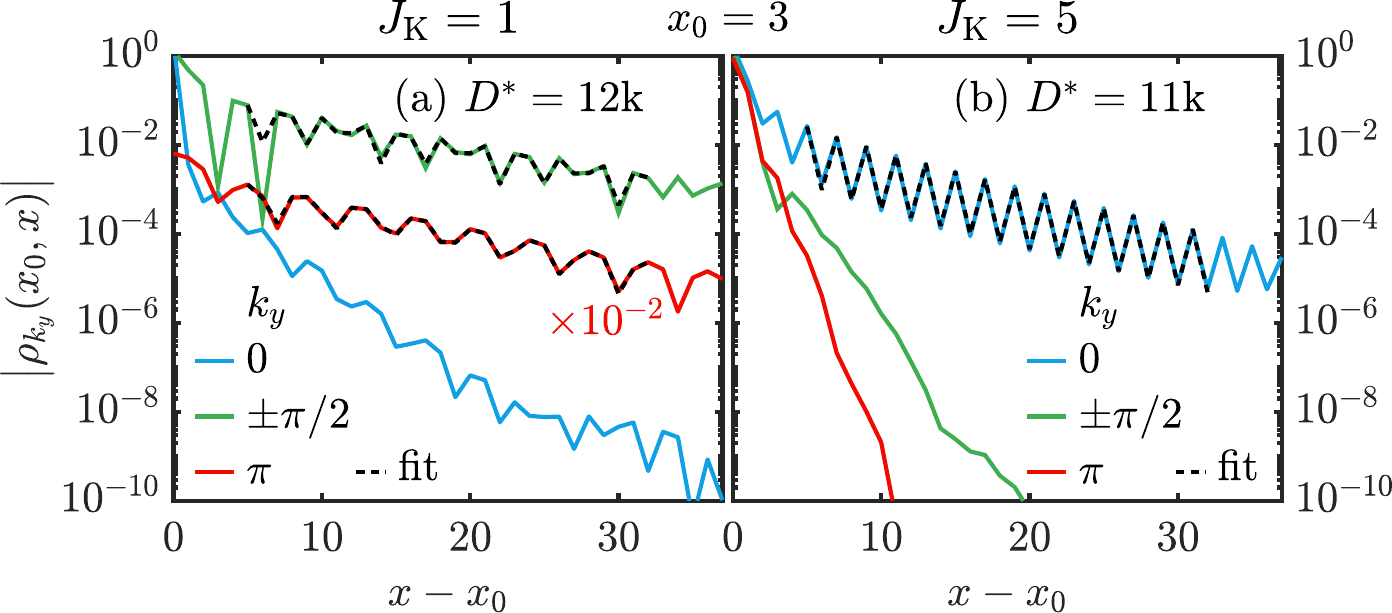} \vspace{-6mm}
 \caption{%
 \label{fig:KondoCyl_ccfit}
Absolute values of the off-diagonal elements of the single-particle density matrix of the Kondo-Heisenberg model on a $40\times4$ cylinder at $25\%$ electron doping, for $J_{\mr{H}}=0.5$ and (a) $J_{\mr{K}} = 1$, (b) $J_{\mr{K}} = 5$. Solid lines are CBE--DMRG data; black dotted lines are fits to Eq.~\eqref{eq:ccfit} to extract $\lambda$ and $k_{\mr{F}x}$.} \vspace{-4mm}
 \end{figure}
\begin{figure}[t!]
 \includegraphics[width=\linewidth]{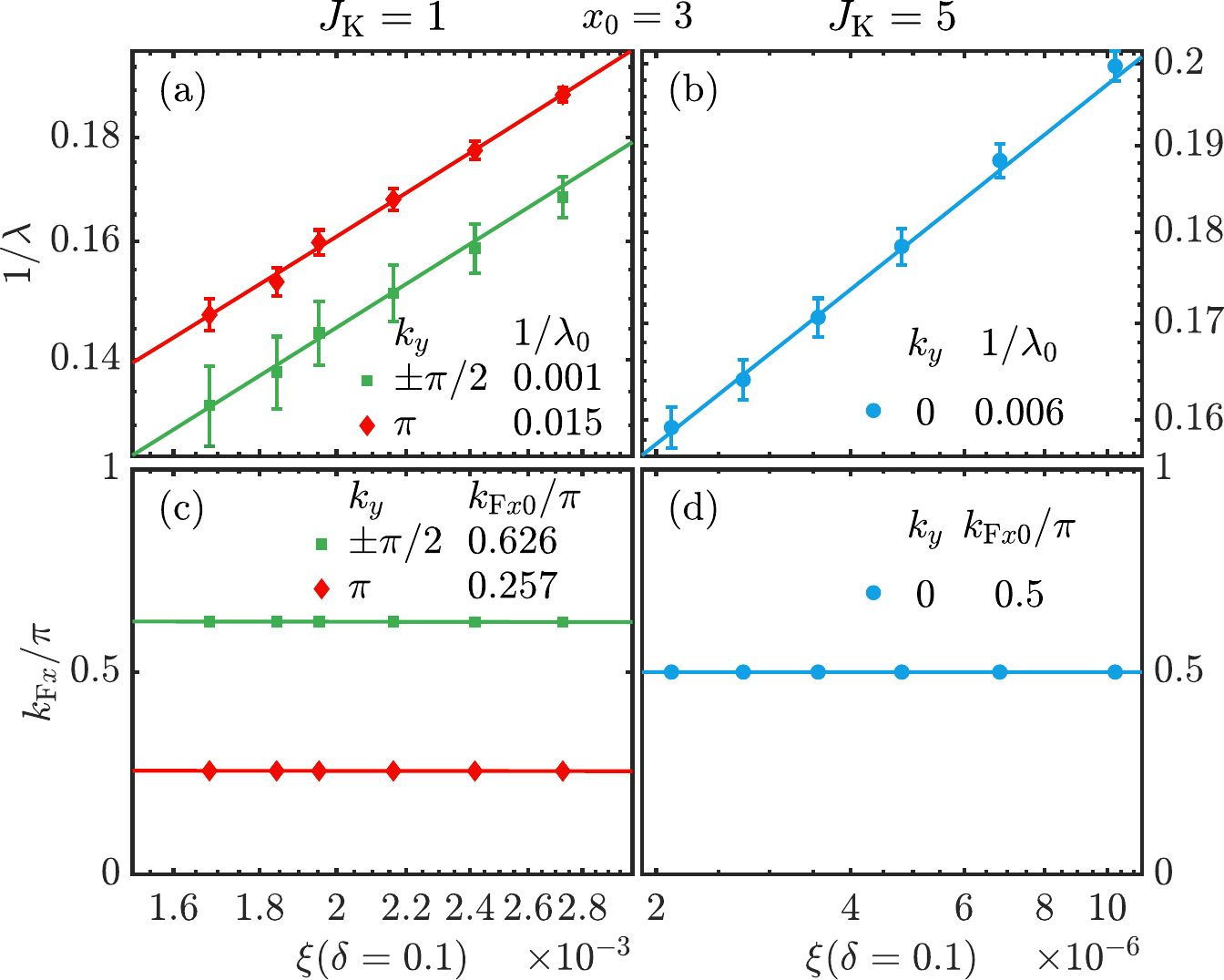} \vspace{-6mm}
 \caption{
 \label{fig:KondoCyl_ccfit_extrap}
Discarded weight extrapolation of (a,b) the correlation length and (c,d) the corresponding Fermi wavevectors, obtained through the fits of $\rho_{k_y}(x_0,x)$ to Eq.~\eqref{eq:ccfit}, as shown in Fig.~\ref{fig:KondoCyl_ccfit}. Error bars indicate $68.2\%$ confidence intervals (i.e.\ one standard deviation) for the fit parameters (below symbol size in (c,d)). 
 } \vspace{-4mm}
 \end{figure}

A jump in $\rho_{k_y}(\alpha)$ suggests that the corresponding $k_y$ value is visited by the Fermi surface, i.e.\ there exists a point on the FS with Fermi wavevector
$\vec{k}_{\mr{F}} = (k_{\mr{F}x}(k_y),k_y)$.
Note, however, that
since we use open boundary conditions,
the eigenbasis of $\rho_{k_y}$ is not the Fourier basis. We can therefore not rely on the eigenbasis of $\rho_{k_y}$ to determine the corresponding $x$-direction Fermi wavevector $k_{\mr{F}x}(k_y)$. Instead, we use the off-diagonal elements $\rho_{x_0x}(k_y)$ in the real space basis, for fixed $x_0 = 3$, and study the behaviour of $\rho_{k_y}(x_0,x)$ as a function of $|x-x_0|$.
The expected behaviour in the case of a Fermi point can be parametrized by the Ansatz
\begin{align}
\label{eq:ccfit}
\rho_{k_y}(x_0,x) \sim \cos\left(k_{\mr{F}x}(k_y)|x-x_0| + \phi\right) \frac{\mr{e}^{-|x-x_0|/\lambda}}{|x-x_0|^{\alpha}} \, .
\end{align}
Here, the exponent in the denominator is given by $\alpha=1$ in case of a Fermi liquid (obtained by Fourier transforming a step function), or takes some non-universal, interaction-dependent value in the case of a Luttinger liquid \cite{Senechal1999}. Because CBE--DMRG approximates the true ground state by a MPS, the correlation length $\lambda$ is finite.  
When $\Dast \to \infty$, or equivalently when $\xi \to 0$, we expect $\lambda\to\infty$.
In Fig.~\ref{fig:KondoCyl_ccfit}, we show that a fit of $\rho_{k_y}(x_0,x)$ to Eq.~\eqref{eq:ccfit} indeed works well for those $\rho_{k_y}$ with gapped spectrum
(green, red curves in Fig.~\ref{fig:KondoCyl_rhok_ev}(a), blue curve in Fig.~\ref{fig:KondoCyl_rhok_ev}(b)).
Note that such fits are not possible for 
the remaining cases.
Figures~\ref{fig:KondoCyl_ccfit_extrap}(a,b) show the behaviour of the inverse correlation length $1/\lambda$ versus discarded weight $\xi$. In the cases where we have identified a possible Fermi wavevector $k_{\mr{F}x}(k_y)$, $1/\lambda$ indeed extrapolates to zero (i.e.\ $\lambda \to \infty$) within our numerical accuracy, consistent with expectations for either a Fermi or Luttinger liquid. 
In Fig.~\ref{fig:KondoCyl_ccfit_extrap}(c,d), we show the corresponding Fermi wavevectors $k_{Fx}(k_y)$ plotted against discarded weight $\xi$. It turns out that $k_{Fx}(k_y)$ is almost independent of $\xi$, which means the determination of $k_{\mr{F}x}(k_y)$ is highly accurate.
%

Our way of extracting Fermi wavevectors from DMRG ground states using the single-particle density matrix is reliable and numerically robust. 
In Fig.~5 
of the main text, we only presented Fermi wavevectors for values of $J_{\mr{K}}$ where we 
were able to
converge the DMRG calculation
with reasonable numerical effort ($D^\ast \leq 12\mr{k}$ on the $40\times4$ cylinder). Closer to the putative KB-QCP, more numerical resources are
needed. These more challenging calculations are beyond the scope of the current work (which mainly focuses on the development of the CBE method)
and are left for the future.
\subsection{Kondo-Heisenberg cylinders: Fermi volume and Luttinger's sum rule}
\label{sec:LuttingersTheorem}%

The FS is especially interesting in the context of Luttinger's sum rule
\cite{Luttinger1960,Oshikawa2000},
\begin{align} 
n_{\mr{eff}} \! =\! 2v_{\mr{FS}} 
\end{align}
(prefactor 2 for spin).
It links the volume enclosed by the FS, $v_{\mr{FS}}$ (measured in terms of Brillouin zone volumes), to the effective number of mobile charge carriers $n_{\mr{eff}}$
(defined modulo 2, i.e.\ excluding filled bands). 
An unambiguous definition of the volume of the FS must include 
a criterion distinguishing its inside and outside. 
The inside of the Fermi
volume is usually defined as those momentum space states which are ``filled",  
having
$n_{\vec{k}} = \sum_\sigma \langle c^{\dagger}_{\vec{k} \sigma} c^{\pdag}_{\vec{k} \sigma} \rangle \simeq 2$.
We point out that the criterion based on $n_{\vec{k}}$ is only stringent in the non-interacting limit where $n_{\vec{k}} \in \{0,2\}$ can only take two values, which is not the case for interacting systems. A stringent criterion for interacting systems can be formulated in terms of single-electron Green's functions 
(see, e.g., Ref.~\onlinecite{Nishikawa2018}, Eq.~(7)), but the computation 
of such dynamical quantities is beyond the scope of this work.
Here, we take the heuristic approach based on $n_{\vec{k}}$.
To make progress on a formula for $v_{\mr{FS}}$ in 2D, we assume that single-electron states in the vicinity of $\vec{k} = (0,0)$ are usually lower in energy than those in the vicinity of $\vec{k} = (\pi,\pi)$. Thus, we consider the states between $\in [-k_{\mr{F}x}(k_y),k_{\mr{F}x}(k_y)]$ filled. For an infinite 2D system, we can now compute
\begin{align}
\label{eq:VFS_infinite}
v_{\mr{FS}} 
= \int_{-\pi}^{\pi} \frac{\mr{d}k_{y}}{2\pi} \int_{-k_{\mr{F}x}(k_y)}^{k_{\mr{F}x}(k_y)} \frac{\mr{d}k_x}{2\pi} 
= \int_{-\pi}^{\pi} \frac{\mr{d}k_{y}}{2\pi} \frac{|k_{\mr{F}x}(k_y)|}{\pi} \, .
\end{align}
Our KH cylinders at hand are however not infinite 2D systems due to the finite circumference of $\eLL_{\!y} = 4$ (the finite length $\eLL_{\!x}$ can in practice chosen large enough to not play a conceptionally problematic role). In this case, we replace
the integral in Eq.~\eqref{eq:VFS_infinite} by a sum to obtain
\begin{align}
\label{eq:VFS_finite}
v_{\mr{FS}} &= \tfrac{1}{\scripteLL_{\!y}} \sum_{k_y} |k_{\mr{F}x}(k_y)| / \pi \, .
\end{align}
Note that we are now faced with the ambiguity of how to define $k_{\mr{F}x}(k_y)$ 
for those $k_y$ values for which no Fermi points exist. The corresponding $k_{\mr{F}x}(k_y)$ could be either $\pi$ or $0$, depending on whether $n_{\vec{k}}$ is filled or  empty for all $k_x$, respectively. For $\eLL_{\!y} \to \infty$, this can be decided based on continuity of $k_{\mr{F}x}(k_y)$. By contrast, for finite $\eLL_{\!y}$, where $k_y$ takes only discrete values, the definition of $k_{\mr{F}x}(k_y)$
has to be based on heuristic arguments. To this end, we
use the eigenvalues of the single-particle density matrix $\rho_{k_y}(\alpha)$ (see Fig.~\ref{fig:KondoCyl_rhok_ev}) as a proxy 
for $n_{\vec{k}}$ 
(in the limit $\eLL_{\!x} \to \infty$, these quantities coincide). If 
the eigenvalues  $\rho_{k_y}(\alpha)$ are close to (or not close to) 2 for \emph{all} $\alpha$, we take that as an indication that all states are filled (or empty), 
and accordingly define $k_{\mr{F}x}(k_y) = \pi$ (or $=0$).

For $J_{\mr{K}} = 1$, only $k_{\mr{F}x}(k_y=0)$ is undecided. Since 
 $\rho_{0}(\alpha)\simeq 2$ (see Fig.~\ref{fig:KondoCyl_rhok_ev}(a), blue dots), we define 
 $k_{\mr{F}x}(0)=\pi$. Together with the Fermi points found at $k_y = \pm\tfrac{\pi}{2},\pi$, we thus find 
$(|k_{\mr{F}x}|,|k_y|) = (\pi, 0)$, $(0.625\pi, \frac{\pi}{2})$ and  $(0.256\pi, \pi)$,
matching the free-electron values at $J_{\mr{K}} = 0$. 
Inserting these into Eq.~\eqref{eq:VFS_finite}, we find $v_{\mr{FS}} = 0.627$ and $n_{\mr{eff}} = 1.25$, consistent with $25\%$ electron doping.
By contrast, for $J_{\mr{K}} \geq 5$,
we find Fermi points only at $(\tfrac{\pi}{2},0)$.
For $k_y = \pm\tfrac{\pi}{2},\pi$, we have to consult $\rho_{k_y}(\alpha)$ shown in Fig.~\ref{fig:KondoCyl_rhok_ev}(b) (green squares and red diamonds).
Since these are well below 2, we define $k_{\mr{F}x} = 0$ for these, so that
$(|k_{\mr{F}x}|,|k_y|) = (0,\frac{\pi}{2})$ and  $(0,\pi)$.
Insertion into Eq.~\eqref{eq:VFS_finite} yields $v_{\mr{FS}} = 0.125$ and $n_{\mr{eff}} = 0.25 = 2.25\,\mr{mod}\,2$ ($n_{\mr{eff}}$ is only defined modulo 2, i.e.\ up to filled bands). 
This is consistent with spins becoming mobile charge carriers by ``binding'' to the electrons~\cite{Si2014} by forming collective Kondo singlets. These collective Kondo singlets break up when approaching the KB--QCP from $J_{\mr{K}} > J_{\mr{K,c}}$ (hence the name 
``Kondo breakdown'') and cease to exist for $J_{\mr{K}} < J_{\mr{K,c}}$. The existence of collective Kondo singlets manifests in a pole in the single-electron 
self-energy. Due to this pole, the Fermi wavevector is shifted, leading to a FS consistent with spins counting as mobile charge
carriers~\cite{Si2014,Oshikawa2000}.
\end{document}